\documentclass[usenatbib]{mn2e}
\usepackage{graphicx}
\usepackage{epstopdf}
\usepackage{natbib}
\usepackage{verbatim}
\usepackage{amssymb,amsmath}
\usepackage{ulem}
\usepackage{color}
\usepackage{comment}
\usepackage[colorlinks=true, citecolor=blue, linkcolor=black]{hyperref}

\pdfoutput=1

\def \ltsim {\lesssim}

\renewcommand{\thefootnote}{\fnsymbol{footnote}}

\newcommand\apj{ApJ}
\newcommand\apjl{ApJ}
\newcommand\apjs{ApJS}
\newcommand\aap{A\&A}
\newcommand\mnras{MNRAS}
\newcommand\pasp{PASP}
\newcommand\nat{Nature}

\title[Illustris Images and Spectra]{Synthetic Galaxy Images and Spectra from the Illustris Simulation}

\author[P. Torrey et al.]
       {\parbox{18cm}{
       Paul~Torrey$^{1,2,3}$\footnotemark[1], 
       Gregory F. Snyder$^{4}$, 
       Mark Vogelsberger$^{2}$, 
       Christopher C. Hayward$^{1,3,5}$\footnotemark[2],
       Shy Genel$^{1}$, 
       Debora Sijacki$^{6}$, 
       Volker Springel$^{5,7}$, 
       Lars Hernquist$^{1}$,\\
       Dylan Nelson$^{1}$,
       Mariska Kriek$^{8}$,
       Annalisa Pillepich$^{1}$,
       Laura V. Sales$^{1}$,
       and Cameron K. McBride$^{1}$
       }\vspace{0.3cm}\\ 
         $^1$ Harvard-Smithsonian Center for Astrophysics, 60 Garden Street,
         Cambridge, MA, 02138, USA\\ 
         $^2 $ MIT Kavli Institute for Astrophysics \& Space Research, Cambridge, MA, 02139, USA\\
         $^3 $ TAPIR 350-17, California Institute of Technology, 1200 E. California Boulevard, Pasadena, CA 91125, USA\\
         $^4 $ Space Telescope Science Institute,	3700 San Martin Drive, Baltimore, MD 21218, USA\\
         $^5$ Heidelberger Institut f\"{u}r Theoretische Studien, Schloss-Wolfsbrunnenweg 35, 69118 Heidelberg, Germany\\
         $^6$ Institute of Astronomy and Kavli Institute for Cosmology, University of Cambridge, Madingley Road, Cambridge CB3 0HA\\
         $^7$ Zentrum f\"{u}r Astronomie der Universit\"{a}t Heidelberg, ARI,  M\"onchhofstr. 12-14, 69120 Heidelberg, Germany\\
         $^8$ Astronomy Department, University of California at Berkeley, Berkeley, CA 94720, USA\\
         }

\begin{document}

\maketitle

\begin{abstract}

We present our methods for generating a catalog of 7,000 synthetic images and 40,000 integrated spectra of redshift $z=0$ galaxies from the Illustris Simulation.
The mock data products are produced by using stellar population synthesis models to assign spectral energy distributions (SED) to each star particle in the galaxies.
The resulting synthetic images and integrated SEDs therefore properly reflect the spatial distribution, stellar metallicity distribution, and star formation history of the galaxies. 
From the synthetic data products it is possible to produce monochromatic or color-composite images, perform SED fitting, classify morphology, determine galaxy structural properties, and evaluate the impacts of galaxy viewing angle.  
The main contribution of this paper is to describe the production, format, and composition of the image catalog that makes up the Illustris Simulation Obsevatory.
As a demonstration of this resource, we derive galactic stellar mass estimates by applying the SED fitting code {\small FAST} to the synthetic galaxy products, and compare the derived stellar masses against the true stellar masses from the simulation.
We find from this idealized experiment that systematic biases exist in the photometrically derived stellar mass values that can be reduced by using a fixed metallicity in conjunction with a minimum galaxy age restriction.
\end{abstract}


\renewcommand{\thefootnote}{\fnsymbol{footnote}}
\footnotetext[1]{E-mail: ptorrey@cfa.harvard.edu}
\footnotetext[2]{Moore Prize Postdoctoral Scholar in Theoretical Astrophysics}
\footnotetext[3]{The synthetic images and spectra presented in this paper are available at http://www.illustris-project.org/galaxy\_obs/}

\section{Introduction}

The formation of dark matter haloes has been studied extensively using 
numerical dark matter only simulations~\citep[e.g.,][]{Millennium,Millennium2,Fosalba2008,Teyssier2009,Bolshoi}.
Extending the insight from dark matter only simulations 
to include a theory of galaxy formation requires a method to link the formation 
of dark matter haloes to observable galaxy properties. 
The most direct method available is hydrodynamical simulations, which model the co-evolution of dark matter and 
baryons~\citep[e.g.][]{Katz1992,
KatzCooling, Weinberg1997, Murali2002, SH03b,Keres05, Ocvirk2008,
Crain2009, Croft2009, SchayeCSFR, Oppenheimer10, Vogelsberger2012}.
By directly including hydrodynamics in structure formation simulations one can probe the
thermal state and column density distribution of the intergalactic medium (IGM) via the Lyman-$\alpha$ forest~\citep{Cen94, Zhang95, Hernquist96, Theuns98}, 
the phase structure and heavy element composition of the circumgalactic medium (CGM) and IGM~\citep{Aguirre01, Cen06, vandeVoort2012a, Shen13},
and the nature and rates of accretion of dark matter and baryons into galaxies~\citep{Keres05, vandeVoort2011a, Nelson2013}.  
Since these simulations can follow the dynamics of both the dark matter and 
baryons down to small spatial scales,
predictions can be made about their 
internal structure including the distribution of gas~\citep{Keres2012, Torrey2012} and the 
formation of stellar disks and bulges~\citep{Abadi2003, Governato2004, Agertz2011, Sales2012, Marinacci2014}.  
As a result of this detailed information, simulations are a valuable tool for interpreting
observational data and placing observed galaxies into a more complete evolution based cosmological context.

Many important galaxy properties -- such as stellar mass -- have substantial uncertainties 
associated with their measurement~\citep[e.g.,][and references therein]{Conroy2009, Conroy2010b}.
These uncertainties can be physical in origin (e.g., 
the slope of the IMF, treatment of the thermally pulsating asymptotic giant branch phase, etc.) 
or due to assumptions made during the spectral energy distribution (SED) fitting 
procedures~\citep{Papovich2001, Wuyts2007, Gallazzi2009, Michalowski2012, Banerji2013}.  
Accurately estimating physical galaxy properties based on broadband photometric data points (or 
even full SEDs) is difficult 
because observed galaxy SEDs contain contributions from stars with a complex distribution 
of stellar ages and metallicities.  
Simplifying assumptions regarding the functional 
form of the star formation history, or a uniform stellar metallicity, are often assumed.
These assumptions can lead to systematic errors on the derived galaxy stellar mass, independent of 
physical uncertainties~\citep[e.g.,][]{Mitchell2013}.
Similarly, derived galaxy structural properties -- such as bulge-to-disk 
decompositions or galaxy sizes -- are often determined inconsistently in 
simulations and observations~\citep{Scannapieco2010}.

A crucial difficulty when relating hydrodynamical simulations and 
observations is the translation of results between the two, a necessary step 
for accurate comparisons~\citep[e.g.,][]{Conroy2010a, Hayward2012, Hayward2013a, Hayward2013b}.  
There are two fundamental ways to make this translation, by either converting observed quantities 
into the space of physical parameters, or by generating `mock' (or synthetic) observations of 
the simulation output. Only the second approach provides a unique mapping, avoiding the uncertainties 
involved in a potentially non-unique inversion. It also allows observational effects to be accounted 
for (e.g. noise, survey geometry, ...), and enables observational tools and data analysis techniques 
to be run on the simulated observations as is.
For comparison to observations of stellar light, this can be achieved in simulations by adopting stellar population synthesis (SPS) models 
to assign light to all star particles within a galaxy based on their age and 
metallicity distributions~\citep{Tinsley1972, Bruzual1983, Buzzoni1989, Bruzual1993, Worthey1994, Maraston1998, SB99, BC03, Thomas2003, Maraston2005}.  

These mock observations can be used to address potential issues in the way 
that SED fitting or photometric analysis is performed and even to directly fit galaxy SEDs rather than relying on SED modelling~\citep{Lanz2014}.
A large body of literature already exists within the idealized merger simulation community in this direction: 
\citet{Lotz2008} used mock broadband images of idealized merger simulations~\citep{Cox2006} to 
determine the conversion from observed galaxy pair counts into merger rates.  
\citet{Wild2009} and~\citet{Snyder2011} used 
mock galaxy SEDs of idealized mergers to determine 
the lifetime and redshift dependent abundance of K+A galaxies, while 
\citet{Snyder2013} used synthetic galaxy spectra to propose a new Mid-IR diagnostic of AGN.
\citet{Wuyts2010} investigated how well the true mass profiles of simulated massive compact 
galaxies could be recovered from their light profiles in different bands. 
\citet{Hayward2011} elucidated the physical meaning of the submillimetre galaxy selection, 
and \citet{Hayward2014a} tested how well the integrated IR luminosity traces the star formation rate.

Similar work exists within the semi analytic community
including the production of mock galaxy luminosity functions~\citep[e.g.,][]{Henriques2011, Somerville2012}
and mock light-cone data~\citep{Kitzbichler2007, Henriques2012b, 
Overzier2013, Bernyk2014} based on dark matter only simulations~\citep{Millennium}.
The Millennium Run Observatory~\citep[MRObs,][]{Overzier2013} built a theoretical virtual observatory 
capable of producing synthetic data products that could then be reduced and analyzed with observational software.
The MRObs synthetic observations covered forty filters where the light was assigned to each galaxy by combining information about 
the star formation history with stellar population synthesis models.  
The resulting data products were formatted in a way that they could be analyzed using standard observational software packages~\citep{Overzier2013}
and distributed using a web based relational SQL database~\citep{Lemson2006}.

In this paper, we describe the production of a mock galaxy image catalog 
based on the recent Illustris simulation~\citep{Vogelsberger2014a, Vogelsberger2014b, Genel2014}.  The Illustris project has two 
basic principles: (i) build a comprehensive physical framework that allows
galaxy populations to regulate their stellar mass growth appropriately and (ii) apply that 
physical framework in hydrodynamical simulations of large cosmological volumes. 
A detailed description of our adopted feedback module, which includes a model for star formation driven winds as well as AGN driven outflows, was 
presented in~\citet{Vogelsberger2013} and demonstrated to appropriately regulate galaxy 
populations across cosmic time in~\citet{Torrey2014}.
Adopting this feedback model ensures that the galaxies within these simulations match 
a wide range of galaxy observable properties such as the redshift $z=0-3$ galaxy stellar mass function.
The main Illustris simulation was run in a periodic box of size $L\sim 100$ Mpc with sufficiently high mass and spatial resolution
(i.e. $M_{{\rm b}}\sim 10^6 M_\odot$ and $\epsilon \ltsim 1$ kpc) to resolve a wide range of galaxy structures~\citep{Vogelsberger2014a, Vogelsberger2014b, Genel2014}.

Using results from the Illustris simulation, we employ stellar population synthesis (SPS)
templates to produce synthetic broadband galaxy images and integrated SEDs.  
These mock images encode information about the spatial distribution and formation history of galactic stellar components.
The synthetic galaxy SEDs and images produced from the Illustris simulation are unique compared 
to previous synthetic data catalogs.
Compared against idealized isolated and merging galaxy simulations~\citep[e.g.,][]{Lotz2008, Snyder2011}, 
the Illustris simulation galaxies are embedded in a proper cosmological context.
The galaxies in the Illustris simulation accrete gas from the surrounding intergalactic medium, undergo major and minor mergers,
and experience feedback as described in~\citet{Vogelsberger2013}.  
All of these processes are reflected in the stellar mass distribution and star formation history, which define the resulting mock images.
Previous generations of synthetic images and spectra have been generated from ``zoom-in" style simulations of individual (or a small number of) objects~\citep[e.g.,][]{Guedes2011, Stinson2013, Snyder2014}.
Like the Illustris simulation, these simulations contain a proper cosmological context for galaxy growth, and they have there been used to study the growth of galactic disks and bulges.
Our large-volume simulation approach is distinct from the ``zoom-in'' style simulations in that it allows us 
to analyze the properties of a large population of galaxies with varied formation histories and environments under uniform assumptions
about the input physics~\citep[e.g.,][]{Pedrosa2014}. 
In contrast to semi-analytic based mock data products~\citep[e.g.,][]{Overzier2013}, our hydrodynamic approach allows us to self-consistently treat the
internal galactic dynamics (e.g., bar formation, merger events, etc.) that drive galaxy morphology evolution. 
The resulting synthetic image catalog has many applications including building mass dependent average galaxy SED 
templates, isolating the origin of shelled galaxies, and understanding the viewing angle dependence of barred galaxies.

In this paper, our main goal is to detail the methods for producing these images, explain the data format, 
and explain how the image catalog will be accessible.
We additionally demonstrate some initial science applications of the resulting synthetic images and spectra.
This paper is outlined as follows.
In section \ref{sec:Methods} we describe the simulation that has been used to produce our mock image catalog 
as well as our methods for producing the mock galaxy image catalog based on its output.
In section \ref{sec:Data} we describe the format for the mock 
galaxy images, present properties of the galaxy SEDs, and demonstrate multi-band image generation.  
In section \ref{sec:Example} we provide a detailed application of this data, by determining stellar 
mass measurements using a popular SED fitting code to compare the 
SED derived stellar masses against the simulation based stellar mass determination.
In section \ref{sec:Discussion} and section \ref{sec:Conclusions} we summarize and conclude.

\begin{table}
\begin{center}
\caption{List of broadband filters used in the Illustris Virtual Observatory, selected as a reasonably complete sampling of the historical, current and future bands used for the observation of galactic stellar light.}
\label{table:BroadbandFilters}
\begin{tabular}{ l c c  }
\hline
Filter                          	& $\lambda _{\rm{eff}}$ (\AA)   		& Field Number	     			\\
				& 							& 		 				\\
\hline
\hline 
                                	&                 						&				       		\\
GALAX FUV		& $ 1513.3 $      					&	1					\\
GALAX NUV		& $ 2300.5 $      					&	2					\\
                                	&               							&						\\
SDSS-u			& $ 3573.1 $      					&	3					\\
SDSS-g			& $ 4724.1 $      					&	4					\\
SDSS-r			& $ 6201.4 $      					&	5					\\
SDSS-i			& $ 7524.7 $      					&	6					\\
SDSS-z			& $ 8917.3 $      					&	7					\\
                                	&               							&						\\
IRAC1			& $ 35667.8 $      					&	8					\\
IRAC2			& $ 45023.2 $      					&	9					\\
IRAC3			& $ 56852.6 $      					&	10					\\
IRAC4			& $ 79040.1$      					&	11					\\
                                	&               							&						\\
Johnson-U		& $ 3650.9 $      					&	12					\\
Johnson-B		& $ 4449.0 $      					&	13					\\
Cousins-R		& $ 6599.4 $      					&	14					\\
Cousins-I			& $ 8061.5 $      					&	15					\\
Johnson-V		& $ 5506.9 $      					&	16					\\
Johnson-J		& $ 12269.0 $      					&	17					\\
Johnson-H		& $ 16466.8 $      					&	18					\\
Johnson-K		& $ 22008.3 $      					&	20					\\
                                	&               							&						\\
2MASS-H			& $ 16567.4 $      					&	19					\\
2MASS-Ks		& $ 21620.0 $      					&	21					\\
                                	&               							&						\\
ACS-F435		& $ 4329.2 $      					&	22					\\
ACS-F606		& $ 5929.7 $      					&	23					\\
ACS-F775		& $ 7712.9 $      					&	24					\\
ACS-F850		& $ 9071.6 $      					&	25					\\
                                	&               							&						\\
f105w			& $ 10539.9 $      					&	26					\\
f125w			& $ 12449.4 $      					&	27					\\
f160w			& $ 15314.6 $      					&	28					\\
                                	&               							&						\\
NIRCAM-F070W${}^{\rm a}$
& $ 6955.5 $      					&	29					\\
NIRCAM-F090W	& $ 9031.3 $      					&	30					\\
NIRCAM-F115W	& $ 11515.3 $      					&	31					\\
NIRCAM-F150W	& $ 15064.8 $      					&	32					\\
NIRCAM-F200W	& $ 19831.5 $      					&	33					\\
NIRCAM-F277W	& $ 27728.0 $      					&	34					\\
NIRCAM-F356W	& $ 35751.1 $      					&	35					\\
NIRCAM-F444W	& $ 44292.7 $      					&	36					\\
\hline
\hline
\end{tabular}
\end{center}
{\footnotesize 
${}^{\rm a}$ Preliminary JWST/NIRCAM filter curves were obtained from \textrm{www.stsci.edu/jwst/instruments/nircam/instrumentdesign/filters}}
\end{table}

\section{Methods}
\label{sec:Methods}
In this section we describe our procedure for producing mock galaxy SEDs and photometric images.  
We provide details on the simulation that forms the basis of our mock image catalog, explain how light is assigned to each galaxy to produce mock images, and outline the resulting data product format.

\subsection{The Illustris Simulation}
All mock images are produced based on galaxies formed in the Illustris simulation.  A detailed description of the Illustris simulation can be found in~\citet{Vogelsberger2014a}, \citet{Vogelsberger2014b}, and \citet{Genel2014}.  We briefly summarize the simulation properties below.

Illustris is a large-volume cosmological hydrodynamical simulation that was run in a periodic box of side length $L=106.5$ Mpc with $N_{{\rm DM}} = 1820^3$ dark matter particles ($m_{{\rm DM}} = 6.3 \times 10^6 M_\odot$) and $N_{{\rm baryon}} \approx 1820^3$ baryon resolution elements ($m_{\rm baryon} \approx 1.3 \times 10^6 M_\odot$).  
Cosmological parameters consistent with the latest Wilkinson Microwave Anisotropy Probe (WMAP)-9 measurements were adopted ($\Omega_M=0.2726$, $\Omega_\Lambda=0.7274$, $\Omega_b=0.0456$, $\sigma_8 = 0.809$, $n_s=0.963$, and $H_0= 70.4$ km s${}^{-1}$ Mpc${}^{-1}$ with $h=0.704$).
The Illustris simulation was run using the moving mesh code {\small AREPO} which, in addition to gravity and hydrodynamics~\citep{AREPO}, includes comprehensive physics and feedback modules allowing for radiative gas cooling~\citep{KatzCooling, WiersmaCooling}, 
heating and ionisation by a UV background~\citep{Faucher2008,Faucher2009,McQuinn2009},
star formation with associated feedback~\citep{SH03}, mass and metal return to the interstellar medium from aging stellar populations~\citep{WiersmaGasReturn}, and active galactic nuclei (AGN) feedback~\citep{Springel2005, Sijacki2006, Sijacki2007, Hayward2014b} as described in detail in~\citet{Vogelsberger2013}.  In practice, the feedback included in the Illustris simulation was tuned to match the redshift $z=0$ galaxy stellar mass function and evolving cosmic star formation rate density~\citep{Vogelsberger2013}, and shown to produce galaxy populations that evolve consistently with observations~\citep{Torrey2014}.

\subsection{Galaxy Image Production from Illustris Data}
The main function of the image pipeline is to produce synthetic images and spectra for galaxies in the Illustris simulation.
Haloes are defined in the Illustris simulation via a Friends-of-Friends (FoF) algorithm while galaxies are found via the {\small SUBFIND} halo finder~\citep{SUBFIND}.
For each halo and galaxy we identify all associated cells and particles (gas, stars, dark matter, and black holes) and produce a `mini-snapshot' file that contains all relevant particle information.
These mini-snapshots files are in principle a redundant replication of the simulation data, however they are produced to make the data for the galaxies more easily accessible (each full Illustris snapshot is \~1.4TB, and therefore it can be rather inefficient to extract individual galaxy data).
We produce two types of mini-snapshot files including either the full FoF groups or only individual subhalos. 
Images and spectra made from the FoF mini-snapshots will contain contributions from neighboring and satellite galaxies.
This can be an advantage when, e.g., visually identifying merger events, but is not favorable when, e.g., trying to derive a galaxy's mass based on its SED (if it contains "extra light" contributions from neighboring galaxies).
These mini-snapshots can be opened and processed very efficiently and are in the natural format for simulation post-processing analysis tools that have been designed to run on individual galaxies.

We employ the radiative transfer code {\small SUNRISE}\footnote{Sunrise is freely available and documented at \textrm{http://code.google.com/p/sunrise/}.}~\citep{SUNRISE, SUNRISE2} to then perform the following steps:
\begin{enumerate}
\item assign full spectra to each star particle based on its age and metallicity (see Section \ref{SEC:LIGHT});
\item account for obscuration and nebular emission from the birth cloud (see Sections \ref{SEC:DUSTOBS} and \ref{SEC:NEBULAREMISSION});
\item generate images of arbitrary field of view and pixel size for several camera positions (see Section \ref{SEC:IMAGES});
\item convolve spectra with pre-tabulated broadband filter transmission functions (see Section \ref{SEC:IMAGES} and Table 1).  
\end{enumerate}
We perform these tasks without applying the dust absorption, scattering, or emission functionalities of the {\small SUNRISE} code.
While these are important considerations for SED modeling, we do not expect the {\small SUNRISE} radiative transfer results to be fully converged for the mass and spatial resolutions employed by the Illustris simulation~\citep{SUNRISE2}.
Instead, we use only a simple empirically motivated dust obscuration correction (described below) and leave a more complete examination of dust and its consequence for mock images to future studies.  
Using {\small SUNRISE} allows us to rely upon an existing, tested, publicly available code, and to prepare our pipeline software for use on future hydro simulations where full dust modeling may be appropriate.
Although AGN are present in the simulation, we do not include their light contributions. 

We post-process all galaxies with more than 500 stellar particles -- roughly corresponding to stellar masses of $M_* > 5 \times 10^8 M_\odot$, 
and run {\small SUNRISE} in two modes depending on the particle count of the system.
For systems with stellar masses $M_* > 10^{10} M_\odot$ ($N_* \gtrsim 10^4$ stellar particles) we calculate both integrated SEDs and spatially resolved photometric maps in 36 broadband filters (see Table~\ref{table:BroadbandFilters} and Section \ref{SEC:IMAGES}).
The simulation volume contains $\sim$7,000 galaxies above this limit at redshift $z=0$.
Systems with stellar masses below this limit are not very well spatially resolved -- making imaging less useful -- so we store only integrated SEDs.
The integrated SEDs do not allow us to study the internal structure of galaxies, but still allow us to identify the impact of metallicity and star formation history on galaxy SEDs which we use in Section \ref{SEC:FAST}.

\begin{figure}
\centerline{\vbox{\hbox{
\includegraphics[width=3.35in]{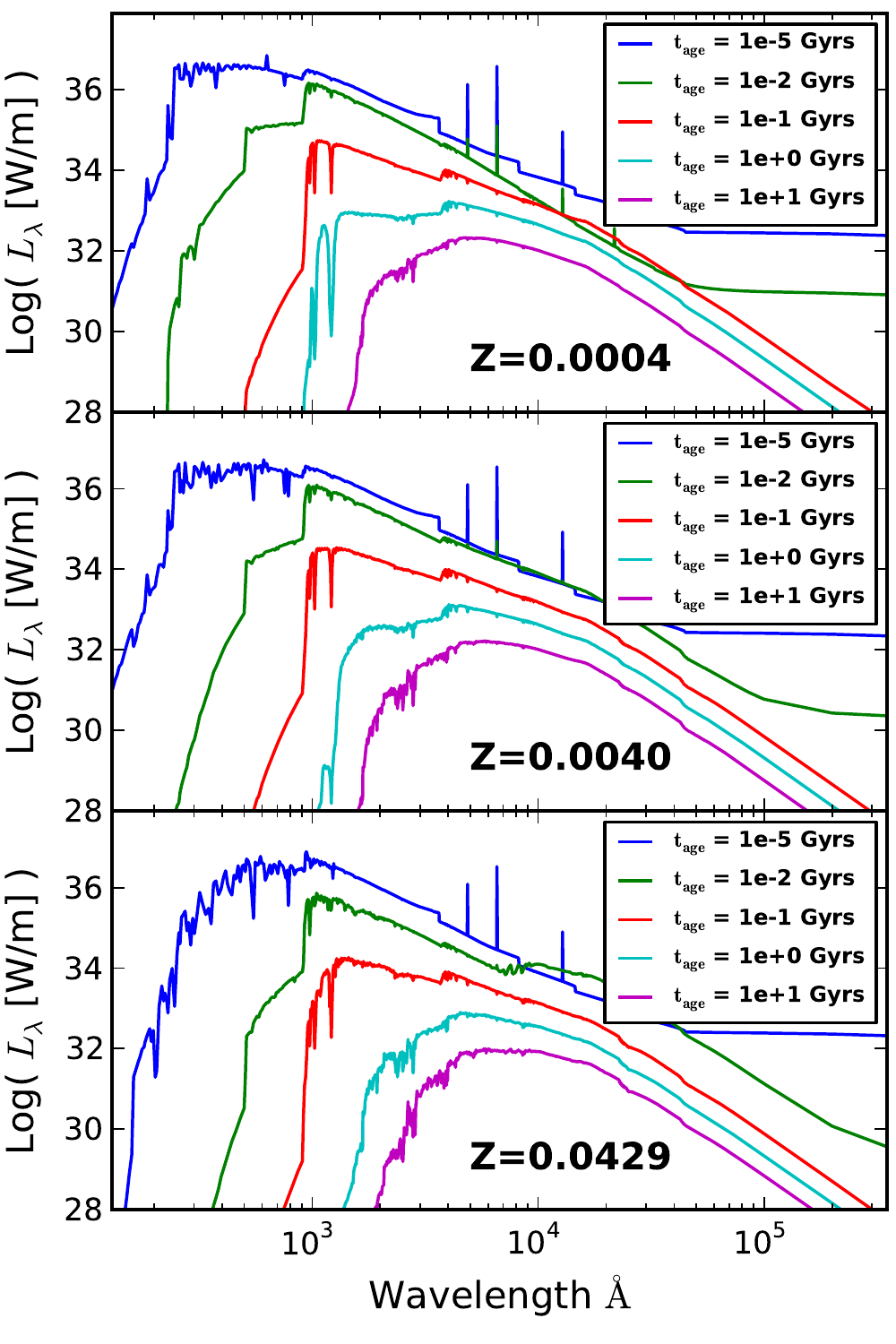}
}}}

\caption{Input SEDs based on the SB99 stellar population synthesis models.  We show SEDs for metallicities of Z=0.0004 (top), 0.004 (middle), and 0.043 (bottom) at several times, as noted in the legend of the figure.  Each star particle in the simulation is assigned a single SED by selecting the nearest SED in age-metallicity space.  All SEDs used in this paper assume Chabrier IMFs, as does the Illustris simulation itself.
}
\label{fig:BC03_SED}
\end{figure}

\begin{figure*}
\centerline{\vbox{\hbox{
\includegraphics[width=7.0in]{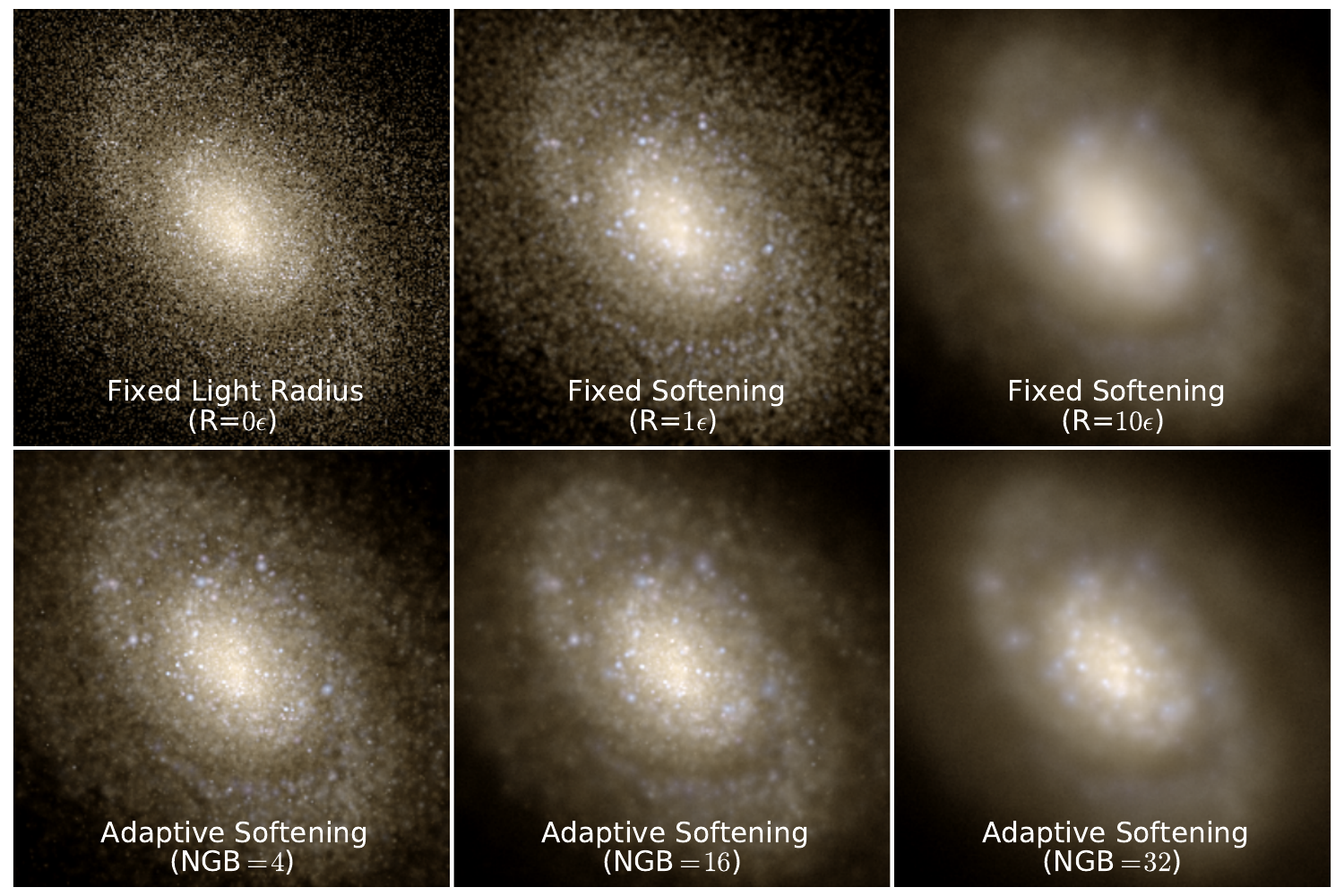}
}}}

\caption{Montage of images based on several options for the smoothing of stellar light for a single galaxy.  The top row shows constant smoothing, with point like source (top left), smoothing set equal to the gravitational softening (top middle), and smoothing set equal to ten times the gravitational softening (top right).  The bottom row shows adaptive smoothing, with the number of nearest neighbors, $N_{{\rm NGB}}$, set to $N_{{\rm NGB}} = 4$, $N_{{\rm NGB}} = 16$, and $N_{{\rm NGB}} = 64$.  We adopt $N_{{\rm NGB}}=16$ for all subsequent figures and analysis.
}
\label{fig:smoothing}
\end{figure*}

\subsubsection{Stellar Light}
\label{SEC:LIGHT}
When a star particle is formed in the simulation it inherits the metallicity of the local ISM gas and records its birth time and initial formation mass.
Star particles then slowly return mass to the ISM to account for mass loss from aging stellar populations.  
We track both the formation mass (a fixed value), as well as the current mass (a time dependent value) for each star particle in the simulation.
For the galaxies, light is then assigned to all constituent star particles based on their mass, age, and metallicity values, using single-age stellar population SED templates.  
We adopt the {\small STARBURST99} (SB99) single-age stellar population models which includes an SED grid resolved with 240 wavelength bins at 308 age and 25 metallicity values~\citep{SB99, SB99_2, SB99_3}, from which a single SED is assigned to each star particle, scaled according to its formation mass.
The fiducial SB99 input SEDs also contain nebular emission and absorption lines according to the model outlined by~\cite{SB99}.
Figure~\ref{fig:BC03_SED} shows the age dependent SEDs for three representative stellar metallicity values in the SB99 templates.  

In our current simulations, the star particles have an initial stellar mass of $M_*\approx 1.3 \times 10^6 M_\odot$ and represent an unresolved full stellar population. 
As a result, it is not immediately obvious how the associated stellar light should be spatially distributed.  
Adopting an appropriate scheme for spreading stellar light is unimportant for integrated quantities (e.g., the integrated galaxy SED) when radiative transfer is not performed but can impact visual interpretation of galactic structure and bias quantitative measures.  
Over smoothing of stellar light will lead to an undesirable modification or deformation of, e.g., galaxy light profiles, whereas 
under smoothing light can have adverse consequences for the resulting galaxy images.
For example, assigning all stellar light to a single pixel (i.e. assuming each simulation star particle is best represented by a point source) will result in a population of easily identified bright pixels in the galactic outskirts.  
This can bias our mock galaxy images toward easy detection of diffuse light components which may in fact have very low surface brightness.  
When using stellar light point sources the peak surface brightness and distribution of surface brightnesses for pixels in an image will be unphysically set by the adopted pixel scale and/or number.  
Applying a finite spatial extent to the light assigned from the star particles ensures that the surface brightness will not be dependent on the image resolution as long as the pixel scale is smaller than the smoothing scale for the star particles.

We consider six choices for spreading the stellar light as described below, and show an example of each in Figure~\ref{fig:smoothing}.
The top left panel of Figure~\ref{fig:smoothing} assigns the light from the star particles as a point source, a method which requires no additional assumptions, but leads to pixelated images where all light from any star particle is always contained within a single pixel.
Image pixelation becomes particularly noticeable in the outskirts of the galaxy, and in general is 
not realistic since each star particle represents a {\it population} of stars, with finite spatial extent.
Artificial pixelation can be reduced by spreading the light from the star particles following some radial distribution, e.g. a Gaussian kernel.
This then requires a smoothing length, for which one natural choice is the gravitational softening length -- the same example image resulting from this choice is shown in the middle panel of Figure~\ref{fig:smoothing}.  
Employing this fixed smoothing length results in structure in the center of the galaxy being under-resolved, while the galaxy outskirts are represented by isolated, circular stellar light distributions.
Increasing the fixed smoothing length further, as shown in the top right panel of Figure~\ref{fig:smoothing}, leads to a smooth distribution of light in the galaxy outskirts, although with ever greater loss of visible galaxy structure, as one might expect.

As an alternative, we find that adaptive light spreading based on the local density of star particles yields a smooth and well-resolved distribution of light from the inner to outer parts of the galaxy.
In practice we set the Gaussian FWHM for each star equal to the distance enclosing its ${\rm N}^{\rm{th}}$ nearest neighbor (in 3D).  
The level of smoothing is then set based on the neighbor number, $N_{{\rm NGB}}$, that we adopt.
The bottom three panels of Figure~\ref{fig:smoothing} show examples of this adaptive light smoothing for three choices of the neighbor number ($N_{{\rm NGB}} = \{4,16,32\}$).  
We have modified {\small SUNRISE} to include the same tree build and neighbor finding steps included in {\small GADGET-2}~\citep{GADGET} to facilitate efficient neighbor finding for all systems.

None of the light distribution options are formally any more or less valid than the others, and 
we cannot rule out that a more physical light assignment procedure may exist. 
In all subsequent analysis in this paper 
we adopt $N_{{\rm NGB}} = 16$ as our fiducial choice for the adaptive light smoothing.

We use $10^8$ photon packets to perform the {\small SUNRISE} Monte Carlo photon propagation scheme.
This results in some residual Monte Carlo noise in low surface brightness features that manifests as fluctuations in the pixel-to-pixel intensity.
These low surface brightness features are in general poorly resolved and the Monte Carlo noise should not be mistaken for detailed information about the stellar mass/light distribution.
We have tested and confirmed that $10^8$ photon packets is sufficient such that all of the results shown in this paper are well-converged and would not change significantly with increased photon packet count.
However, caution should still be exercised when examining the detailed structure of low surface brightness features.

\subsubsection{Dust Obscuration}
\label{SEC:DUSTOBS}
A proper treatment of dust and its modification of simulated galaxy spectra requires knowledge of the distribution of gas and dust on small ($\ll$ 1 kpc) spatial scales.
To most accurately determine the effects of dust attenuation, full radiative transfer should be performed on the simulated galaxies, but the spatial resolution of state-of-the-art cosmological simulations is not sufficient for such calculations~\citep{SUNRISE2}.
In principle, one can estimate the impact of dust obscuration by calculating the column density of dust (via the gas density, temperature, and metallicity spatial distribution), and thus the optical depth, along lines of sight to each star in 
the galaxies~\citep[e.g.,][]{Hopkins2005,Robertson2007,Wuyts2009a,Wuyts2009b}.
However, in the Illustris simulation we are limited to $\sim$ kpc scale resolution, with a pressurized, effective equation-of-state treatment of the complex structure of the ISM~\citep{SH03}.  
As a result, the gas disks in our simulated galaxies are thicker (in terms of disk scale height) and more homogeneous than the ISM in real galaxies.  
A naive calculation of the optical depths to stars within galaxies based on the simulated gas distribution will therefore not necessarily lead to accurate dust obscuration measurements.  
In principle, a more robust dust correction can be constructed via a sub-grid dust-obscuration model that accounts for the unresolved properties of the gas disks.
Such a model is beyond the scope of this paper, but is being considered in Snyder et al., (in prep).

Here, we instead consider the simple empirical dust model of \citet[][hereafter CF00]{CF00}.
The CF00 model accounts for the effects of dust by applying an effective absorption of $\tau = 1.0 \left( \lambda /0.55 \mu m \right)^{-0.7}$ to the standard SB99 SED.
The normalisation of the absorption curve is lowered by a factor of three for stars older than $t_{{\rm age}} > 3 \times 10^7$ yrs, which are assumed to have left their birth clouds.
The only parameter in this model is stellar age; no knowledge of the local ISM conditions (including the local gas density or metallicity) is taken into account.  
Since most galaxies have only a very low fraction of star particles with ages $t_{{\rm age}} < 3 \times 10^7$ yrs, applying this CF00 model typically results in a more-or-less uniform screen being placed in front of the fiducial galaxy image. 
As a result, the CF00 models have a spatial distribution of light which is similar to the fiducial non-dusty images (albeit, with an intensity offset determined by the CF00 attenuation curve).

\subsubsection{Nebular Emission}
\label{SEC:NEBULAREMISSION}
To appropriately include the impact of the unresolved birth cloud on the emergent SED from young star particles, we adopt a model that takes into account nebular emission around young stars.
This is potentially important because line emission from gas in HII regions (e.g., H$\alpha$) can contribute substantially to the flux in certain broadband filters.
Line contamination in broadband filters can modify inferred ages and stellar masses, especially for high redshift galaxies where the average star formation rates and nebular emission line emission are much higher than in the local universe~\citep{Zackrisson2008, Schaerer2009, Ono2010, Stark2013}.  
We calculate SEDs for the galaxy population including a model~\citep[for a complete description see][]{SUNRISE2} that includes the environment around young stellar populations.
To account for this in our mock galaxy SEDs, {\small SUNRISE} reassigns the intrinsic SEDs of young star particles assuming a partially obscured young stellar spectra with added contributions from HII regions~\citep{Dopita2005, Dopita2006a, Dopita2006b} as calculated in ~\citet{Groves2008} via the {\small MAPPINGS-III} 1-D photoionisation and radiative transfer code~\citep{Binette1985, SutherlandDopita, Groves2004a, Groves2004b}.  
We assume a photodissociation region covering fraction of $f_{{\rm PDR}} = 0.2$ and cluster masses of $10^6 M_\odot$.
The {\small MAPPINGS} model replaces SB99 emission for all young ($t_{age} < 10^7$ yrs) star particles.
As in the case of the CF00 model, the spatial distribution of light for our simulated galaxies with and without the {\small MAPPINGS} model is similar and so we store only integrated quantities for this model.
In summary, our treatment models the partial obscuration of young stars by their birth clouds, where some of the obscured light is re-emitted in nebular emission lines.

\begin{figure}
\centerline{\vbox{\hbox{
\includegraphics[width=3.5in]{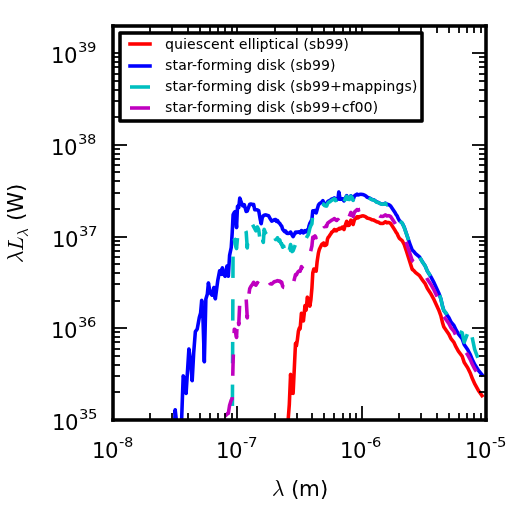}
}}}

\caption{Integrated SEDs are shown for a quiescent elliptical (red line) and star forming disk (blue line) of similar stellar mass.  Due to the younger stellar population, the star forming disk has substantially more UV emission.  We also show the star forming disk's SED when using the {\small MAPPINGS} HII region model (cyan), and the CF00 empirical dust correction (magenta).  Both of these models reduce the UV luminosity substantially. }
\label{fig:gal_sed}
\end{figure}

\subsubsection{Image Production and Broadband Definitions}
\label{SEC:IMAGES}
Images are produced by projecting the stellar light using pinhole cameras with a resolution of $N_{{\rm pixels}} \times N_{{\rm pixels}}$ pixels.  
We use a fiducial value of $N_{{\rm pixels}} = 256$ for the full image catalog.
However, we have used higher pixel resolution ($N_{{\rm pixels}} = 512$) for many of the individual images included in this paper.
Each galaxy is imaged with four cameras, which see the galaxy from four different viewing angles.
The cameras sit at the vertices of a tetrahedron and are pointed toward the center of the tetrahedron.
We move the galaxy's position such that the galaxy's potential minimum sits at the center of the tetrahedron, which 
results in the four cameras being randomly aligned with respect to the rotation axis of each galaxy. 
The first camera (CAMERA0, in our later notation) corresponds to the positive $z$ direction.
Each camera is placed 50 Mpc away from the galaxy's center, regardless of the galaxy's intrinsic redshift, and 
the image field-of-view is set to be 10 times the stellar half mass radius for the galaxy.

The first imaging step is the production of a mock integral field unit (IFU) data cube 
-- with full spectra being associated with each pixel.  
Integrated global SEDs are produced by adding the light contributions from all IFU pixels.  As such, we 
note that any light contributions that fall outside of the initially selected field of view (10 times the 
stellar half mass radius) will not be included.  We note that we have examined global SEDs
computed using larger fields of view, and found them to be nearly indistinguishable from those 
produced using our fiducial field of view choice.
We convolve the IFU mock data cubes with 36 broadband filter transmission 
functions -- as listed in Table~\ref{table:BroadbandFilters} -- to produce an 
equal number of broadband images for the galaxy.
The resulting broadband images can be used to generate RGB images 
based on various filter combinations, and to facilitate quantitative comparisons 
with observations made with particular filters.

The images generated from this process are idealized images, in the sense that they 
contain no sky noise, no camera point-spread-function (PSF) blurring, and lack 
galactic and extragalactic image contamination (from the foreground/background).
We do not include these ``image realism" contributions because (i) they can 
be added to the idealized images in ``post processing" and (ii) the best choice 
for the, e.g., camera PSF blurring will depend on the specific instrument and/or application.
In future work we will include such non-intrinsic contributions before analyzing the photometrically inferred galaxy properties.

The post processing of every galaxy takes roughly $\sim$ 30 minutes running 
on 4 processors -- leading to a total compute time of roughly 80,000 CPU hours 
for the redshift $z=0$ population.

\section{Results: Galaxy Images and SEDs}
\label{sec:Data}
The redshift $z = 0$ galaxy population has been run through the {\small SUNRISE} pipeline producing a catalog of $\sim$ 40,000 galaxy spectra and $\sim$ 7,000 galaxy images with associated photometric data.  
This data is available through a web-based query and download service.\footnote{http://www.illustris-project.org/galaxy\_obs/} 
The image database is searchable based on galaxy properties (e.g., stellar mass, halo mass, etc.) as well as unique identification numbers, which are provided for the example galaxy images used in this section.

Photometric data for a galaxy is contained in a single, multi-extension FITS file containing 
four image extensions corresponding to the four camera views.
Each extension contains an array of dimensionality $N_{{\rm bands}} \times N_{{\rm pixels}} \times N_{{\rm pixels}}  $.  
The ordering of the broadband filters can be found in Table~\ref{table:BroadbandFilters} as well as in the first field in the {\small FILTERS} extension.  
Image units, pixel count, and other relevant details are stored in the camera extension header as outlined in Table~\ref{table:FITScontents}.
This initial release closely follows the {\small SUNRISE} output format, for which additional information and documentation was presented by \citet{SUNRISE,SUNRISE2}.

\begin{table*}
\begin{center}
\caption{FITS file content overview.  Bracket quantities indicated multiple fields.}    
\label{table:FITScontents}
\begin{tabular}{ l c l l l }
\hline
Extension Name                          						& Extension								& 	Notes	\\
\hspace*{3em} Field Name						& 	Number				 		 			&					\\
\hline
\hline 
                                								&                 								       	&												\\

{\small CAMERA[0,1,2,3]-PARAMETERS}			&	2-5								&	 		\\
\hspace*{1.5em} Header Info:						&											&												\\
\hspace*{3em} cameradist						&											&								Distance from camera to galaxy [kpc]			\\
\hspace*{3em} linear\_fov							&											&								Image FOV [kpc]							\\
\hspace*{3em} CAMPOS[X,Y,Z]					&											&								Camera Position [kpc]						\\
\hspace*{3em} CAMDIR[X,Y,Z]						&											&								Camera Viewing Direction 					\\
\hspace*{3em} CAMUP[X,Y,Z]						&											&								Camera Up Direction 						\\

                                								&                 								       	&												\\

{\small INTEGRATED\_QUANTITIES}				&	7								&						\\
\hspace*{1.5em} Header Info:						&											&												\\
\hspace*{3em} TTYPE[1-12]						&											&	Names of included data fields			\\
\hspace*{3em} TUNIT[1-12]						&											&	Units for included data fields			\\
\hspace*{3em} L\_bol\_grid						&											&	Bolometric Luminosity [W]			\\
\hspace*{1.5em} Data Fields:						&											&												\\
\hspace*{3em} lambda							&											&	SED $\lambda$ array [m]			\\
\hspace*{3em} L\_lambda						&											&	SED $L_\lambda$ array [W/m]			\\

{\small FILTERS}								& 13      										&															\\
\hspace*{1.5em} Header Info:						&											&															\\
\hspace*{3em} TTYPE[1-20]						&											&	Names of included data fields					\\
\hspace*{3em} TUNIT[1-20]						&											&	Units for included data fields					\\

\hspace*{1.5em} Data Fields:						&											&															\\
\hspace*{3em} Filter Name List						&											&	String array giving the broadband filter names						\\
\hspace*{3em} Effective Wavelengths				&											&	$\lambda_{{\rm eff}}$ for each filter [m]							\\
\hspace*{3em} AB\_mag\_nonscatter[0,1,2,3]			&											&	Absolute magnitudes in each band								\\

                                								&                 								       	&															\\
{\small CAMERA[0,1,2,3]-BROADBAND}				& 14-17    					&					 	\\
\hspace*{9em} {\small -NONSCATTER}				& 			&					 	\\
\hspace*{1.5em} Header Info:						&											&															\\
\hspace*{3em} IMUNIT							&											&	Units for included images					\\

\hspace*{1.5em} Data Fields:						&											&															\\
\hspace*{3em} data							&											&	36 x N${}_{\rm pixel}$ x N${}_{\rm pixel}$  array;  Contains all broadband images						\\

\hline
\hline
\end{tabular}
\end{center}

\end{table*}

\subsection{Integrated SEDs}
We store integrated global SEDs for each galaxy under the {\small IntegratedQuantities} field.  
Example SEDs are shown in Figure~\ref{fig:gal_sed} for one star-forming blue galaxy (blue line, ID=350781), and one quiescent red galaxy (red line, ID=138415).  
Both galaxies have stellar masses of $M_*=10^{11.0} M_\odot$.
The galaxies have star formation rates of $5.7 M_\odot / $yr (blue) and $0.0 M_\odot / $yr (red).
While the blue galaxy still has a reservoir of star-forming gas, the red galaxy has quenched its star formation via AGN feedback.
The two galaxies have noticeably different SED slopes with the star-forming galaxy having substantially more UV emission.
Moreover, since a somewhat significant amount of light in the near IR comes from young stellar populations, an offset in the near IR SED can also be seen.

The fiducial SEDs do not take into account dust, which implies that the UV luminosities are overestimated due to missing SED attenuation.  
This is demonstrated in Figure~\ref{fig:gal_sed}, where the cyan and magenta lines show the star-forming galaxy's SED when applying the {\small MAPPINGS} HII region model and the CF00 empirical dust correction, respectively.
In both cases, these models reduce the UV luminosity by approximating the obscuration of young stellar spectra.
At optical wavelengths, the CF00 model depresses the SED by $\sim20\%$, while the {\small MAPPINGS} model is nearly coincident with the pure SB99 spectra.   
However, since both galaxies are fairly massive and dominated by older stellar populations which contribute to the near IR light, the offset in the near IR portion of the SED is significantly less noticeable than in the UV.

\begin{figure}
\centerline{\vbox{\hbox{
\includegraphics[width=3.5in]{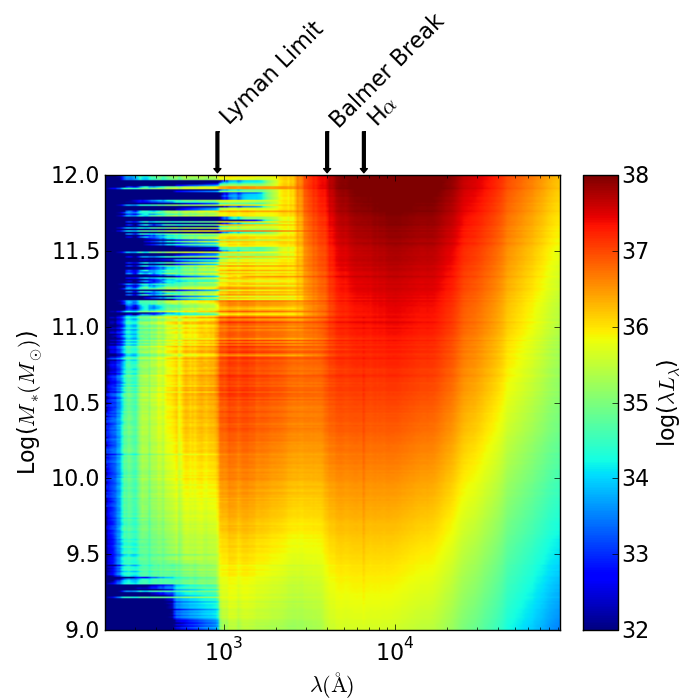}
}}}

\caption{The mass-binned median integrated spectra from the synthetic image catalog is shown.  Each row represents the median galaxy spectra from a bin of width $\Delta M_* = 0.01$ dex.  The intensity of the NIR SED increases smoothly with increasing galaxy stellar mass, while the low star formation rates of the least massive galaxies and quenching of massive galaxies is indicated via low levels of UV emission.   }
\label{fig:two_d_mean_spectra}
\end{figure}

To present a more complete overview of the synthetic spectra properties, we show all 40,000 Illustris galaxies by binning in narrow (0.01 dex) stellar mass bins and calculate the median galaxy spectra at each wavelength as a function of galaxy stellar mass in Figure~\ref{fig:two_d_mean_spectra}.
This figure shows the mean galaxy SED template that applies to our redshift $z=0$ synthetic galaxy spectra as a function of stellar mass.
The intensity of the median NIR SED increases smoothly with increasing galaxy stellar mass.
Little UV emission from both low- and high-mass galaxies can be identified.
The least massive galaxies produce little UV emission owing to their low star formation rates derived from their position on the star formation main sequence.
The high mass galaxies have low star formation rates owning to quenching from AGN feedback in our model.
This view of the data also allows us to pick out several features of the mock spectra easily.  
Examples include the Lyman limit at 912 \AA, Balmer Break at 4000 \AA, and H$\alpha$ emission line at 6563 \AA -- which are all included in the input SB99 template SEDs.
Departures from the median trends identified in Figure~\ref{fig:two_d_mean_spectra} can be substantial in the UV part of the spectrum owing to fluctuations in the star formation rates of galaxies -- even at similar stellar masses.
This variability is pronounced for high-mass systems (i.e. $M_*\gtrsim^{11} M_\odot$) where quenched and non-quenched galaxies sit in the same stellar mass bins.
On the other hand, the variability of the median spectra are low for the most massive systems in the NIR region owing to their NIR luminosities being dominated by old stellar populations.

The median galaxy SED templates for the Illustris galaxy population presented in Figure~\ref{fig:two_d_mean_spectra} do somewhat depend on the adopted choice for the stellar population synthesis model. 
We have compared the findings of Figure~\ref{fig:two_d_mean_spectra} with the BC03-based outputs and found that qualitatively the mass binned median galaxy spectra generated using the BC03 stellar population synthesis models look very similar to those generated using the SB99 procedure.
The main distinction between these models can be seen in their UV spectra.
Whereas the SB99 models have little or no UV emission from old stellar populations, the BC03 catalogs contain non-negligible UV emission from old 
stars.
This causes the UV brightness of massive (quenched) galaxies to be substantially brighter for the BC03 model.
In principle, this difference also contributes to the UV emission of intermediate mass galaxies (i.e. $M_* \sim 10^{10.5} M_\odot$).
However, since the intermediate mass galaxies have substantial ongoing star formation dominating their UV spectra, any contribution to the UV spectra from old stellar populations is not easily identified.

\subsection{Galaxy Images}
Spatially resolved galaxy images are stored in 36 bands, from which 
surface brightness maps and color images can be directly produced.
Color galaxy images produced from our image pipeline were first presented in Figure 13 of~\citet{Vogelsberger2014b} which shows 42 blue star forming galaxies and 42 red, passive galaxies as viewed through the SDSS-g, -r, -i bands using the~\citet{Lupton2004} asinh scaling.
Here we present additional applications of the broadband images by demonstrating: (1) the viewing angle dependence of a select set of galaxy types, (2) the visual appearance of a simulated galaxy in different filters, and (3) the time evolution of several elliptical galaxies exhibiting stellar shells.

\subsubsection{Multiple Galaxy Viewing Angles}
\begin{figure}
\centerline{\vbox{\hbox{
\includegraphics[width=3.5in]{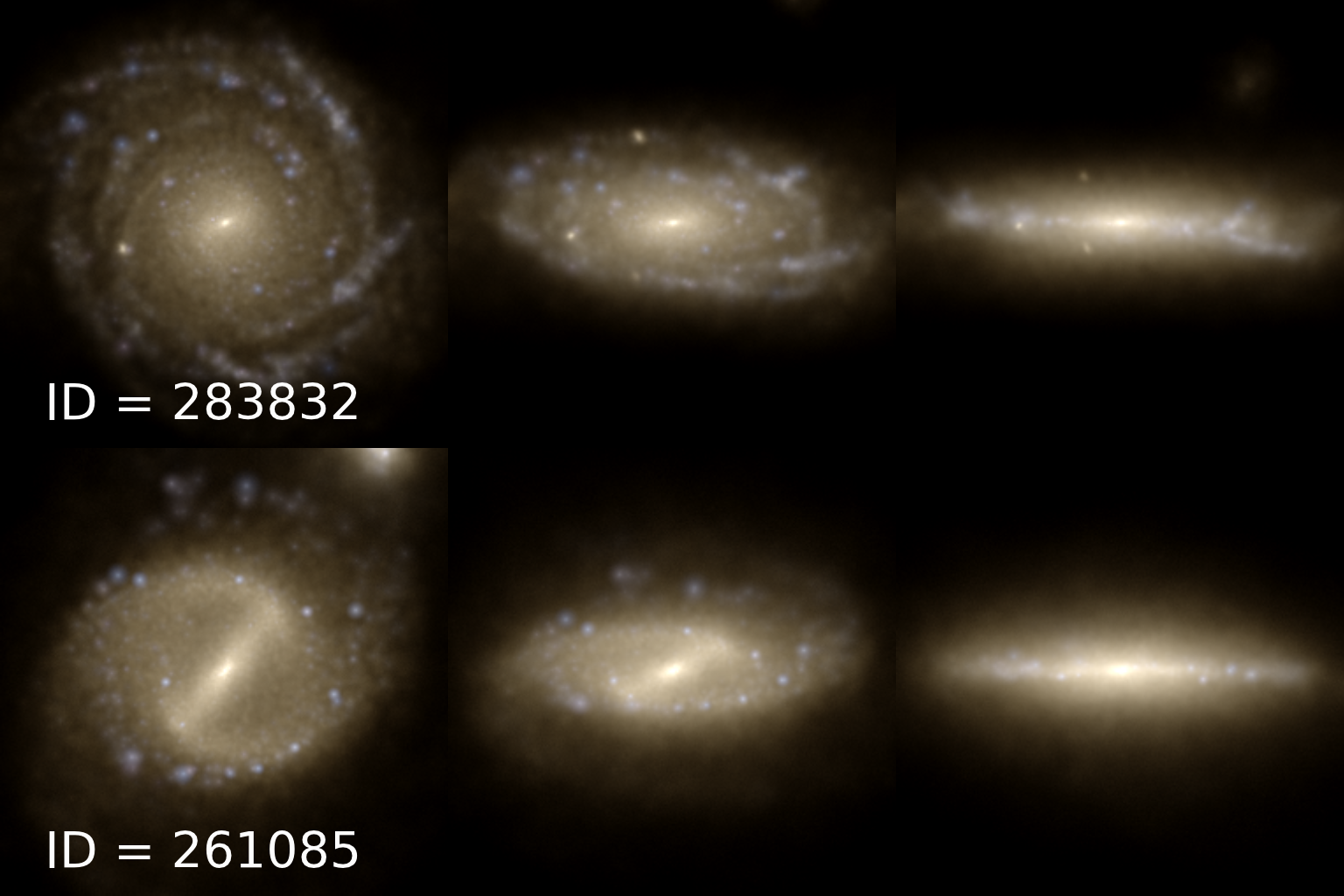}}}}
\caption{Montage of two disk galaxies, respectively without a bar (top) and with a bar (bottom), viewed from three different viewing angles. The degree to which the observable strength of the bar depends on viewing angle can be quantified, and be used as a correction for visual classification studies. }
\label{fig:multi_angle_disks}
\end{figure}

Figure~\ref{fig:multi_angle_disks} shows RGB color images of two disk galaxies made using the SDSS-g, -r, -i bands as seen from three 
different directions to demonstrate the impact of viewing angle on perceived galactic morphology.  
The left (right) most image is aligned (anti-aligned) with the angular momentum vector for the stars in the galaxy.  
The central images uniformly transition between these two polar views.
We note that the full galaxy image catalog contains only four viewing angles for a galaxy, without a guarantee of any of the images being exactly (anti-)aligned with the rotation axes.  
The images presented here have been constructed after manually rotating the orientation of each galaxy to highlight the viewing angle dependence.
While the top galaxy (ID=283832) is relatively axisymmetric, the bottom galaxy has a prominent bar (ID=261085).  
The observable strength of the bar feature is a function of viewing angle, with the bar appearing strongest when viewed face-on and being completely unidentifiable when the galaxy is seen edge-on.  
These mock images could then be used to derive quantitative correction factors for visually classified bar strengths, or 
to determine what fraction of strongly barred systems would have been classified as having weak bars, or no bars altogether, based on viewing angle.

Similar examples of the impact of galaxy viewing angle can be constructed for merging/interacting galaxies~\citep[e.g.,][]{Lotz2008}, bulge-to-disk identification~\citep[e.g.,][]{Scannapieco2010}, and measured isophotal shape of elliptical systems~\citep[e.g.,][]{Carter1979}.

\begin{figure}
\centerline{\vbox{\hbox{
\includegraphics[width=3.5in]{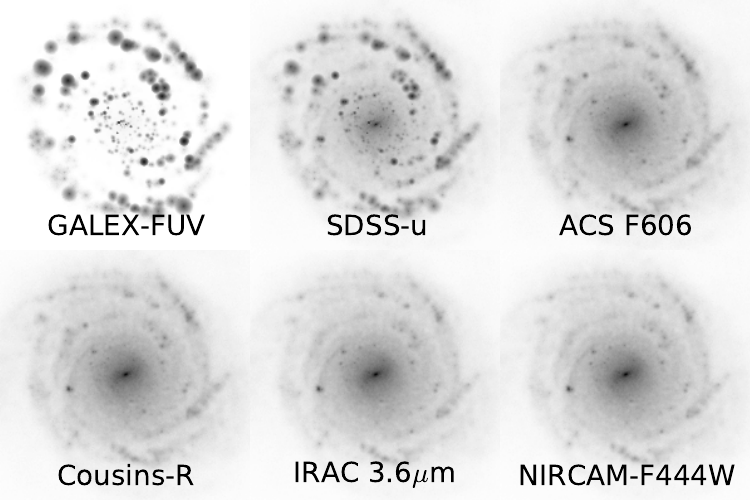}}}}
\caption{Montage of a sample simulated disk galaxy in 6 bands as labeled within the figure.  Each image has been scaled independently using an ``asinh" scaling, with the non-linear transition being set to the mean pixel intensity for that band.  Bands increase in wavelength from top left to lower right.  UV bands are strongly impacted by localized emission from young star particles.  Longer-wavelengths bands trace stellar mass faithfully. }
\label{fig:bw_all_bands}
\end{figure}

\subsubsection{Multi-wavelength Images}
Figure~\ref{fig:bw_all_bands} shows an example galaxy as seen through 12 different broadband filters, as labeled within the figure.  
Each image has been scaled independently using an ``asinh" scaling, with the non-linear transition set to the mean pixel intensity for that band.  
Bands increase in wavelength from top left to lower right (effective wavelengths associated with the bands can be found in Table~\ref{table:BroadbandFilters}).  
The UV bands reveal the emission from young star particles -- tracing the location of ongoing star formation in this system.
Longer-wavelength bands trace stellar mass more efficiently and exhibit less band-to-band variation.
We note that this particular galaxy (ID=283832) was selected because it possesses a large stellar disk ($M_* = 10^{11.5} M_\odot$), with moderate levels of ongoing star formation, enabling us to make a clear demonstration of the wavelength-dependent image variation.
This is the largest and most massive disk galaxy in the simulation volume; only a small fraction of the galaxies in the simulation are as well-resolved as the one presented here.

\subsubsection{Image Masking}
\begin{figure*}
\centerline{\vbox{\hbox{
\includegraphics[width=7.0in]{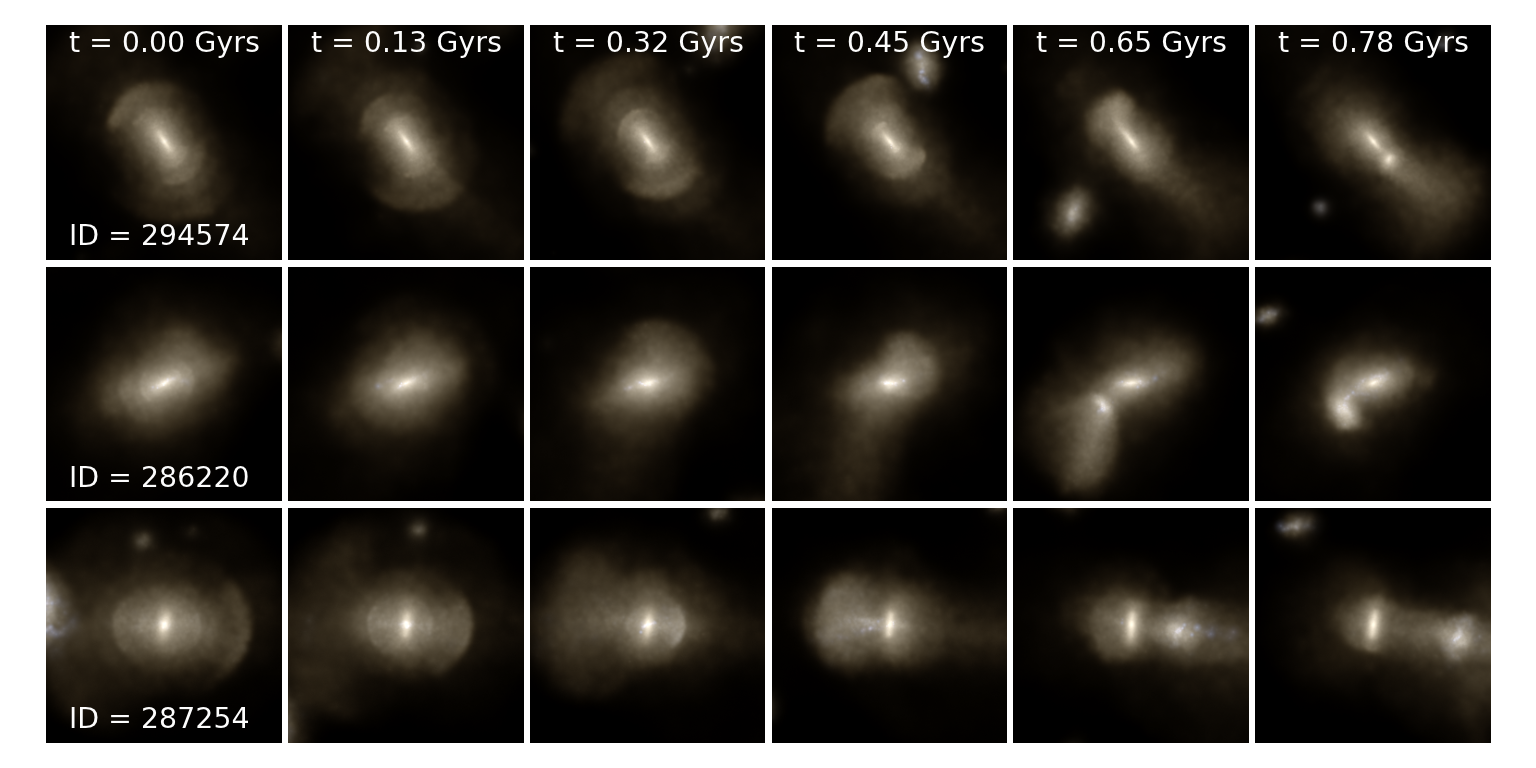}}}}
\caption{The time evolution of three galaxies with stellar shells at redshift $z=0$ is shown, for a fixed field of view of 120 physical kpc.  Lookback time for each column is noted in the top row. The stellar shells can be seen to form during merger events in the past $\sim 1$ Gyr.   }
\label{fig:shell_evolution}
\end{figure*}

Figure~\ref{fig:shell_evolution} shows a time series of RGB images from the rest frame SDSS-g, -r, -i bands for three particular galaxies that we visually identify to possess stellar shells at redshift $z=0$. 
We trace these systems back in time using merger trees, identifying the most massive progenitor galaxy at each previous snapshot.
We generate images for a system at every snapshot using a fixed field of view of 120 physical kpc.
We find that merger events occur for all three systems within the past $\sim$Gyr, which are then responsible for forming the observed present-day stellar shells.
This is consistent with the established theory of stellar shell formation via major~\citep{Hernquist1992} and minor~\citep{Quinn1984, Dupraz1986, Hernquist1988, Kojima1997} mergers.

\begin{figure*}
\centerline{\vbox{\hbox{
\includegraphics[width=7.0in]{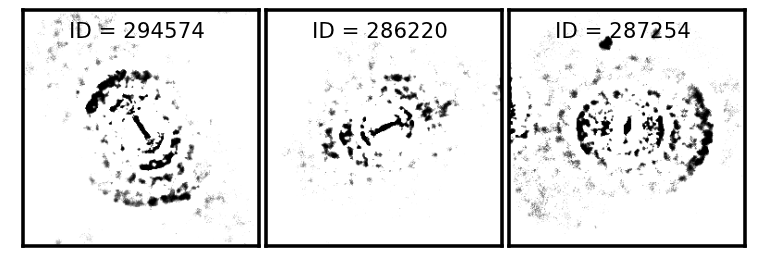}}}}
\caption{ Johnson-K band images after applying an unsharp mask to highlight non-smooth (high frequency) components in the stellar light distribution for the redshift $z=0$ galaxies shown in Figure~\ref{fig:shell_evolution}. }
\label{fig:unsharp_mask_shell_evolution}
\end{figure*}

To identify the shell locations, we make monochromatic images of these systems in the Johnson-K band (as a good representation of the stellar mass) and apply an unsharp mask procedure.
In detail, we convolve the original image with a two dimensional gaussian with a standard deviation of 5 pixels.
We then subtract the gaussian blurred image from the original image, and discard all negative pixels to specifically highlight stellar light excesses relative to the blurred image.  
The resulting masked image is shown in Figure~\ref{fig:unsharp_mask_shell_evolution} using a logarithmic stretch.
All high frequency image contributions -- including noise from the discrete particle representation of stellar mass -- are amplified in the masked image.  
As a result, the shells are more easily distinguished from the smoothly varying light component.
Using the masked images, we identify the stellar shells down to fairly low galactocentric radii with $\sim$3-5 shells being visually apparent in the systems at redshift z=0.  
Stellar shells have been observed at low galactocentric radius using similar image masking techniques~\citep[see][for a detailed description]{Canalizo2007}.  
We have confirmed the presence of low radii stellar shells by examination of a radial velocity versus radius phase diagram.

The presence of multiple systems with layered stellar shells in our cosmological simulation is of interest since the majority of previous theoretical work studying the evolution and characteristics of stellar shells formed via mergers has been done either with idealized merger simulations~\citep{Quinn1984, Hernquist1992, Dupraz1986, Hernquist1988, Kojima1997, Canalizo2007} or with dark matter only cosmological simulations~\citep[e.g.,][]{Cooper2011}.
Since these systems are realized here within the full cosmological context, we can address some previously inaccessible questions,
including:
(i) the origin of very tightly bound shells, (ii) the observability lifetime of shells and other tidal features following merger events,
and (iii) the predicted frequency of visible shells in the galaxy population.
A deeper exploration into these topics using the Illustris simulation and images presented here is deferred to a separate, targeted study.

\section{Example Application: Derived Stellar Masses}
\label{sec:Example}
\label{sec:StellarMass}

An examination of the distribution of galaxies of various morphological types as a function of mass, the morphology-environment relation, and the role of mergers in driving morphological evolution will be considered elsewhere (Snyder et al., in prep.).
Here we investigate a related problem which can be attacked with the image pipeline catalog:  the accuracy of stellar mass estimates based on broadband photometric fitting.
The simple question we ask is:  How well do broadband fitting routines recover the stellar mass from the simulations?  
Similar topics have been addressed using semi-analytic models~\citep[e.g.,][]{Mitchell2013} and idealized hydrodynamical simulations ~\citep{Wuyts2009a, Michalowski2014, Hayward2014c}.
 
For observational data, stellar masses are often determined by applying one of a few commonly adopted functional forms for the star formation history and finding the best-fitting stellar mass, galaxy age, star formation timescale, dust attenuation parameters, and metallicity to the observed SED.
One common assumption is an exponentially declining star formation history which is characterized by two free parameters: the galaxy formation time and the star formation timescale.
The simplest method for constraining these parameters is to perform a chi-squared minimization of the observed SED or broadband photometry relative to a grid of mock SEDs which uniformly cover the parameter space.  
This is the general approach that is taken in, e.g., {\small FAST}~\citep{FAST}.

To perform this exercise we assign light to a galaxy based on a stellar population synthesis template (e.g., the SB99 model), and then use the same (or a similar) stellar population synthesis model to find a best fit to the SED.
One might therefore worry that this is a circular argument, which will not shed any light on the validity of ``true" galaxy mass estimates.  
However, the star formation history and heavy element composition for our simulated galaxies is known and properly reflected in the integrated galaxy SED, while the observational models must assume a functional form for the star formation history and metallicity.  
The discrepancy leads, in fact, to the two recommendations for improved stellar mass determination, described below.

To make the comparison as clear and idealized as possible, we neglect dust in the simulated galaxy SEDs, and force {\small FAST} to neglect dust obscuration as well (i.e. set $A_V=0$).
Neglecting dust removes a large age/dust degeneracy which is present in real extragalactic data and mass measurements.
Furthermore, we ignore dust re-radiation at long wavelengths -- which would add uncertainty into both our models and model SED fitting.
Finally, we have not included any photometric and/or distance related errors in the galaxy's positioning.  
We assume that we know a galaxy's distance exactly (without error), and assign apparent magnitudes in each band appropriately.
Owing to these simplifying assumptions, the comparisons that we perform below are highly idealized.
We expect that introducing any additional SED complications (dust, distance errors, etc.) would bring the derived stellar masses into worse agreement than what we present below because there would be additional characteristics of the intrinsic galaxy SEDs that are not fully accounted for by {\small FAST}.
Because of these simplifying assumptions, we will show below that we can obtain relatively good mass estimates for our mock galaxies based on their mock SEDs using only 5 bands.
However, we find that restricting the allowable age range and metallicity values that {\small FAST} is permitted to sample can improve the derived stellar mass errors.

\subsection{Comparison: {\small FAST} vs. {\small SUBFIND} stellar masses}
\label{SEC:FAST}
We employ the code {\small FAST}~\citep{FAST} to fit stellar population synthesis models to the artificially generated image pipeline broadband photometry.  
We assume an exponentially declining star formation history ($\psi \propto \exp \left( - t / \tau \right)$) to determine stellar masses based on the stellar-light-only galaxy photometry from the image pipeline.  
For the initial comparison, we allow {\small FAST} to select from the full range of input SED possibilities.  
Specifically, the galaxy age is taken in the range $10^7 < t_{{\rm age}} < 10^{10.5}$ years with logarithmic steps of $\Delta t_{{\rm age}} = 0.1$ dex, 
the star formation timescale is chosen in the range $10^7 < \tau < 10^{10}$ years with logarithmic steps of $\Delta \tau = 0.1$ dex, and 
the metallicity is selected from $Z=[0.004, \;0.008, \;0.02, \;0.05]$.

The mock galaxy photometry uses the SB99 templates while {\small FAST} uses the BC03 templates.\footnote{The SB99 and BC03 model templates agree well enough in the optical bands that this does not introduce a major source of error into our analysis.  However, since the old stellar populations have different UV emission in the two models, very poor fits can be identified when including, e.g., the GALEX bands.}
We ignore dust attenuation in the mock photometry and restrict {\small FAST} fits to neglect attenuation as well.  

{\small FAST} input files are generated by calculating rest-frame broadband fluxes for the five SDSS bands (u, g, r, i, z) converted to units of flux density assuming the galaxies are at redshift $z=0.01$ using the same cosmology employed in the Illustris simulation.\footnote{We note that it is slightly inconsistent to use rest-frame broadband fluxes for input into {\small FAST} (which assumes redshifted fluxes), but we have checked that the neglected k-correction error is minimal for such a low redshift.}  
Although {\small FAST} has the ability to determine spectroscopic/photometric redshifts, in our initial comparison we supply correct redshifts to {\small FAST} to avoid errors.
We compare the stellar mass derived from {\small FAST} (hereafter {\small FAST} mass) against the simulation determined stellar mass found in the {\small SUBFIND} galaxy catalog (hereafter {\small SUBFIND} mass) for all galaxies with integrated spectra in the image pipeline.

\begin{figure*}
\includegraphics[width=3.25in]{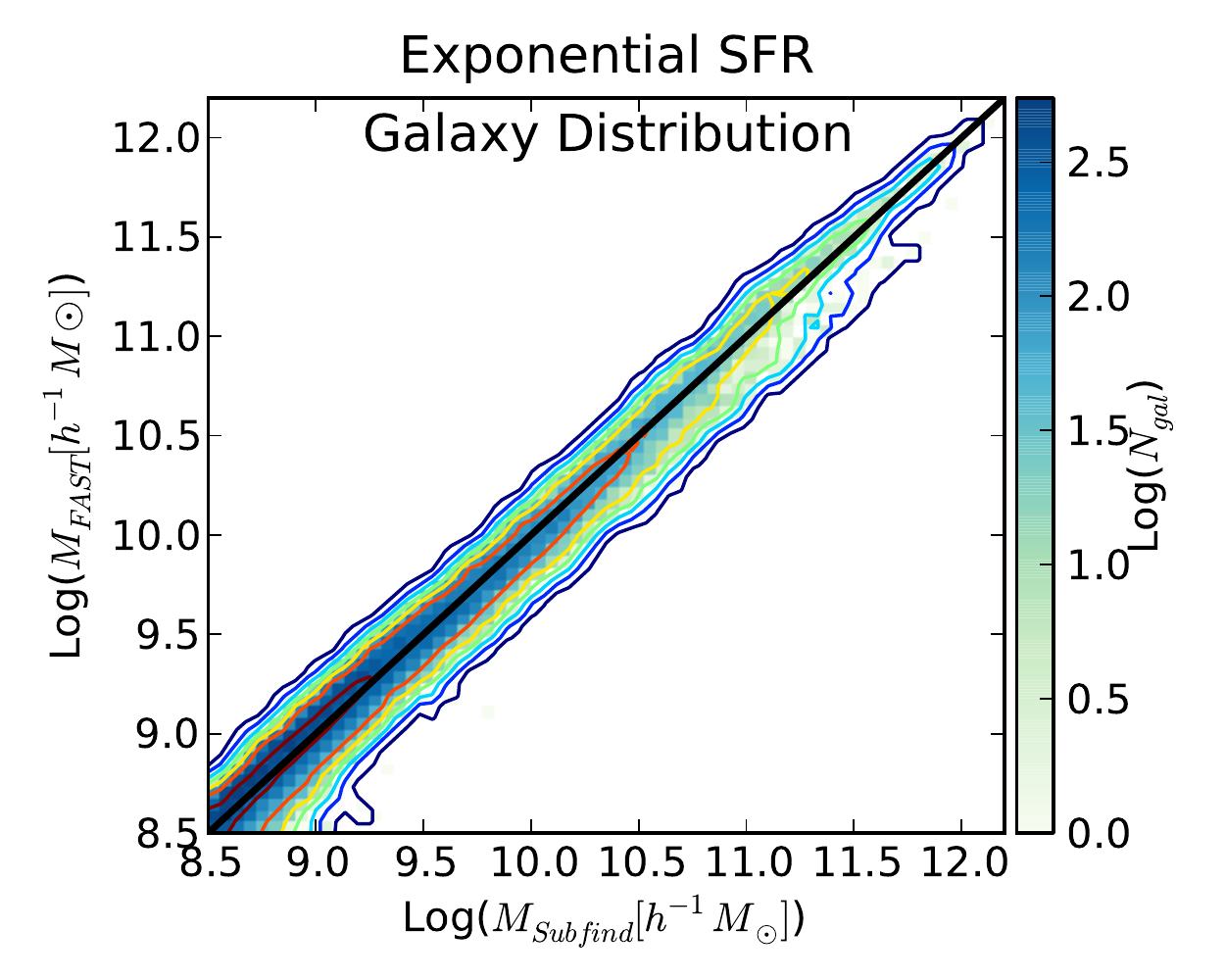}
\includegraphics[width=3.25in]{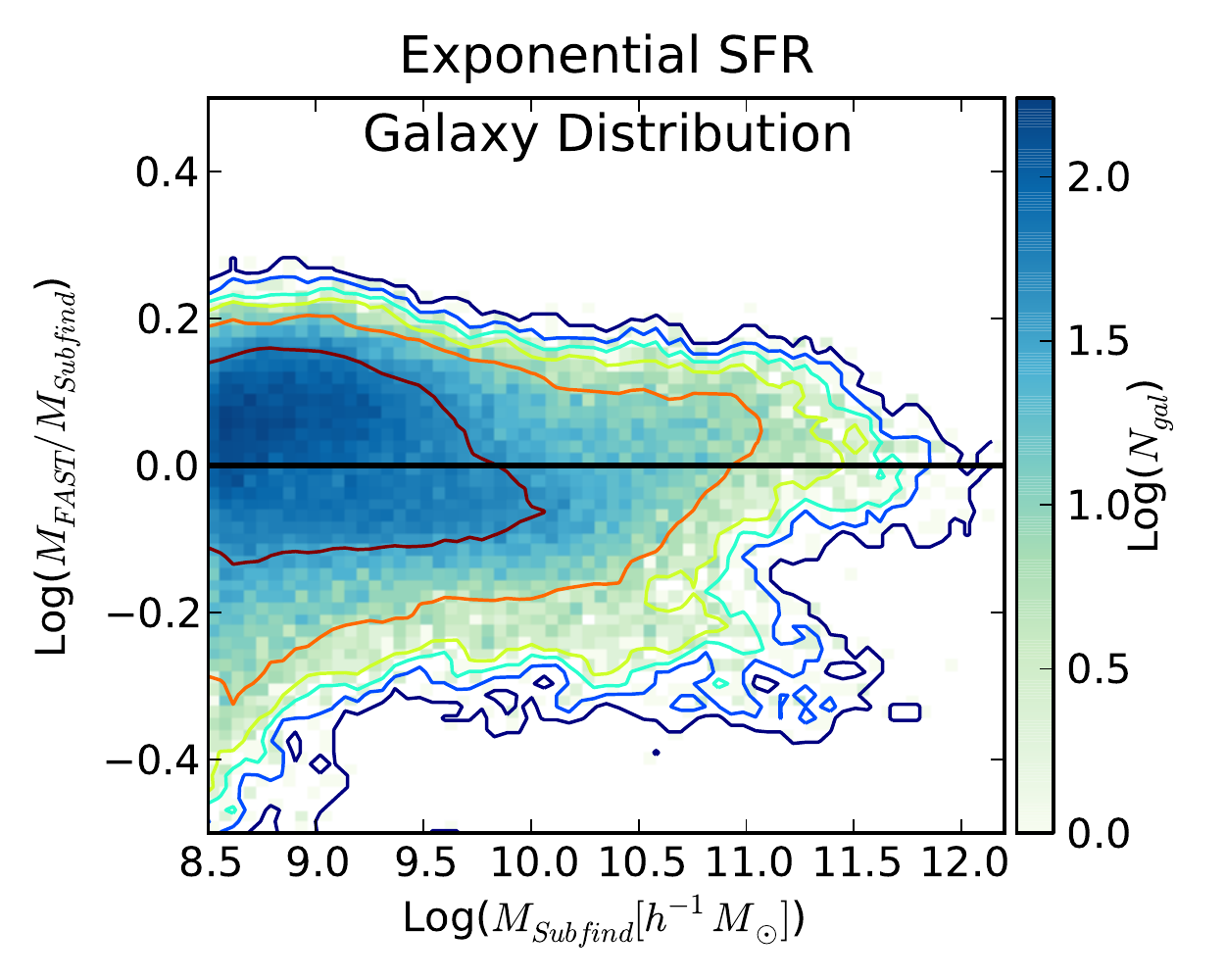}
\caption{Two-dimensional histograms of the distribution of stellar mass derived from {\small FAST} and intrinsic stellar mass from {\small SUBFIND} (left), and the ratio of these two values (right), as a function of galaxy mass.  In both cases, the color denotes the number of galaxies in each pixel, as noted in the colorbar.  Solid line contours show iso-density surfaces to highlight the overall galaxy distribution.  The general agreement between the derived stellar mass and the true simulation mass is generally very good, with $>95\%$ of systems obtaining the correct mass within a factor of two.    }
\label{fig:fast_masses}
\end{figure*}

\begin{figure}
\includegraphics[width=3.25in]{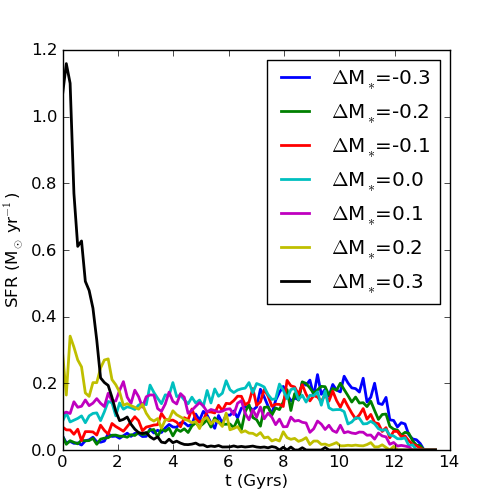}
\caption{ The average (median) star formation histories of galaxies in the mass bin $10^9 < M_* [M_\odot] < 10^{10}$ are shown for galaxies.  
The lines indicates the median star formation history for 50 galaxies with a mass error $\Delta$M${}_*$= Log(M${}_{{\rm fast}}$/ M${}_{{\rm subfind}}$) closest to the prescribed value in the plot legend.  
Galaxies with underestimated stellar masses have systematically older stellar populations.   }
\label{fig:delta_sfr_hist}
\end{figure}

The left panel of Figure~\ref{fig:fast_masses} shows a two dimensional histogram indicating the relationship between {\small FAST} photometrically derived stellar mass and the {\small SUBFIND} tabulated stellar mass for all 41,517 galaxies with more than 500 star particles at redshift $z=0$.  The black line indicates a perfect 1:1 relationship (i.e. the {\small FAST} masses exactly equal the {\small SUBFIND} masses).  There is a tight relationship between these two quantities that nearly follows the ideal 1:1 scaling, with a slight offset and scatter.  Out of the 41,517 galaxies for which we perform this fitting, 1755 (4\%) disagree by more than a factor of two and 10 (0.02\%) disagree by more than a factor of five.
We calculated the average (median) star formation rate histories for galaxies in a mass bin ($10^9 < M_* [M_\odot] < 10^{10}$) for several Log(M${}_{{\rm fast}}$/ M${}_{{\rm subfind}}$) values.  
The resulting plot is shown in Figure~\ref{fig:delta_sfr_hist} and can be used to partially determine what drives the {\small FAST} mass estimate.
The median star formation histories evolve with this ``mass error'' parameter.  
Galaxies with the lowest mass error parameter,  which correspond to galaxies whose masses have been most severely underestimated, have the oldest stellar populations and earliest peak in their SFR histories.  
Galaxies with old stellar populations will have the largest mass-to-light ratios, which drives their systematically underestimated stellar masses if FAST does not choose a suitably large galaxy age.  
Moving toward larger (and positive) mass error parameters we find that the average SFR history becomes biased toward later times and the present day star formation rate monotonically increases.  
For the most extreme positive mass error parameters, we find that they are heavily peaked toward very recent star formation.
The most extreme objects have either very early or late formation histories which are not captured in the FAST best fit parameters.

\begin{figure*}
\includegraphics[width=2.30in]{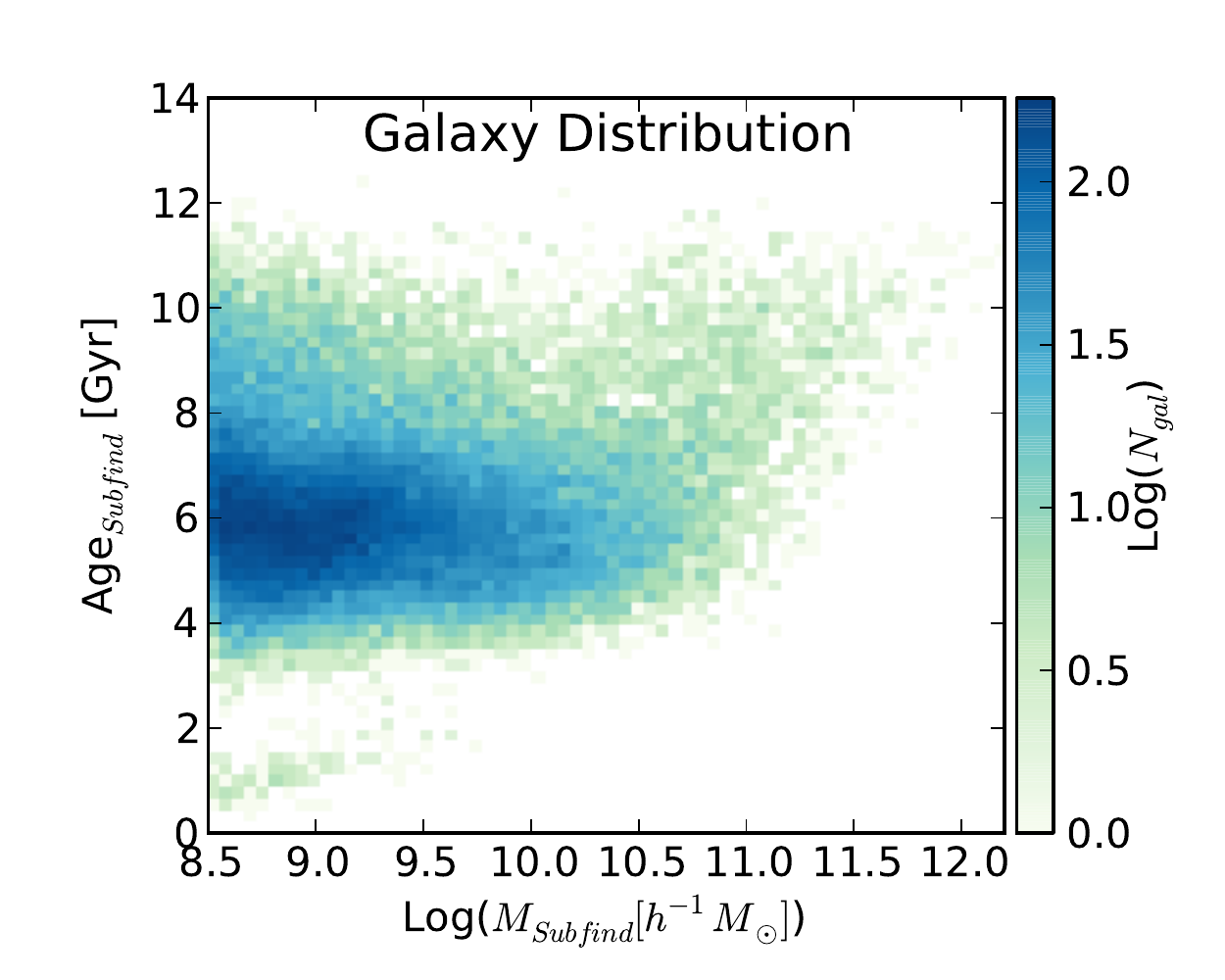}
\includegraphics[width=2.30in]{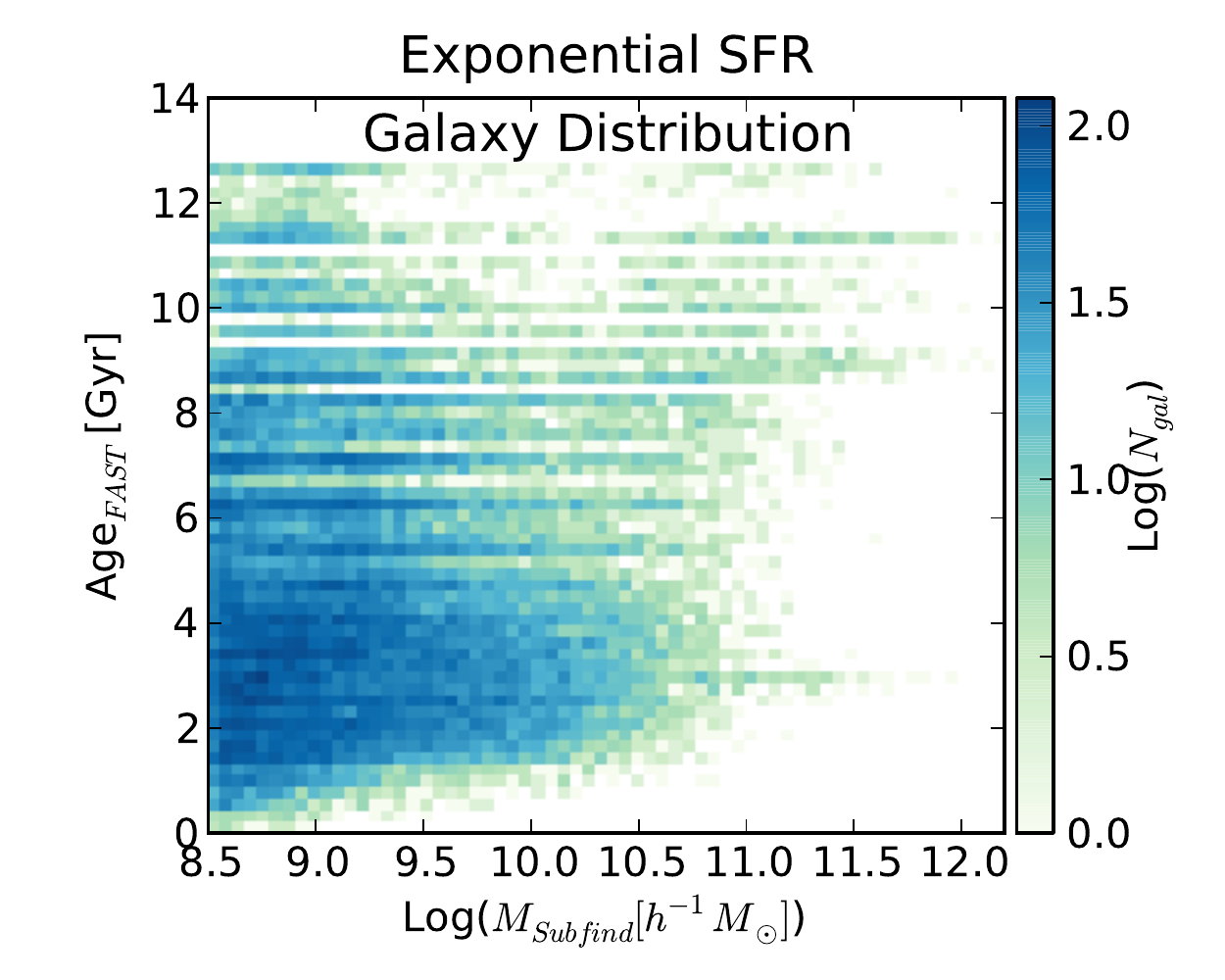}
\includegraphics[width=2.30in]{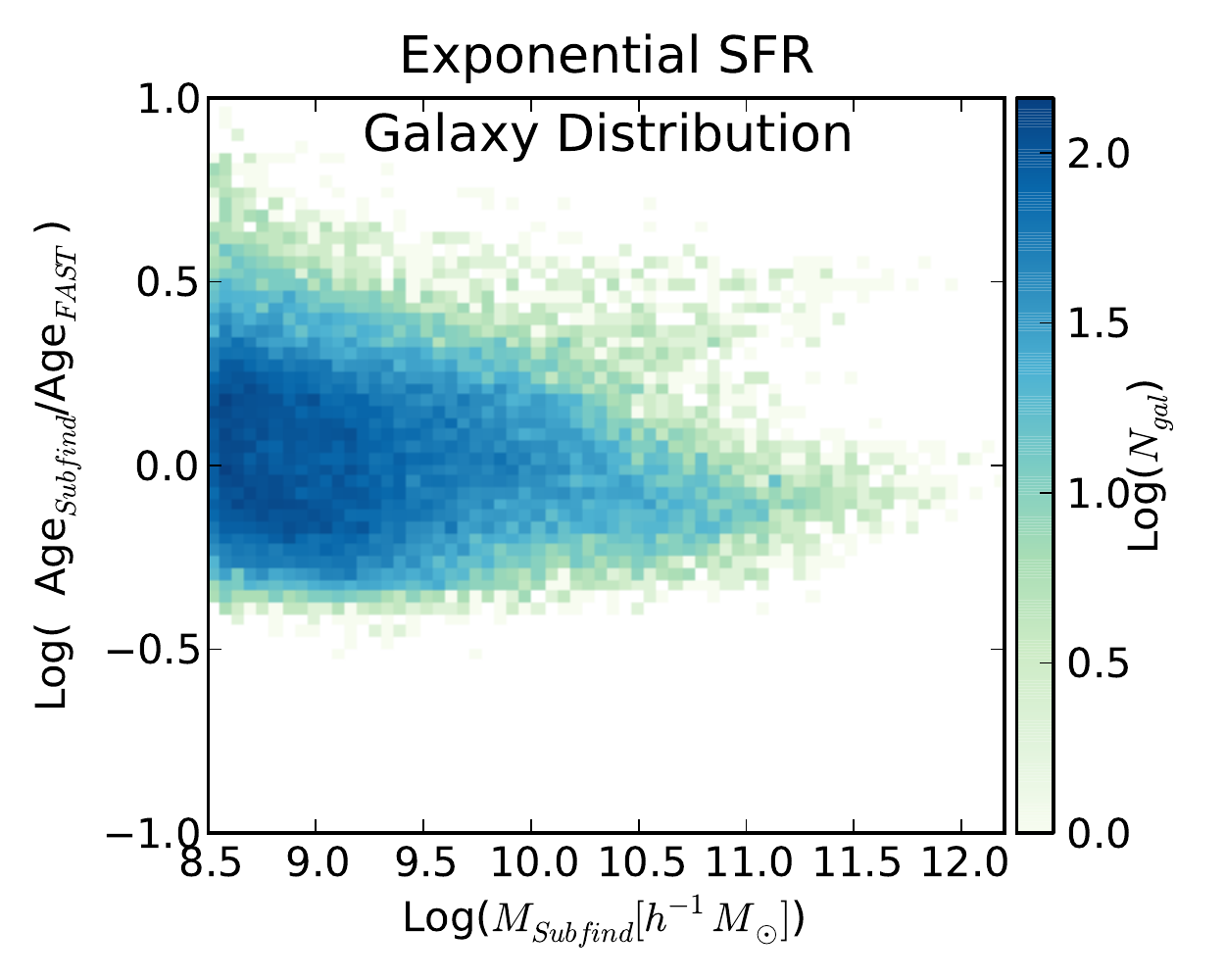}
\caption{(left) The distribution of galaxy formation times defined as the time of stellar half mass assembly as a function of galaxy stellar mass.  (center) The distribution of galaxy age parameters as derived from {\small FAST} defined as the time of onset of star formation for each galaxy. (right) The ratio of the half mass assembly time to the {\small FAST} assigned stellar age as a function of galaxy mass. 
} 
\label{fig:fast_sub_ages}
\end{figure*}

The left panel of Figure~\ref{fig:fast_sub_ages} shows the stellar half mass assembly time.  
This can be compared against the center panel of Figure~\ref{fig:fast_sub_ages} which shows the stellar half mass assembly time based on the assumed {\small FAST} star formation history.
In general the two distributions look very different.
The {\small FAST} age values span a larger range than found in {\small SUBFIND}.
We plot the ratio of the {\small SUBFIND} age to the {\small FAST} age in the right panel of Figure~\ref{fig:fast_sub_ages} as a function of galaxy mass.
There is significant scatter in this relation that is fairly mass dependent, with massive galaxies being assigned more accurate age estimates than low mass systems.
This mis-estimation of galaxy formation times can cause problems for the inferred mass-to-light ratios.

\begin{figure*}
\includegraphics[width=3.45in]{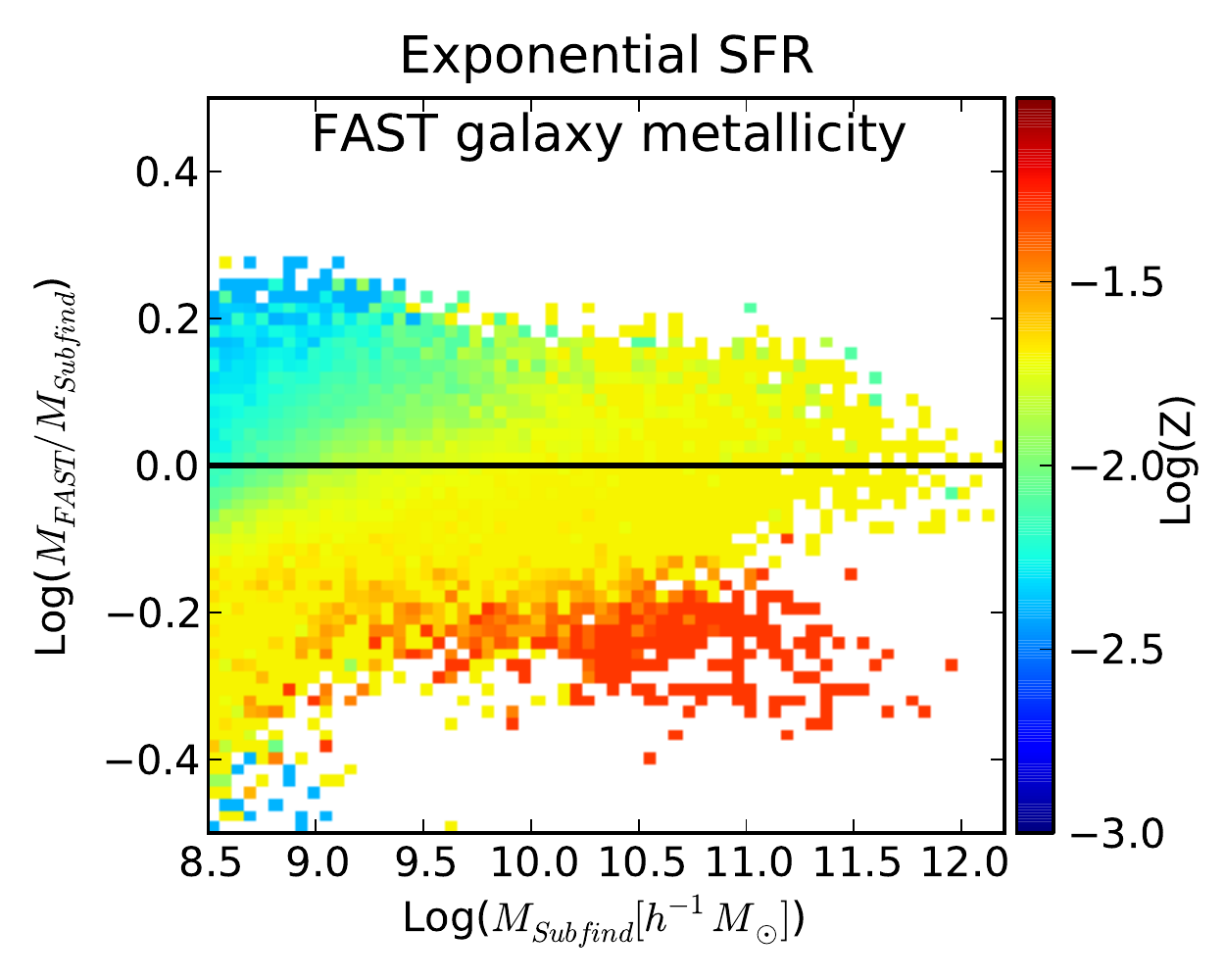}
\includegraphics[width=3.45in]{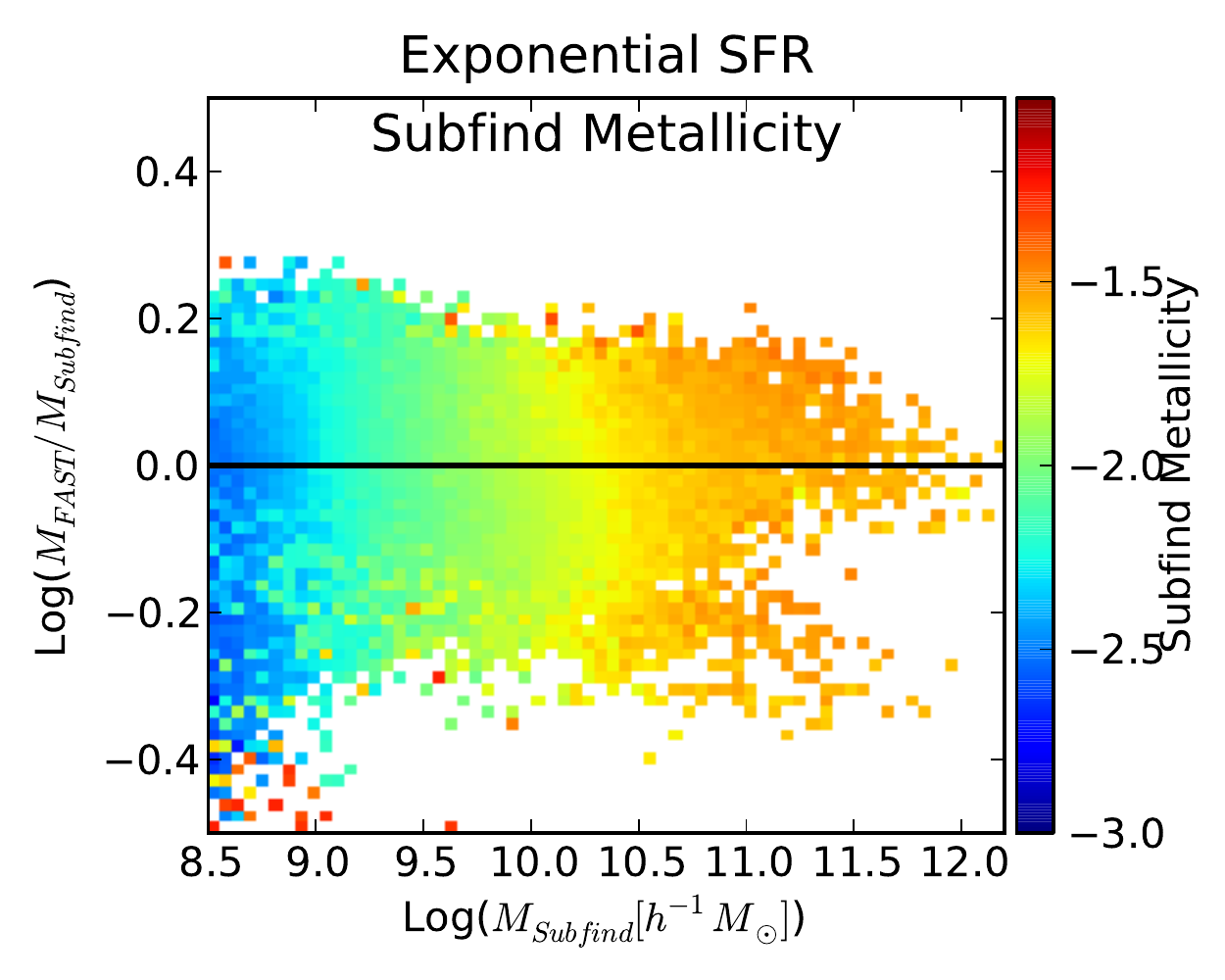}
\caption{Two-dimensional histograms depicting the average metallicity template applied by {\small FAST} (left) and average galaxy metallicity based on the simulation data (right).  In both cases, color denotes an average (mean) metallicity for all galaxies in each pixel, with the legend indicating the quantitative metallicity values.  While the simulation galaxy metallicity smoothly varies with mass, the {\small FAST} metallicities exhibit discontinuous jumps as a function of mass due to the sparsely spaced metallicity templates. }
\label{fig:fast_metallicities}
\end{figure*}

The right panel of Figure~\ref{fig:fast_masses} shows the ratio of the {\small FAST} masses to the {\small SUBFIND} masses as a function of the {\small SUBFIND} mass, which helps to identify some residual trends in the derived {\small FAST} masses.  
There is a mass dependence in the average {\small FAST} mass offset and scatter which can be ascertained from the distribution of galaxies outlined by the contour lines.  
We find that a principle origin of the mass dependent offset and scatter is the discrete and sparsely sampled metallicity values in the {\small FAST} BC03 stellar population synthesis templates.  
The standard BC03 templates used in {\small FAST} draw from discrete metallicities values of $Z=[0.004, \;0.008, \;0.02, \;0.05]$ without interpolation, and the result is a discrete jump in the derived stellar masses depending on which template metallicity is applied.
The left panel of Figure~\ref{fig:fast_metallicities} shows the average (mean) template metallicity that {\small FAST} applied to galaxies in each pixel.  
Since {\small FAST} is forced to pick one of these sparsely populated metallicity values, the derived galaxy mass offsets correlate directly with different {\small FAST} metallicity values.  
The right panel of Figure~\ref{fig:fast_metallicities} shows the mass weighted average stellar metallicity for these same galaxies.  
We can gather from this plot that, even in this highly idealized experiment, {\small FAST} is not able to accurately determine a galaxy's metallicity when using only 5 broadband filters.
Moreover, allowing for {\small FAST} to assign metallicity values to each galaxy independently results in galaxies of similar intrinsic mass being assigned factor of $\sim 2$ different stellar mass estimates.
We have further checked  -- and we discuss in section \ref{sec:fixed_metallicity} -- that if we restrict {\small FAST} to using only a single metallicity ($Z=0.02$) template, all derived galaxy masses fall along a single continuous relation, albeit with an offset from the desired $M_{{\rm FAST}} = M_{{\rm SUBFIND}}$ relation.

\subsection{Alternative Assumed Star Formation Histories}
Here we re-compute {\small FAST} masses by employing delayed or single-burst star formation histories.  Both the delayed and single burst star formation histories are described by two parameters:  a galaxy formation time and star formation timescale, $\tau$.  The galaxy formation time gives the lookback time when the galaxy started forming stars (i.e. for all times before this, the star formation rate is assumed to be zero).  In the case of the delayed star formation history ($\psi \propto t \exp \left( -t /\tau \right)$), the star formation timescale modulates the decay timescale for the late time star formation rate decline.  In the case of the single burst star formation history, the star formation rate is assumed to be constant after the galaxy formation time for one star formation timescale, after which it drops back to zero ($\psi = C$ for $t_{{\rm age}} - \tau < t < t_{{\rm age}}$).

\begin{figure*}
\includegraphics[width=3.45in]{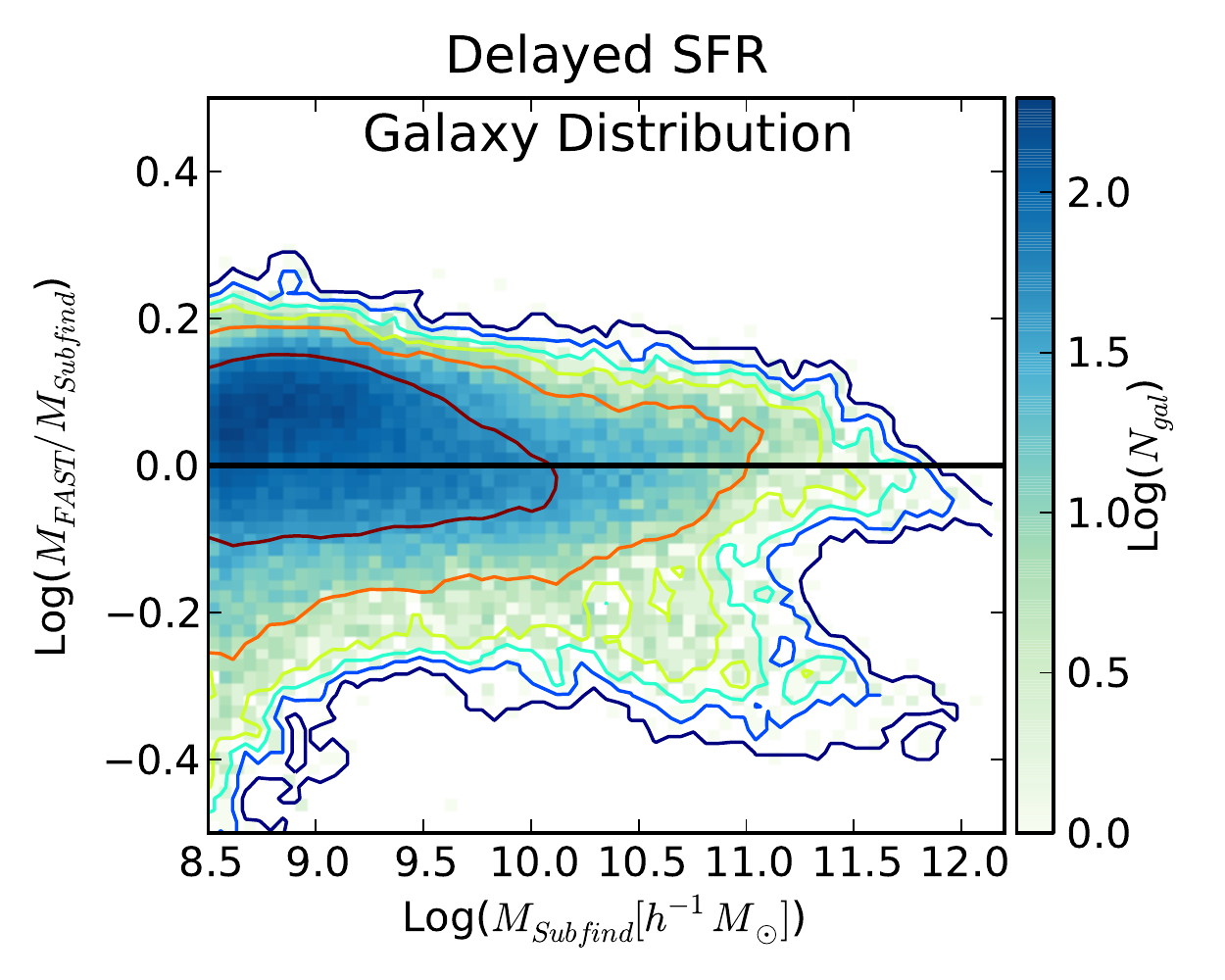}
\includegraphics[width=3.45in]{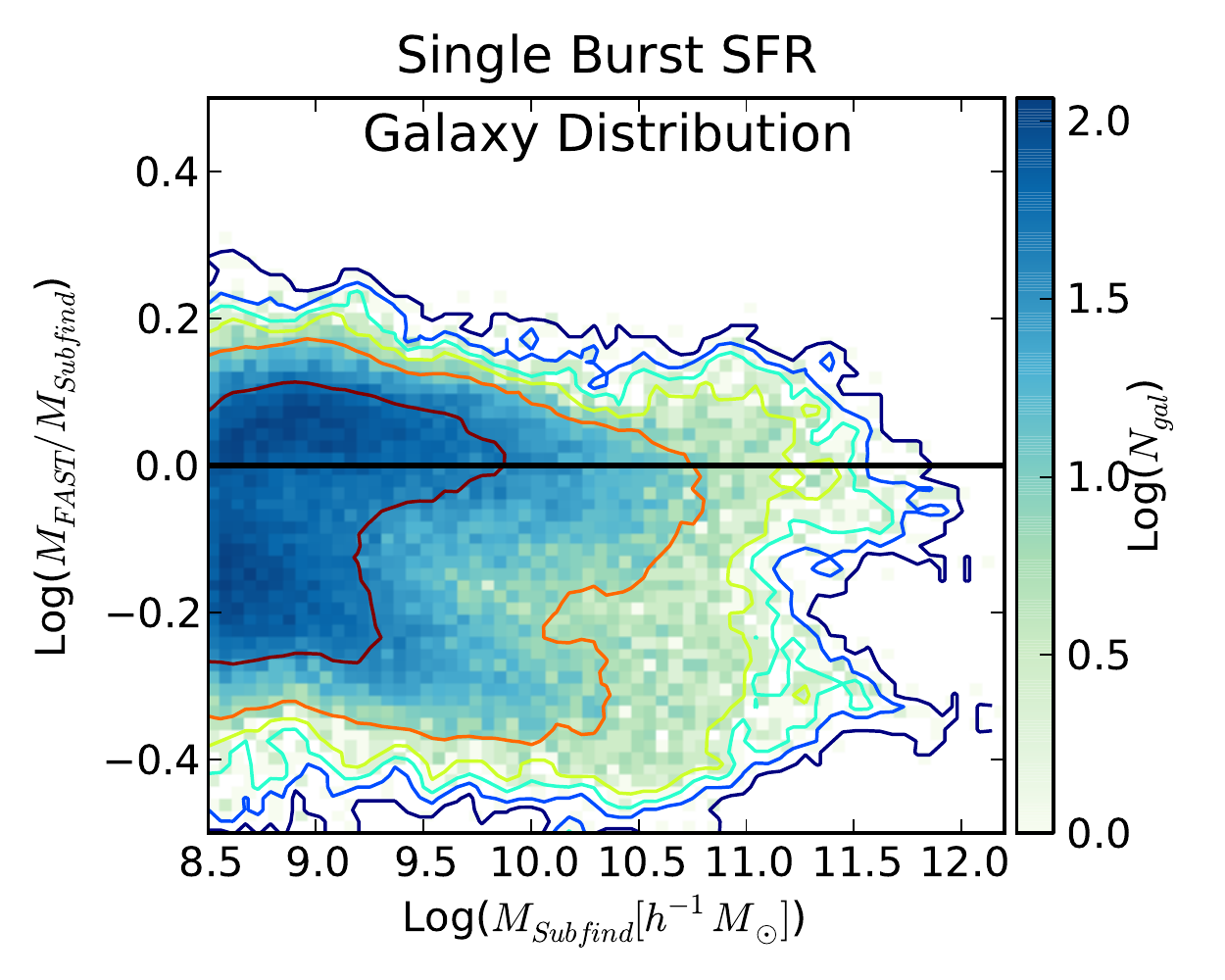}
\caption{Two dimensional histograms depicting the distribution of derived {\small FAST} masses versus intrinsic masses for a delayed star formation history (left) and single burst star formation history (right).  The distribution of derived galaxy masses for the delayed star formation history shows only subtle differences from the exponential star formation history.  The distribution of derived galaxy masses for the single burst star formation history is noticeably different from the exponential or delayed star formation histories. }
\label{fig:fast_alt_sfr}
\end{figure*}

\begin{figure*}
\includegraphics[width=3.45in]{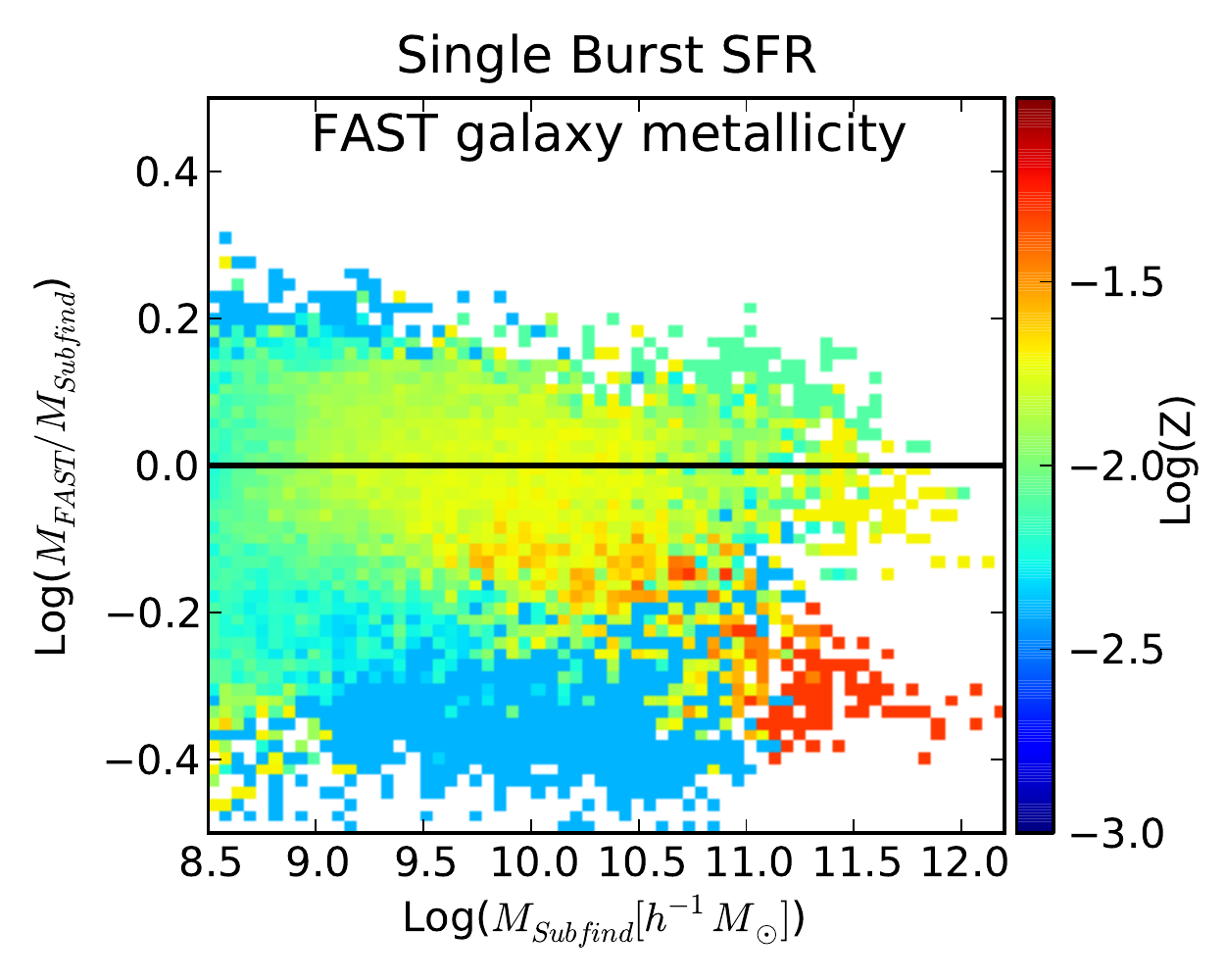}
\includegraphics[width=3.45in]{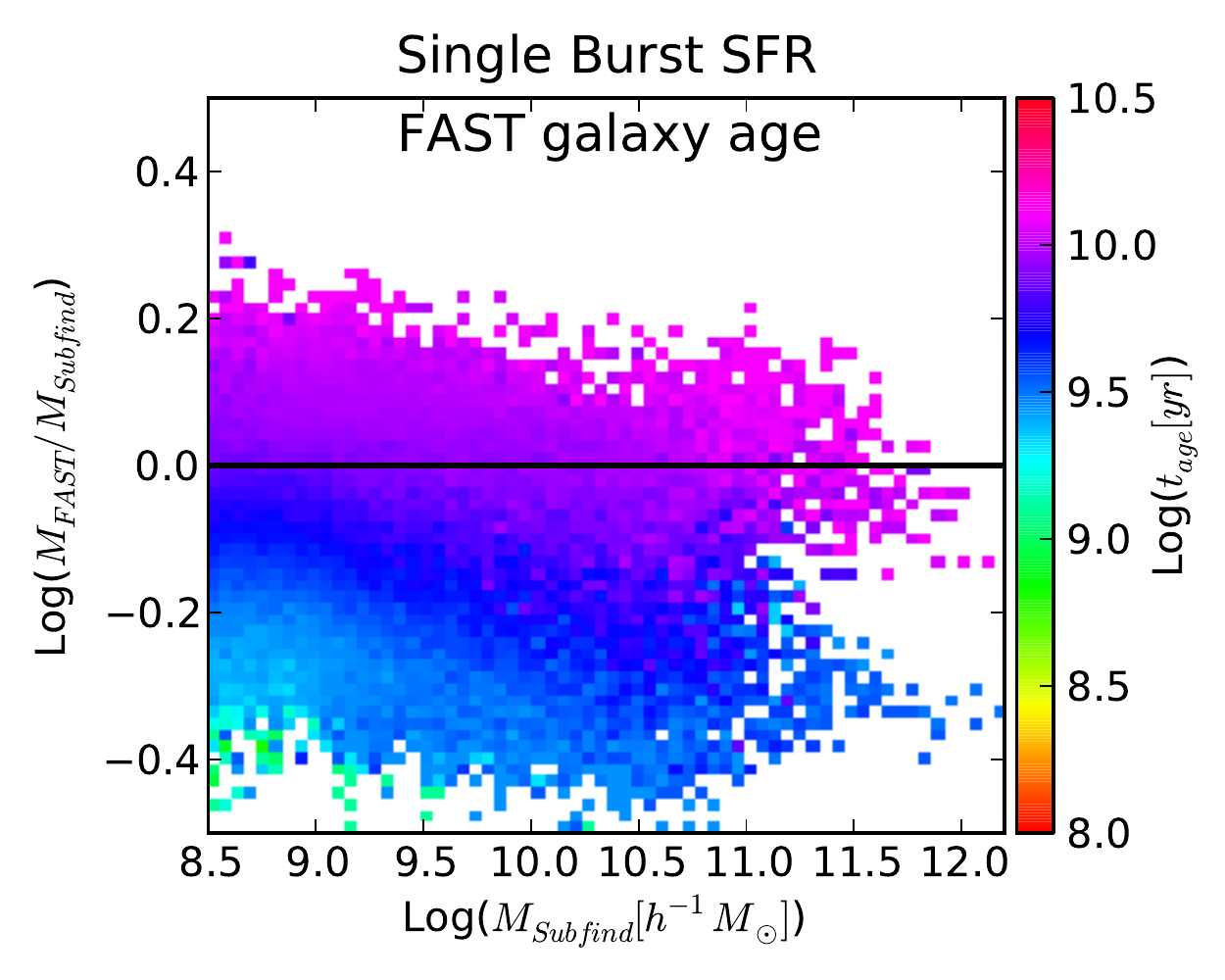}
\caption{Two dimensional histograms depicting the distribution of average metallicity (left) and average galaxy age (right) for the single burst star formation history.  The sparse metallicity template spacing is only partially responsible for the structure seen in the distribution of galaxies in Figure~\ref{fig:fast_alt_sfr}.  There is a large gradient of best fit ages, and galaxies that are assigned to be very young are also assigned masses that are too small.   }
\label{fig:fast_tru_t_and_z}
\end{figure*}

Figure~\ref{fig:fast_alt_sfr} shows the ratio of the {\small FAST} mass to the {\small SUBFIND} mass as a function of the {\small SUBFIND} mass for the delayed star formation history (left) and single burst star formation history (right).  The distribution of galaxies in the delayed star formation history space looks similar to what was found for the previously assumed exponential star formation history.  
By examining a number of ``best fit" star formation histories from the exponential and delayed models, we find that {\small FAST} prefers very similar (and young) star formation history in many cases, explaining why the distribution of derived galaxy masses is so similar.

The distribution of derived galaxy masses for the single burst star formation rate history is more complex.  
Figure~\ref{fig:fast_tru_t_and_z} show the average metallicity (left) and age (right).   
There are three metallicities assigned to galaxies, causing mass estimate continuity issues as discussed above.  
The metallicity alone cannot explain the large spread in the galaxy population at ${\rm M}_{{\rm SUBFIND}} \approx 10^9 {\rm M} _\odot$.  
The right panel of Figure~\ref{fig:fast_tru_t_and_z} shows that there is a large age gradient, where the location along this gradient determines a galaxy's {\small FAST} mass relative to the intrinsic {\small SUBFIND} mass.  
Galaxies that are assigned old ages (and long star formation timescales) yield larger derived mass estimates (owing to their larger mass-to-light ratios).  
The origin of the large age spread for the low-mass galaxy population likely arises from the diversity of recent star formation histories of these low mass systems, which can substantially impact the best fit SED that {\small FAST} selects.
It has been found, for instance, that when bursty systems are fit with single-component SFHs, their stellar masses can be significantly underestimated~\citep[e.g.,][]{Michalowski2012, Michalowski2014}.  
Regardless of the underlying cause, we conclude that when {\small FAST} assigns a very young age to a galaxy it is likely that the galaxy's mass will be underestimated.

\subsection{Holding the Metallicity Fixed}
\label{sec:fixed_metallicity}
\begin{figure*}
\includegraphics[width=3.45in]{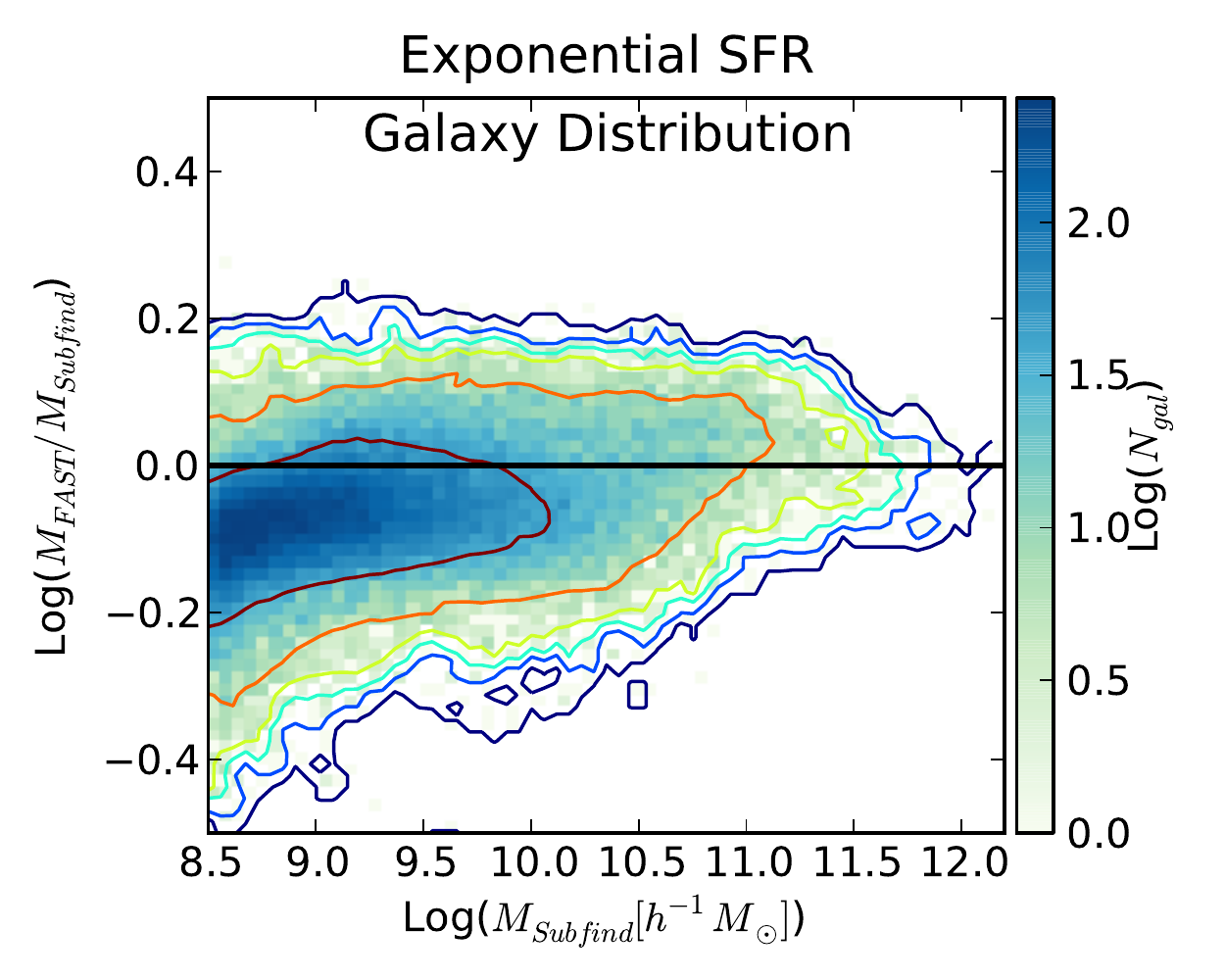}
\includegraphics[width=3.45in]{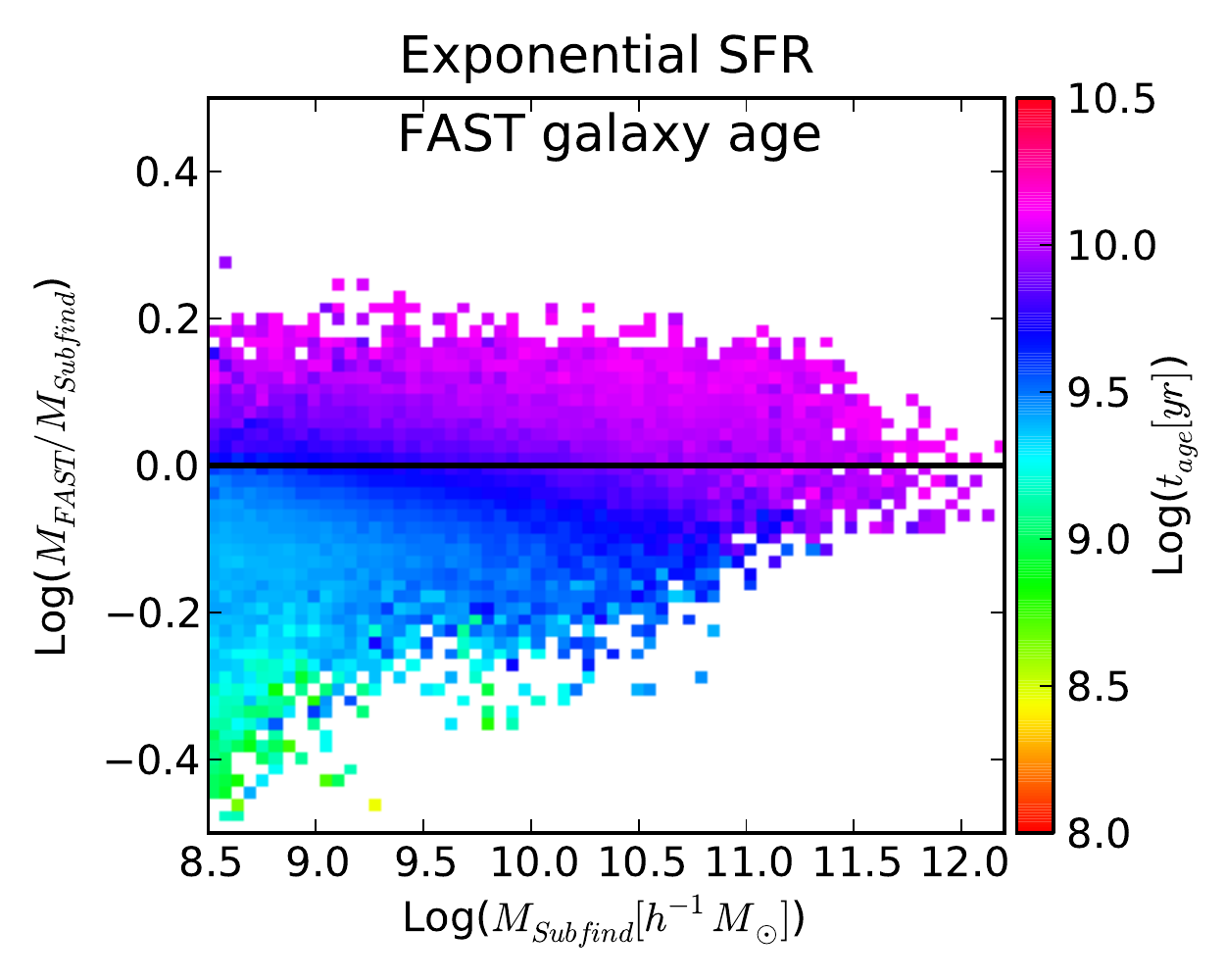}
\caption{Two-dimensional histograms depicting the distribution of derived {\small FAST} masses versus intrinsic masses for an exponential star formation history with a fixed metallicity assumption ($Z=0.02$) in {\small FAST}  (left) and average galaxy age based on the {\small FAST} best fit (right).  In the right plot, color denotes an average (mean) age for all the galaxies in a given pixel, with the legend indicating the quantitative values.  There is a clear relationship between the ``mass error" and the assumed galaxy age.}
\label{fig:fast_fixed_metallicity}
\end{figure*}

In the previous subsections, we have found that allowing {\small FAST} to pick a galaxy's metallicity during the SED fitting process can lead to mass-dependent errors in the derived galaxy masses.
Here we re-compute {\small FAST} masses while holding the assumed metallicity fixed at $Z=0.02$ (i.e. solar metallicity).
This is a commonly adopted assumption when deriving stellar masses for real observational data~\citep[e.g.,][]{Wuyts2012}.
The ratio of the {\small FAST} mass to the {\small SUBFIND} mass as a function of the {\small SUBFIND} mass for a fixed metallicity with an assumed exponential star formation rate history is shown in the left panel of Figure~\ref{fig:fast_fixed_metallicity}.
There is an offset between the derived mass and the intrinsic mass that evolves with intrinsic mass.
This trend is stronger than what was found when the metallicity was allowed to vary (Figure~\ref{fig:fast_masses}).
At the low mass end, the systematic error in the estimated mass is $\sim 0.1$ dex, while the scatter about this value is roughly $\sim 0.2$ dex.

Employing a fixed metallicity provides a more continuous and less broad distribution of mass estimates with respect to the intrinsic mass.  
In particular, holding the metallicity fixed in the {\small FAST} calculation reduces some of the variation in the mass ratios previously found in Section \ref{SEC:FAST}.
However, the estimated mass error shows a clear mass dependent offset from the intrinsic mass and there is still substantial scatter in the derived masses.  
The right hand panel of Figure~\ref{fig:fast_fixed_metallicity} shows the average age that {\small FAST} has assigned to each galaxy.
For all exercises up until this point, we have allowed the {\small FAST} age selection to be selected from $10^7 < t_{{\rm age}} < 10^{10.5}$ years.
Examining {\small FAST}'s selection for the derived galaxy age value, there is a very clear residual trend where {\small FAST} underestimates the mass of a galaxy when it assigns a young age (i.e. $t_{{\rm age}}<2 \times 10^9$ years) to that galaxy.
This is consistent with recent findings in the literature that young galaxy age assignment values can lead to underestimates in the stellar mass~\citep[e.g.,][]{Wuyts2012, Maraston2013, Price2014}.

\subsection{Restricting the Galaxy Age}
\begin{figure*}
\includegraphics[width=3.45in]{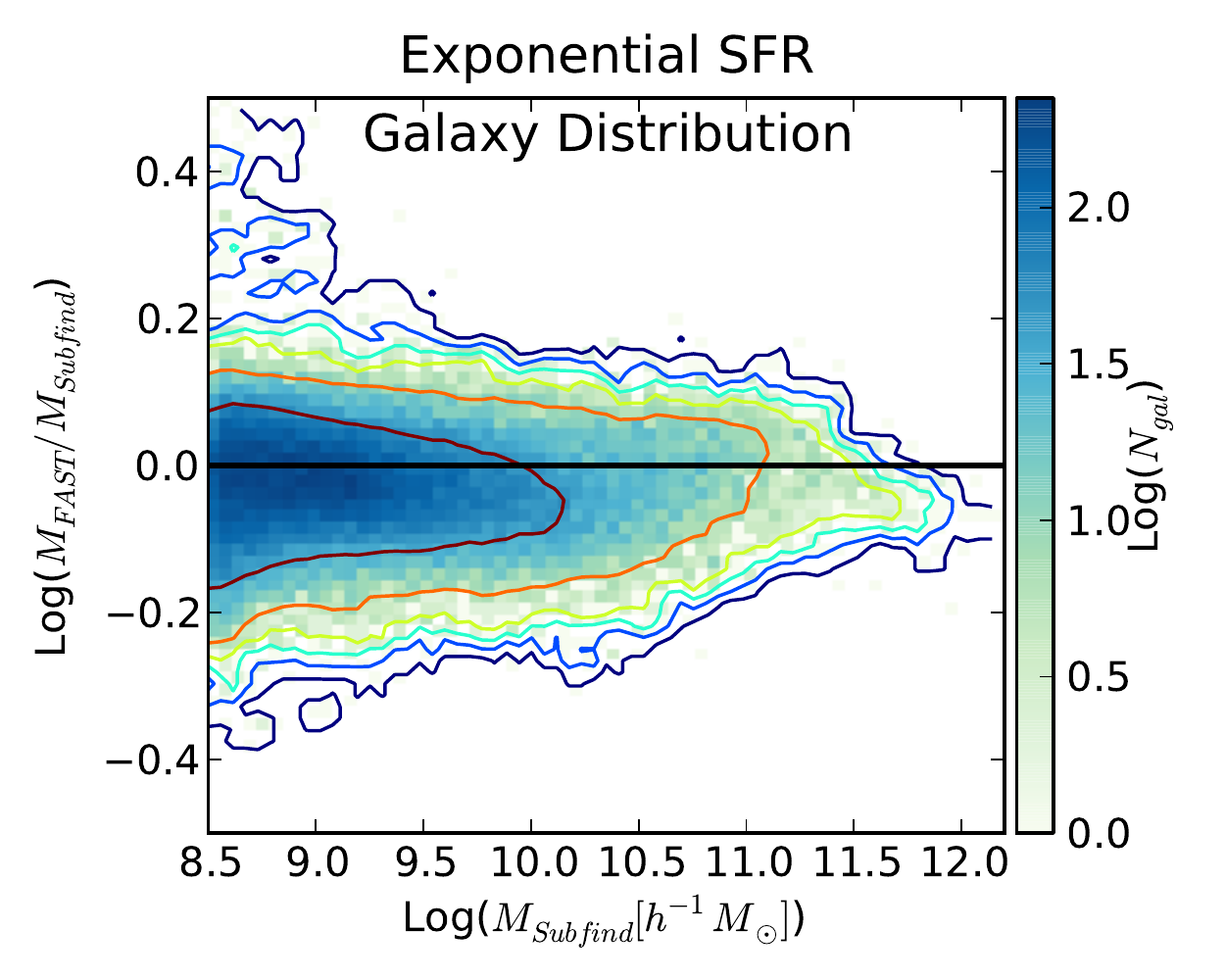}
\includegraphics[width=3.45in]{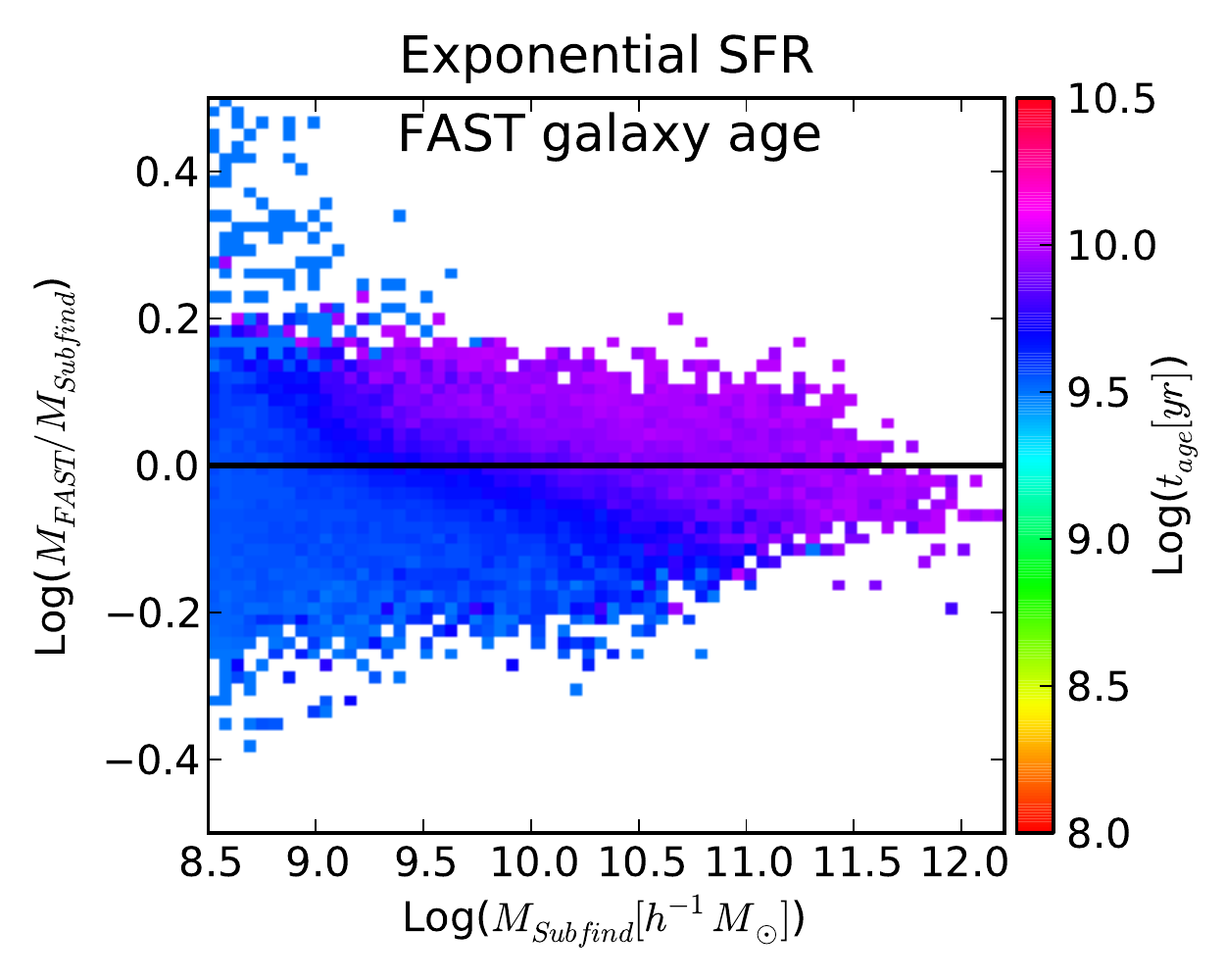}
\caption{Two-dimensional histograms depicting the distribution of derived {\small FAST} masses versus intrinsic masses for an exponential star formation history with a fixed metallicity assumption ($Z=0.02$) and restricted galaxy age range ($10^{9.5} < t_{{\rm age}} < 10^{10.5}$) in {\small FAST} (left) and average galaxy age based on the {\small FAST} best fit (right).  This case has the lowest systematic offset, mass dependence, and scatter in the estimated mass error of all of the examples presented in this paper.  In the right plot, color denotes an average (mean) age for all the galaxies in each pixel, with the legend indicating the quantitative values.  Restricting the galaxy ages reduces the overall spread in derived galaxy mass estimates while reducing the correlation between the ``mass error" and the assumed galaxy age in {\small FAST} -- though one does still remain.} 
\label{fig:fast_age_restriction}
\end{figure*}

In an attempt to improve the accuracy of the estimated masses, we restrict the galaxy age range from which {\small FAST} can select to $10^{9.5} < t_{{\rm age}} < 10^{10.5}$ years.  
The resulting ratio of the {\small FAST} mass to the {\small SUBFIND} mass as a function of the {\small SUBFIND} mass (still for a fixed metallicity with an assumed exponential star formation rate history) is shown in Figure~\ref{fig:fast_age_restriction}.
This case has the lowest systematic offset, mass dependence, and scatter in the estimated mass error of all of the cases presented in this paper.
The origin of this reduced bias is fairly clear:  a relatively low mass of recently formed stars (say, $10\%$ of a galaxy's total stellar mass) can contribute substantially to the integrated SED -- especially at short wavelengths -- forcing {\small FAST} to assume a relatively young age for galaxy as a whole when only a single exponential star formation rate history is assumed. 
By restricting the minimum galaxy age to be $10^{9.5}$ years, we effectively prevent low galaxy ages (with correspondingly low mass-to-light ratios) from being assigned.
We would therefore encourage caution when allowing a single galaxy age parameter to vary in an unconstrained
manner, when determining the stellar masses of real systems.
For redshift $z=0$ we find that a lower age bound of $t_{{\rm age}} =10^{9.5}$ years works well for our simulated galaxies.
However, this value will need to be lowered for high redshift galaxies. 
It is also possible that this issue could be solved by adopting multi-component star formation histories \citep[e.g., an exponential with superimposed bursts][]{Michalowski2012, Michalowski2014}.

\subsection{Spatially Resolved Derived Galaxy Properties}

\begin{figure*}
\includegraphics[width=2.15in]{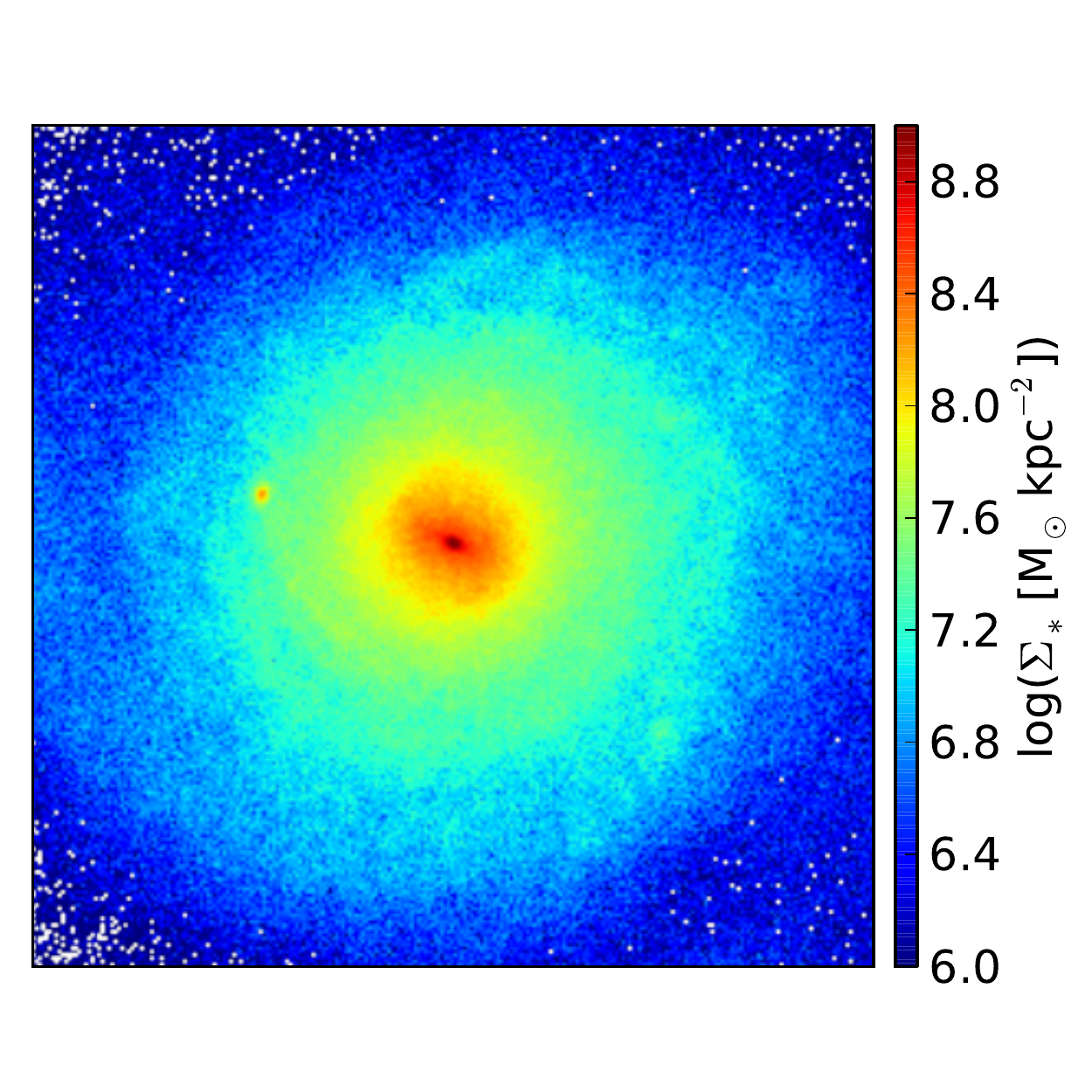}
\includegraphics[width=2.15in]{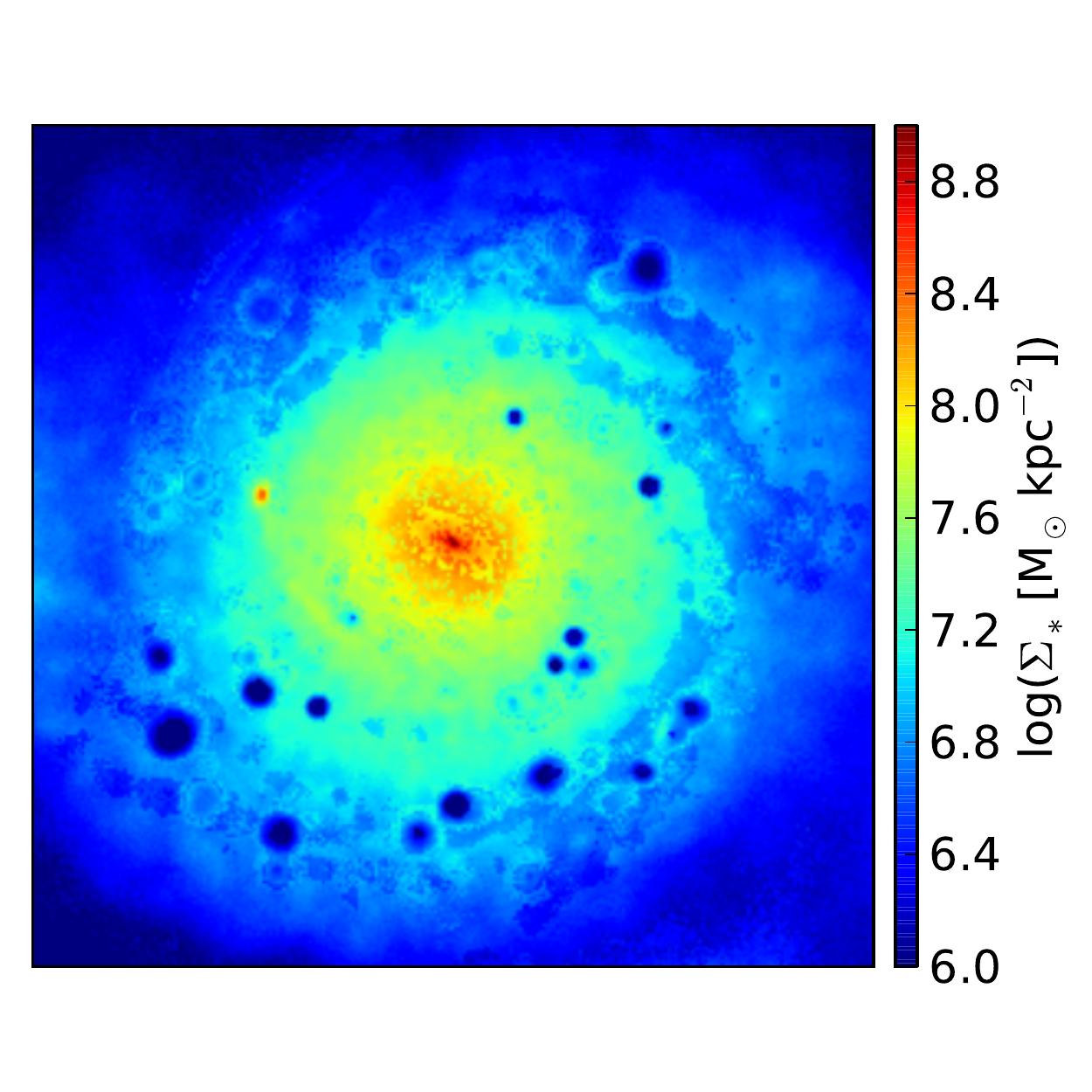}
\includegraphics[width=2.15in]{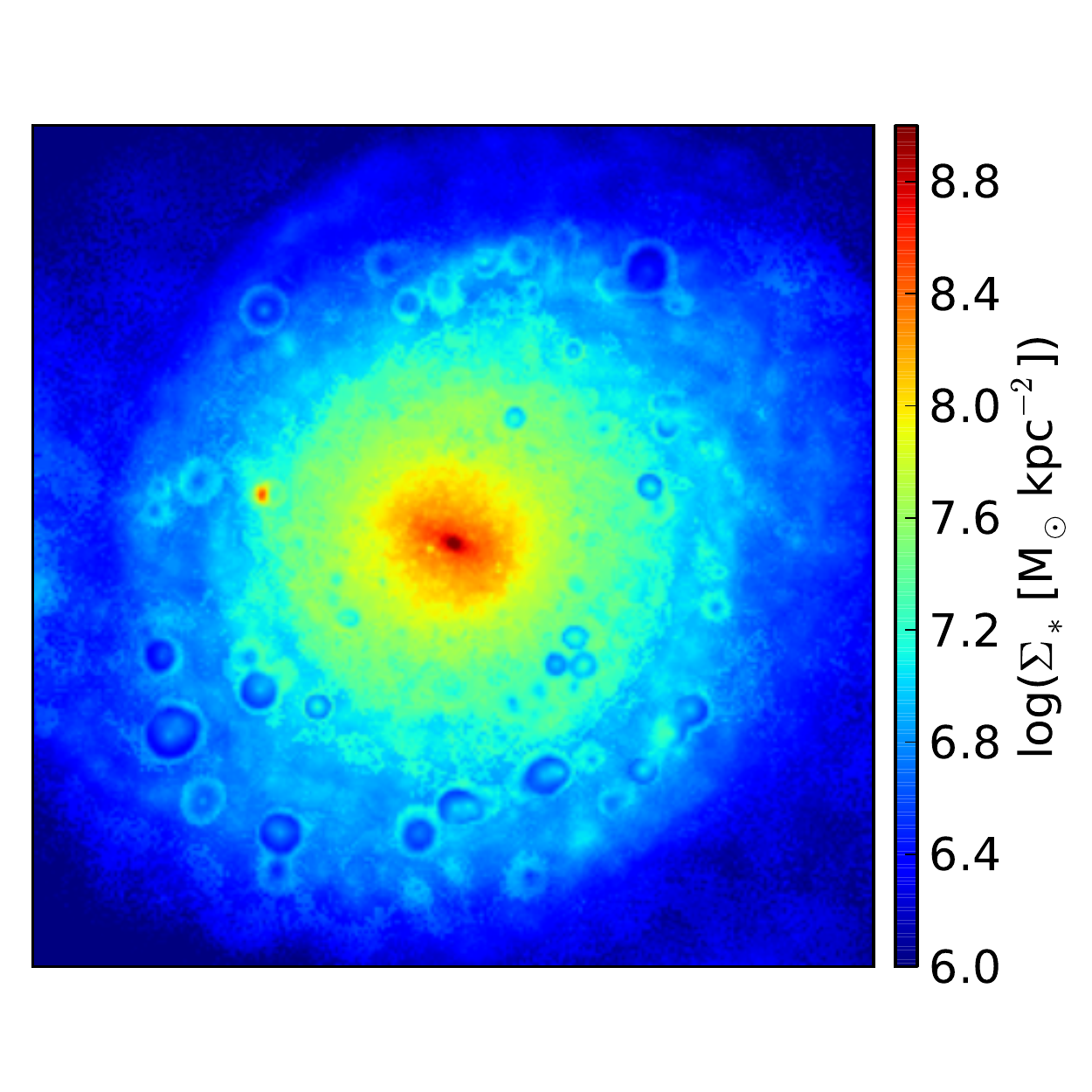}
\caption{Spatially resolved maps of the stellar mass surface density as determined from (left) the simulation directly, (center) {\small FAST} without any age or metallicity restrictions, (right) {\small FAST} with fixed metallicity and minimum age.} 
\label{fig:gal_1_fast}
\end{figure*}

In this subsection we apply the {\small FAST} broadband fitting procedure to spatially resolved galaxy images.
By performing the fitting procedure in each pixel we estimate the resolved surface density distribution which contains information about the dynamical state of the system as predicted by the simulation.
We first perform this analysis on the idealized image of a single redshift $z=0$ galaxy without applying any restrictions on the {\small FAST} best fit parameters.
The central panel of Figure~\ref{fig:gal_1_fast} shows the stellar mass surface density as derived from applying {\small FAST} to each pixel of the image.
For reference, the projected stellar surface density taken directly from the simulation for this galaxy is shown in the left panel of Figure~\ref{fig:gal_1_fast}.
This galaxy (ID=283832) possesses a radially declining surface density profile as seen in the left panel of Figure~\ref{fig:gal_1_fast}.
The stellar surface density profile is recovered in the {\small FAST} analysis, with some detailed differences.  
We find ``holes" in the {\small FAST} stellar mass surface density profiles which are present in pixels which have significant contributions from young star particles.
These pixels are assigned young ages owing to the influence of the luminous young stellar population despite the fact that the majority of the mass in these areas is characterized by an older average age.
The low mass-to-light ratios associated with the young stellar ages result in an underestimation of the stellar mass surface density in these regions.

We next restrict the {\small FAST} best fit parameters based on the minimum age and fixed metallicity criteria that we prescribe in the previous section.
The right panel of Figure~\ref{fig:gal_1_fast} shows the stellar mass surface density map derived from {\small FAST} when we restrict the minimum stellar age to $10^{9.5}$ years and fix the stellar metallicity to solar.
The derived stellar mass surface density map is qualitatively similar to that shown in the central panel of Figure~\ref{fig:gal_1_fast}, with the surface density holes being somewhat less pronounced.
Because there are no pixels in this particular galaxy image which are mass dominated by young stellar populations we find that applying the minimum stellar age objectively improves the derived stellar mass surface density map.
Applying a fixed metallicity impacts the derived stellar mass surface density maps at the $\sim$0.1 dex level, which is subdominant compared to the adopted age.

\section{Discussion}
\label{sec:Discussion}

The Illustris Simulation Observatory presented in this paper is a large mock galaxy image catalog constructed directly from a hydrodynamical simulation.  
In the past, large mock galaxy catalogs have been produced by applying semi-analytic models of galaxy formation to the output of dark matter only simulations~\citep[e.g.,][]{Kitzbichler2007, Henriques2012b, Overzier2013}.  
The approach used in the MRObs is to assign light to a bulge and disk component of a galaxy separately based on assumptions about their radial surface brightness profiles by using the bulge-to-total ratio, bulge half light radius, and disk scale height~\citep{Overzier2013}.  
These galactic light components are then included in images by assuming a sky position angle and inclination.
After projecting all galaxies from a conic observational volume onto an image, point spread functions taken from the appropriate instruments are applied to yield mock light cones that compare against real observations favorably~\citep{Overzier2013}.
While these models have the advantage that they can be run on very large simulated volumes -- allowing them to produce mock light-cones and mock survey data -- they can only indirectly take into account the role that galactic dynamics (e.g., galaxy bar formation, major/minor mergers, disk instabilities, etc.) play in morphological evolution.

In contrast, hydrodynamic simulations model gas processes and internal galactic dynamics self-consistently down to the gravitational softening limit.
The resulting sample of simulated galaxies can therefore be directly converted into synthetic images which encode information about the stellar age distribution, stellar metallicity distribution, and stellar spatial distribution.
No further assumptions need to be made in order to generate a galaxy image.  
The resulting light distribution from galaxies may therefore end up as disk-dominated, bulge-dominated, irregular, or highly asymmetric as dictated by the formation history and internal galactic dynamics of each system.
In this paper we have limited our focused to the construction of idealized images.
In a companion paper, Snyder et al., (in prep) outlines our steps for adding image realism to the idealized galaxy images presented in this paper.
Specifically, after creating the idealized images one can in post-processing rebin the images to an appropriate pixel scale, convolve with a point spread function, add background images or noise at a level consistent with a given instrument and exposure time.
Analysis of the idealized images presented in this paper after being put through such post-processing steps will be soon presented in a companion paper (Snyder et al., in prep).

A series of important limitations need to be considered in order to properly exploit the Illustris output. An incomplete list of these is as follows: 
\begin{enumerate}
\item The redshift $z=0$ gravitational softening length in the simulation is $\epsilon=0.5 \;h^{-1}$ kpc, making small systems and small-scale features within large systems poorly resolved.  
While we do observe barred galaxies formed during merger/flyby events, force softening can suppress disk instabilities and bar formation in non-interacting galaxies~\citep{Kaufmann2007}.
This can lead to a low fraction of barred systems, and potentially modulate the efficiency of stellar bulge formation (to the extent that bars help drive bulge formation).
\item The mass resolution of our models may impact results, particularly for low mass galaxies.
Given our gas/star particle mass resolution of $M_* \approx 1.3 \times 10^6 M_\odot$, young stellar populations will be stochastically sampled with a small number of particles.
This has been shown in the past to potentially bias non-parametric fitting parameters, such as the Gini coefficient~\citep{Lotz2008, SUNRISE2}.
\item Our current models make use of a pressurized equation of state to handle the multi-phase ISM in a sub-grid fashion~\citep{SH03}.  
Features that are expected to exist in a true multiphase ISM, such as spiral arm structure driven by pressure waves, will not be properly captured.
Some simulated galaxies -- such as the disk galaxies presented in Figure~\ref{fig:multi_angle_disks} -- do show signs of spiral features.  
However, this spiral arm structure is driven by swing amplification~\citep{Goldreich1965, Julian1966} of perturbations to the halo potential caused by the representation of dark matter by discrete simulation particles~\citep{Donghia2013}.
\item Our assumed physics (and specifically our feedback implementation) can influence galaxy morphology.
We find, for instance, a number of galaxies with enhanced rings of star formation in our simulation, which are likely caused by the interaction of star formation driven winds with the dense galactic ISM.
The presence of these rings and/or the specific feedback formulation used in our model may influence visual and automated galaxy morphology classification or galaxy size growth.  
\item Our mock images and SEDs do not contain consider detailed radiation transfer. 
The kilo parsec scale spatial resolution of the Illustris simulation is slightly too coarse to carry out self-consistent dust modeling.
We have applied simple empirical corrections based on the CF00 model to the galaxy SEDs.  
Analysis of any systems spatial light distributions and/or SED properties where dust is expected to play a major role should therefore use caution.
\end{enumerate}

Keeping in mind these caveats, we still expect the Illustris Simulation Observatory to be a useful and reliable resource in many situations, including: (i) addressing issues that cannot be accessed with real data (e.g. the viewing angle dependence of galaxy morphology and time evolution of galactic features), (ii) studying situations where the internal galaxy dynamics governing the spatial distribution of stars is of the essence (e.g. in the formation and evolution of stellar shells or the viewing angle dependence of bar classification, as explored briefly in this paper), and (iii) where the full hydro simulation approach of the Illustris simulation is likely robust (e.g. for the formation of bars via mergers/interactions, the observability timescales for galaxy pairs and tidal features, the puffing up of massive galaxies via dry mergers, the formation of disk galaxies via in-situ star formation, etc.).
Moreover, using an even-handed determination of galaxy morphology based on synthetic images will help us further identify the environments and mass ranges where our simulated galaxy populations are successfully or poorly replicating observations (Snyder et al., in prep).

We expect that the Illustris Simulation Observatory will be an evolving resource, not only for the current Illustris simulation, but also for next generation simulations with higher resolution, larger volume, and more sophisticated physics implementations.

\section{Conclusions}
\label{sec:Conclusions}
The main goal of this paper is the presentation of a large synthetic galaxy image catalog based on a high-resolution full-volume cosmological hydrodynamical simulation.
We described our method for building this image catalog, in particular, the assignment of stellar light to each star particle within a galaxy.
We have made $\sim$7,000 galaxy images in 36 broadbands plus integrated SEDs for $\sim$40,000 systems.
We have demonstrated that these broadband images can be used to generate high quality galaxy images.
Our intention when constructing this mock galaxy image database is that it may serve as a useful tool for observers in the interpretation of extragalactic data.  
In principle, the FITS file format of our mock galaxy images will make it easy to process these images with existing image analysis tools, similar to what was done in~\citet{Overzier2013}. 
This should facilitate galaxy selection from our catalog using selection criteria matched to observational targets.  
It also implies that spatially resolved galaxy properties (e.g., half-light-radii, bulge-to-disk ratios, etc.) can be derived using un-modified photometric image decomposition or fitting techniques.

As a first application, we showed that the SED fitting code {\small FAST} is able to derive the ``correct'' galaxy mass within a factor of 2 for $>$95\% of galaxies in our idealized sample.
We also uncovered some residual trends in the stellar mass estimates and argued that sparsely populated stellar population synthesis template metallicity values and low galaxy age assignments are responsible for the majority of the error in the photometrically derived galaxy mass values.  
Although our idealized comparison cannot address systematic uncertainties such as the dependence of the mass-to-light ratio for a galaxy population on the shape of the IMF, we have used our mock image pipeline to show that, overall, {\small FAST} is able to derive reasonably accurate galaxy stellar masses for complex star formation histories using a relatively simple assumption for those histories.
We concluded that, for redshift $z=0$, employing a lower age bound of $t_{{\rm age}} =10^{9.5}$ years with a fixed solar metallicity produces the smallest errors in the stellar mass determination.

Several additional applications of the mock image pipeline are currently under way.  
We are undertaking automated non-parametric measurements of galaxy morphology for direct comparison with survey data (Snyder et al., in prep). 
Among other goals, this will allow us to quantitatively assess the extent to which the morphological mix of the simulated galaxies match reality.
Visual classification of the galaxy population for galaxies with $>10^5$ stellar particles is also planned.  
This selection criterion results in a sample of $\sim$1000 galaxies with stellar masses $M_* \gtrsim 5 \times 10^{10} M_\odot$, which we have verified contains a mix of both disks and elliptical systems.
Visual classification is an alternative to the automated non-parametric classifications, and enables us to directly address whether, for example, the disk-to-elliptical fraction for simulated galaxies is comparable to the values measured from SDSS in {\small GalaxyZoo}~\citep{Lintott2008}.

By making a fair comparison between cosmological simulations and large galaxy surveys, we can begin to connect theoretical ideas on the origins of galaxy structure with actual observations in ever more accurate ways.
The role that simulations play in shaping our understanding of galaxy formation will continue to increase as the models become increasingly more detailed and physically accurate.
Comparing the detailed structure and properties of individual galaxies -- within the context of their larger populations -- is an important point of overlap between theory and observation, and mock catalogs based on realistic, synthetic observations of simulated systems are a fundamental requisite for maximizing the insight gained from that overlap.

\section*{Acknowledgements} 
We thank Dave Sanders and Arjun Dey for helpful suggestions on this work.  
We thank Patrik Jonsson for his significant efforts to write, develop, and support the {\small SUNRISE} code.  
GS acknowledges support from the HST grants program, number HST- AR-12856.01-A. Support for program \#12856 (PI Lotz) was provided by NASA through a grant from the Space Telescope Science Institute, which is operated by the Association of Universities for Research in Astronomy, Inc., under NASA contract NAS 5-26555. 
CCH is grateful to the Klaus Tschira Foundation for financial support and acknowledges the hospitality of the Aspen Center for Physics, which is supported by the National Science Foundation Grant No. PHY-1066293.
VS acknowledges support by the DFG Research Centre SFB-881 The Milky Way System through project A1, and by the European Research Council under ERC-StG EXAGAL-308037. 
LH acknowledges support from NASA grant NNX12AC67G and NSF grant AST-1312095.


\begin{thebibliography}{136}
\expandafter\ifx\csname natexlab\endcsname\relax\def\natexlab#1{#1}\fi

\bibitem[{{Abadi} {et~al.}(2003){Abadi}, {Navarro}, {Steinmetz}, \&
  {Eke}}]{Abadi2003}
{Abadi}, M.~G., {Navarro}, J.~F., {Steinmetz}, M., \& {Eke}, V.~R. 2003, \apj,
  597, 21

\bibitem[{{Agertz} {et~al.}(2011){Agertz}, {Teyssier}, \& {Moore}}]{Agertz2011}
{Agertz}, O., {Teyssier}, R., \& {Moore}, B. 2011, \mnras, 410, 1391

\bibitem[{{Aguirre} {et~al.}(2001){Aguirre}, {Hernquist}, {Schaye}, {Katz},
  {Weinberg}, \& {Gardner}}]{Aguirre01}
{Aguirre}, A., {Hernquist}, L., {Schaye}, J., {et~al.} 2001, \apj, 561, 521

\bibitem[{{Banerji} {et~al.}(2013){Banerji}, {Glazebrook}, {Blake}, {Brough},
  {Colless}, {Contreras}, {Couch}, {Croton}, {Croom}, {Davis}, {Drinkwater},
  {Forster}, {Gilbank}, {Gladders}, {Jelliffe}, {Jurek}, {Li}, {Madore},
  {Martin}, {Pimbblet}, {Poole}, {Pracy}, {Sharp}, {Wisnioski}, {Woods},
  {Wyder}, \& {Yee}}]{Banerji2013}
{Banerji}, M., {Glazebrook}, K., {Blake}, C., {et~al.} 2013, \mnras, 431, 2209

\bibitem[{{Bernyk} {et~al.}(2014){Bernyk}, {Croton}, {Tonini}, {Hodkinson},
  {Hassan}, {Garel}, {Duffy}, {Mutch}, \& {Poole}}]{Bernyk2014}
{Bernyk}, M., {Croton}, D.~J., {Tonini}, C., {et~al.} 2014, ArXiv e-prints

\bibitem[{{Binette} {et~al.}(1985){Binette}, {Dopita}, \&
  {Tuohy}}]{Binette1985}
{Binette}, L., {Dopita}, M.~A., \& {Tuohy}, I.~R. 1985, \apj, 297, 476

\bibitem[{{Boylan-Kolchin} {et~al.}(2009){Boylan-Kolchin}, {Springel}, {White},
  {Jenkins}, \& {Lemson}}]{Millennium2}
{Boylan-Kolchin}, M., {Springel}, V., {White}, S.~D.~M., {Jenkins}, A., \&
  {Lemson}, G. 2009, \mnras, 398, 1150

\bibitem[{{Bruzual} \& {Charlot}(2003)}]{BC03}
{Bruzual}, G., \& {Charlot}, S. 2003, \mnras, 344, 1000

\bibitem[{{Bruzual A.}(1983)}]{Bruzual1983}
{Bruzual A.}, G. 1983, \apj, 273, 105

\bibitem[{{Bruzual A.} \& {Charlot}(1993)}]{Bruzual1993}
{Bruzual A.}, G., \& {Charlot}, S. 1993, \apj, 405, 538

\bibitem[{{Buzzoni}(1989)}]{Buzzoni1989}
{Buzzoni}, A. 1989, \apjs, 71, 817

\bibitem[{{Canalizo} {et~al.}(2007){Canalizo}, {Bennert}, {Jungwiert},
  {Stockton}, {Schweizer}, {Lacy}, \& {Peng}}]{Canalizo2007}
{Canalizo}, G., {Bennert}, N., {Jungwiert}, B., {et~al.} 2007, \apj, 669, 801

\bibitem[{{Carter}(1979)}]{Carter1979}
{Carter}, D. 1979, \mnras, 186, 897

\bibitem[{{Cen} \& {Fang}(2006)}]{Cen06}
{Cen}, R., \& {Fang}, T. 2006, \apj, 650, 573

\bibitem[{{Cen} {et~al.}(1994){Cen}, {Miralda-Escud{\'e}}, {Ostriker}, \&
  {Rauch}}]{Cen94}
{Cen}, R., {Miralda-Escud{\'e}}, J., {Ostriker}, J.~P., \& {Rauch}, M. 1994,
  \apjl, 437, L9

\bibitem[{{Charlot} \& {Fall}(2000)}]{CF00}
{Charlot}, S., \& {Fall}, S.~M. 2000, \apj, 539, 718

\bibitem[{{Conroy} \& {Gunn}(2010)}]{Conroy2010b}
{Conroy}, C., \& {Gunn}, J.~E. 2010, \apj, 712, 833

\bibitem[{{Conroy} \& {Wechsler}(2009)}]{Conroy2009}
{Conroy}, C., \& {Wechsler}, R.~H. 2009, \apj, 696, 620

\bibitem[{{Conroy} {et~al.}(2010){Conroy}, {White}, \& {Gunn}}]{Conroy2010a}
{Conroy}, C., {White}, M., \& {Gunn}, J.~E. 2010, \apj, 708, 58

\bibitem[{{Cooper} {et~al.}(2011){Cooper}, {Mart{\'{\i}}nez-Delgado}, {Helly},
  {Frenk}, {Cole}, {Crawford}, {Zibetti}, {Carballo-Bello}, \&
  {GaBany}}]{Cooper2011}
{Cooper}, A.~P., {Mart{\'{\i}}nez-Delgado}, D., {Helly}, J., {et~al.} 2011,
  \apjl, 743, L21

\bibitem[{{Cox} {et~al.}(2006){Cox}, {Jonsson}, {Primack}, \&
  {Somerville}}]{Cox2006}
{Cox}, T.~J., {Jonsson}, P., {Primack}, J.~R., \& {Somerville}, R.~S. 2006,
  \mnras, 373, 1013

\bibitem[{{Crain} {et~al.}(2009){Crain}, {Theuns}, {Dalla Vecchia}, {Eke},
  {Frenk}, {Jenkins}, {Kay}, {Peacock}, {Pearce}, {Schaye}, {Springel},
  {Thomas}, {White}, \& {Wiersma}}]{Crain2009}
{Crain}, R.~A., {Theuns}, T., {Dalla Vecchia}, C., {et~al.} 2009, \mnras, 399,
  1773

\bibitem[{{Croft} {et~al.}(2009){Croft}, {Di Matteo}, {Springel}, \&
  {Hernquist}}]{Croft2009}
{Croft}, R.~A.~C., {Di Matteo}, T., {Springel}, V., \& {Hernquist}, L. 2009,
  \mnras, 400, 43

\bibitem[{{D'Onghia} {et~al.}(2013){D'Onghia}, {Vogelsberger}, \&
  {Hernquist}}]{Donghia2013}
{D'Onghia}, E., {Vogelsberger}, M., \& {Hernquist}, L. 2013, \apj, 766, 34

\bibitem[{{Dopita} {et~al.}(2005){Dopita}, {Groves}, {Fischera}, {Sutherland},
  {Tuffs}, {Popescu}, {Kewley}, {Reuland}, \& {Leitherer}}]{Dopita2005}
{Dopita}, M.~A., {Groves}, B.~A., {Fischera}, J., {et~al.} 2005, \apj, 619, 755

\bibitem[{{Dopita} {et~al.}(2006{\natexlab{a}}){Dopita}, {Fischera},
  {Sutherland}, {Kewley}, {Tuffs}, {Popescu}, {van Breugel}, {Groves}, \&
  {Leitherer}}]{Dopita2006a}
{Dopita}, M.~A., {Fischera}, J., {Sutherland}, R.~S., {et~al.}
  2006{\natexlab{a}}, \apj, 647, 244

\bibitem[{{Dopita} {et~al.}(2006{\natexlab{b}}){Dopita}, {Fischera},
  {Sutherland}, {Kewley}, {Leitherer}, {Tuffs}, {Popescu}, {van Breugel}, \&
  {Groves}}]{Dopita2006b}
---. 2006{\natexlab{b}}, \apjs, 167, 177

\bibitem[{{Dupraz} \& {Combes}(1986)}]{Dupraz1986}
{Dupraz}, C., \& {Combes}, F. 1986, \aap, 166, 53

\bibitem[{{Faucher-Gigu{\`e}re} {et~al.}(2009){Faucher-Gigu{\`e}re}, {Lidz},
  {Zaldarriaga}, \& {Hernquist}}]{Faucher2009}
{Faucher-Gigu{\`e}re}, C.-A., {Lidz}, A., {Zaldarriaga}, M., \& {Hernquist}, L.
  2009, \apj, 703, 1416

\bibitem[{{Faucher-Gigu{\`e}re} {et~al.}(2008){Faucher-Gigu{\`e}re},
  {Prochaska}, {Lidz}, {Hernquist}, \& {Zaldarriaga}}]{Faucher2008}
{Faucher-Gigu{\`e}re}, C.-A., {Prochaska}, J.~X., {Lidz}, A., {Hernquist}, L.,
  \& {Zaldarriaga}, M. 2008, \apj, 681, 831

\bibitem[{{Fosalba} {et~al.}(2008){Fosalba}, {Gazta{\~n}aga}, {Castander}, \&
  {Manera}}]{Fosalba2008}
{Fosalba}, P., {Gazta{\~n}aga}, E., {Castander}, F.~J., \& {Manera}, M. 2008,
  \mnras, 391, 435

\bibitem[{{Gallazzi} \& {Bell}(2009)}]{Gallazzi2009}
{Gallazzi}, A., \& {Bell}, E.~F. 2009, \apjs, 185, 253

\bibitem[{{Genel} {et~al.}(2014){Genel}, {Vogelsberger}, {Springel}, {Sijacki},
  {Nelson}, {Snyder}, {Rodriguez-Gomez}, {Torrey}, \& {Hernquist}}]{Genel2014}
{Genel}, S., {Vogelsberger}, M., {Springel}, V., {et~al.} 2014, ArXiv e-prints
  1405.3749

\bibitem[{{Goldreich} \& {Lynden-Bell}(1965)}]{Goldreich1965}
{Goldreich}, P., \& {Lynden-Bell}, D. 1965, \mnras, 130, 125

\bibitem[{{Governato} {et~al.}(2004){Governato}, {Mayer}, {Wadsley}, {Gardner},
  {Willman}, {Hayashi}, {Quinn}, {Stadel}, \& {Lake}}]{Governato2004}
{Governato}, F., {Mayer}, L., {Wadsley}, J., {et~al.} 2004, \apj, 607, 688

\bibitem[{{Groves} {et~al.}(2008){Groves}, {Dopita}, {Sutherland}, {Kewley},
  {Fischera}, {Leitherer}, {Brandl}, \& {van Breugel}}]{Groves2008}
{Groves}, B., {Dopita}, M.~A., {Sutherland}, R.~S., {et~al.} 2008, \apjs, 176,
  438

\bibitem[{{Groves} {et~al.}(2004{\natexlab{a}}){Groves}, {Dopita}, \&
  {Sutherland}}]{Groves2004a}
{Groves}, B.~A., {Dopita}, M.~A., \& {Sutherland}, R.~S. 2004{\natexlab{a}},
  \apjs, 153, 9

\bibitem[{{Groves} {et~al.}(2004{\natexlab{b}}){Groves}, {Dopita}, \&
  {Sutherland}}]{Groves2004b}
---. 2004{\natexlab{b}}, \apjs, 153, 75

\bibitem[{{Guedes} {et~al.}(2011){Guedes}, {Callegari}, {Madau}, \&
  {Mayer}}]{Guedes2011}
{Guedes}, J., {Callegari}, S., {Madau}, P., \& {Mayer}, L. 2011, \apj, 742, 76

\bibitem[{{Hayward} {et~al.}(2013{\natexlab{a}}){Hayward}, {Behroozi},
  {Somerville}, {Primack}, {Moreno}, \& {Wechsler}}]{Hayward2013a}
{Hayward}, C.~C., {Behroozi}, P.~S., {Somerville}, R.~S., {et~al.}
  2013{\natexlab{a}}, \mnras

\bibitem[{{Hayward} {et~al.}(2012){Hayward}, {Jonsson}, {Kere{\v s}},
  {Magnelli}, {Hernquist}, \& {Cox}}]{Hayward2012}
{Hayward}, C.~C., {Jonsson}, P., {Kere{\v s}}, D., {et~al.} 2012, \mnras, 424,
  951

\bibitem[{{Hayward} {et~al.}(2011){Hayward}, {Kere{\v s}}, {Jonsson},
  {Narayanan}, {Cox}, \& {Hernquist}}]{Hayward2011}
{Hayward}, C.~C., {Kere{\v s}}, D., {Jonsson}, P., {et~al.} 2011, \apj, 743,
  159

\bibitem[{{Hayward} {et~al.}(2013{\natexlab{b}}){Hayward}, {Narayanan},
  {Kere{\v s}}, {Jonsson}, {Hopkins}, {Cox}, \& {Hernquist}}]{Hayward2013b}
{Hayward}, C.~C., {Narayanan}, D., {Kere{\v s}}, D., {et~al.}
  2013{\natexlab{b}}, \mnras, 428, 2529

\bibitem[{{Hayward} \& {Smith}(2014)}]{Hayward2014c}
{Hayward}, C.~C., \& {Smith}, D.~J.~B. 2014, arXiv:1409.6332

\bibitem[{{Hayward} {et~al.}(2014{\natexlab{a}}){Hayward}, {Torrey},
  {Springel}, {Hernquist}, \& {Vogelsberger}}]{Hayward2014b}
{Hayward}, C.~C., {Torrey}, P., {Springel}, V., {Hernquist}, L., \&
  {Vogelsberger}, M. 2014{\natexlab{a}}, \mnras, 442, 1992

\bibitem[{{Hayward} {et~al.}(2014{\natexlab{b}}){Hayward}, {Lanz}, {Ashby},
  {Fazio}, {Hernquist}, {Mart{\'{\i}}nez-Galarza}, {Noeske}, {Smith}, {Wuyts},
  \& {Zezas}}]{Hayward2014a}
{Hayward}, C.~C., {Lanz}, L., {Ashby}, M.~L.~N., {et~al.} 2014{\natexlab{b}},
  \mnras, 445, 1598

\bibitem[{{Henriques} {et~al.}(2011){Henriques}, {Maraston}, {Monaco},
  {Fontanot}, {Menci}, {De Lucia}, \& {Tonini}}]{Henriques2011}
{Henriques}, B., {Maraston}, C., {Monaco}, P., {et~al.} 2011, \mnras, 415, 3571

\bibitem[{{Henriques} {et~al.}(2012){Henriques}, {White}, {Lemson}, {Thomas},
  {Guo}, {Marleau}, \& {Overzier}}]{Henriques2012b}
{Henriques}, B.~M.~B., {White}, S.~D.~M., {Lemson}, G., {et~al.} 2012, \mnras,
  421, 2904

\bibitem[{{Hernquist} {et~al.}(1996){Hernquist}, {Katz}, {Weinberg}, \&
  {Miralda-Escud{\'e}}}]{Hernquist96}
{Hernquist}, L., {Katz}, N., {Weinberg}, D.~H., \& {Miralda-Escud{\'e}}, J.
  1996, \apjl, 457, L51

\bibitem[{{Hernquist} \& {Quinn}(1988)}]{Hernquist1988}
{Hernquist}, L., \& {Quinn}, P.~J. 1988, \apj, 331, 682

\bibitem[{{Hernquist} \& {Spergel}(1992)}]{Hernquist1992}
{Hernquist}, L., \& {Spergel}, D.~N. 1992, \apjl, 399, L117

\bibitem[{{Hopkins} {et~al.}(2005){Hopkins}, {Hernquist}, {Martini}, {Cox},
  {Robertson}, {Di Matteo}, \& {Springel}}]{Hopkins2005}
{Hopkins}, P.~F., {Hernquist}, L., {Martini}, P., {et~al.} 2005, \apjl, 625,
  L71

\bibitem[{{Jonsson}(2006)}]{SUNRISE}
{Jonsson}, P. 2006, \mnras, 372, 2

\bibitem[{{Jonsson} {et~al.}(2010){Jonsson}, {Groves}, \& {Cox}}]{SUNRISE2}
{Jonsson}, P., {Groves}, B.~A., \& {Cox}, T.~J. 2010, \mnras, 403, 17

\bibitem[{{Julian} \& {Toomre}(1966)}]{Julian1966}
{Julian}, W.~H., \& {Toomre}, A. 1966, \apj, 146, 810

\bibitem[{{Katz} {et~al.}(1992){Katz}, {Hernquist}, \& {Weinberg}}]{Katz1992}
{Katz}, N., {Hernquist}, L., \& {Weinberg}, D.~H. 1992, \apjl, 399, L109

\bibitem[{{Katz} {et~al.}(1996){Katz}, {Weinberg}, \&
  {Hernquist}}]{KatzCooling}
{Katz}, N., {Weinberg}, D.~H., \& {Hernquist}, L. 1996, \apjs, 105, 19

\bibitem[{{Kaufmann} {et~al.}(2007){Kaufmann}, {Mayer}, {Wadsley}, {Stadel}, \&
  {Moore}}]{Kaufmann2007}
{Kaufmann}, T., {Mayer}, L., {Wadsley}, J., {Stadel}, J., \& {Moore}, B. 2007,
  \mnras, 375, 53

\bibitem[{{Kere{\v s}} {et~al.}(2005){Kere{\v s}}, {Katz}, {Weinberg}, \&
  {Dav{\'e}}}]{Keres05}
{Kere{\v s}}, D., {Katz}, N., {Weinberg}, D.~H., \& {Dav{\'e}}, R. 2005,
  \mnras, 363, 2

\bibitem[{{Kere{\v s}} {et~al.}(2012){Kere{\v s}}, {Vogelsberger}, {Sijacki},
  {Springel}, \& {Hernquist}}]{Keres2012}
{Kere{\v s}}, D., {Vogelsberger}, M., {Sijacki}, D., {Springel}, V., \&
  {Hernquist}, L. 2012, \mnras, 425, 2027

\bibitem[{{Kitzbichler} \& {White}(2007)}]{Kitzbichler2007}
{Kitzbichler}, M.~G., \& {White}, S.~D.~M. 2007, \mnras, 376, 2

\bibitem[{{Klypin} {et~al.}(2011){Klypin}, {Trujillo-Gomez}, \&
  {Primack}}]{Bolshoi}
{Klypin}, A.~A., {Trujillo-Gomez}, S., \& {Primack}, J. 2011, \apj, 740, 102

\bibitem[{{Kojima} \& {Noguchi}(1997)}]{Kojima1997}
{Kojima}, M., \& {Noguchi}, M. 1997, \apj, 481, 132

\bibitem[{{Kriek} {et~al.}(2009){Kriek}, {van Dokkum}, {Labb{\'e}}, {Franx},
  {Illingworth}, {Marchesini}, \& {Quadri}}]{FAST}
{Kriek}, M., {van Dokkum}, P.~G., {Labb{\'e}}, I., {et~al.} 2009, \apj, 700,
  221

\bibitem[{{Lanz} {et~al.}(2014){Lanz}, {Hayward}, {Zezas}, {Smith}, {Ashby},
  {Brassington}, {Fazio}, \& {Hernquist}}]{Lanz2014}
{Lanz}, L., {Hayward}, C.~C., {Zezas}, A., {et~al.} 2014, \apj, 785, 39

\bibitem[{{Leitherer} {et~al.}(2010){Leitherer}, {Ortiz Ot{\'a}lvaro},
  {Bresolin}, {Kudritzki}, {Lo Faro}, {Pauldrach}, {Pettini}, \&
  {Rix}}]{SB99_3}
{Leitherer}, C., {Ortiz Ot{\'a}lvaro}, P.~A., {Bresolin}, F., {et~al.} 2010,
  \apjs, 189, 309

\bibitem[{{Leitherer} {et~al.}(1999){Leitherer}, {Schaerer}, {Goldader},
  {Gonz{\'a}lez Delgado}, {Robert}, {Kune}, {de Mello}, {Devost}, \&
  {Heckman}}]{SB99}
{Leitherer}, C., {Schaerer}, D., {Goldader}, J.~D., {et~al.} 1999, \apjs, 123,
  3

\bibitem[{{Lemson} \& {Springel}(2006)}]{Lemson2006}
{Lemson}, G., \& {Springel}, V. 2006, in Astronomical Society of the Pacific
  Conference Series, Vol. 351, Astronomical Data Analysis Software and Systems
  XV, ed. C.~{Gabriel}, C.~{Arviset}, D.~{Ponz}, \& S.~{Enrique}, 212

\bibitem[{{Lintott} {et~al.}(2008){Lintott}, {Schawinski}, {Slosar}, {Land},
  {Bamford}, {Thomas}, {Raddick}, {Nichol}, {Szalay}, {Andreescu}, {Murray}, \&
  {Vandenberg}}]{Lintott2008}
{Lintott}, C.~J., {Schawinski}, K., {Slosar}, A., {et~al.} 2008, \mnras, 389,
  1179

\bibitem[{{Lotz} {et~al.}(2008){Lotz}, {Jonsson}, {Cox}, \&
  {Primack}}]{Lotz2008}
{Lotz}, J.~M., {Jonsson}, P., {Cox}, T.~J., \& {Primack}, J.~R. 2008, \mnras,
  391, 1137

\bibitem[{{Lupton} {et~al.}(2004){Lupton}, {Blanton}, {Fekete}, {Hogg},
  {O'Mullane}, {Szalay}, \& {Wherry}}]{Lupton2004}
{Lupton}, R., {Blanton}, M.~R., {Fekete}, G., {et~al.} 2004, \pasp, 116, 133

\bibitem[{{Maraston}(1998)}]{Maraston1998}
{Maraston}, C. 1998, \mnras, 300, 872

\bibitem[{{Maraston}(2005)}]{Maraston2005}
---. 2005, \mnras, 362, 799

\bibitem[{{Maraston} {et~al.}(2013){Maraston}, {Pforr}, {Henriques}, {Thomas},
  {Wake}, {Brownstein}, {Capozzi}, {Tinker}, {Bundy}, {Skibba}, {Beifiori},
  {Nichol}, {Edmondson}, {Schneider}, {Chen}, {Masters}, {Steele}, {Bolton},
  {York}, {Weaver}, {Higgs}, {Bizyaev}, {Brewington}, {Malanushenko},
  {Malanushenko}, {Snedden}, {Oravetz}, {Pan}, {Shelden}, \&
  {Simmons}}]{Maraston2013}
{Maraston}, C., {Pforr}, J., {Henriques}, B.~M., {et~al.} 2013, \mnras, 435,
  2764

\bibitem[{{Marinacci} {et~al.}(2014){Marinacci}, {Pakmor}, \&
  {Springel}}]{Marinacci2014}
{Marinacci}, F., {Pakmor}, R., \& {Springel}, V. 2014, \mnras, 437, 1750

\bibitem[{{McQuinn} {et~al.}(2009){McQuinn}, {Lidz}, {Zaldarriaga},
  {Hernquist}, {Hopkins}, {Dutta}, \& {Faucher-Gigu{\`e}re}}]{McQuinn2009}
{McQuinn}, M., {Lidz}, A., {Zaldarriaga}, M., {et~al.} 2009, \apj, 694, 842

\bibitem[{{Micha{\l}owski} {et~al.}(2012){Micha{\l}owski}, {Dunlop},
  {Cirasuolo}, {Hjorth}, {Hayward}, \& {Watson}}]{Michalowski2012}
{Micha{\l}owski}, M.~J., {Dunlop}, J.~S., {Cirasuolo}, M., {et~al.} 2012, \aap,
  541, A85

\bibitem[{{Micha{\l}owski} {et~al.}(2014){Micha{\l}owski}, {Hayward}, {Dunlop},
  {Bruce}, {Cirasuolo}, {Cullen}, \& {Hernquist}}]{Michalowski2014}
{Micha{\l}owski}, M.~J., {Hayward}, C.~C., {Dunlop}, J.~S., {et~al.} 2014,
  ArXiv e-prints 1405.2335

\bibitem[{{Mitchell} {et~al.}(2013){Mitchell}, {Lacey}, {Baugh}, \&
  {Cole}}]{Mitchell2013}
{Mitchell}, P.~D., {Lacey}, C.~G., {Baugh}, C.~M., \& {Cole}, S. 2013, \mnras,
  435, 87

\bibitem[{{Murali} {et~al.}(2002){Murali}, {Katz}, {Hernquist}, {Weinberg}, \&
  {Dav{\'e}}}]{Murali2002}
{Murali}, C., {Katz}, N., {Hernquist}, L., {Weinberg}, D.~H., \& {Dav{\'e}}, R.
  2002, \apj, 571, 1

\bibitem[{{Nelson} {et~al.}(2013){Nelson}, {Vogelsberger}, {Genel}, {Sijacki},
  {Kere{\v s}}, {Springel}, \& {Hernquist}}]{Nelson2013}
{Nelson}, D., {Vogelsberger}, M., {Genel}, S., {et~al.} 2013, \mnras, 429, 3353

\bibitem[{{Ocvirk} {et~al.}(2008){Ocvirk}, {Pichon}, \&
  {Teyssier}}]{Ocvirk2008}
{Ocvirk}, P., {Pichon}, C., \& {Teyssier}, R. 2008, \mnras, 390, 1326

\bibitem[{{Ono} {et~al.}(2010){Ono}, {Ouchi}, {Shimasaku}, {Dunlop}, {Farrah},
  {McLure}, \& {Okamura}}]{Ono2010}
{Ono}, Y., {Ouchi}, M., {Shimasaku}, K., {et~al.} 2010, \apj, 724, 1524

\bibitem[{{Oppenheimer} {et~al.}(2010){Oppenheimer}, {Dav{\'e}}, {Kere{\v s}},
  {Fardal}, {Katz}, {Kollmeier}, \& {Weinberg}}]{Oppenheimer10}
{Oppenheimer}, B.~D., {Dav{\'e}}, R., {Kere{\v s}}, D., {et~al.} 2010, \mnras,
  406, 2325

\bibitem[{{Overzier} {et~al.}(2013){Overzier}, {Lemson}, {Angulo}, {Bertin},
  {Blaizot}, {Henriques}, {Marleau}, \& {White}}]{Overzier2013}
{Overzier}, R., {Lemson}, G., {Angulo}, R.~E., {et~al.} 2013, \mnras, 428, 778

\bibitem[{{Papovich} {et~al.}(2001){Papovich}, {Dickinson}, \&
  {Ferguson}}]{Papovich2001}
{Papovich}, C., {Dickinson}, M., \& {Ferguson}, H.~C. 2001, \apj, 559, 620

\bibitem[{{Pedrosa} {et~al.}(2014){Pedrosa}, {Tissera}, \& {De
  Rossi}}]{Pedrosa2014}
{Pedrosa}, S.~E., {Tissera}, P.~B., \& {De Rossi}, M.~E. 2014, \aap, 567, A47

\bibitem[{{Price} {et~al.}(2014){Price}, {Kriek}, {Brammer}, {Conroy},
  {F{\"o}rster Schreiber}, {Franx}, {Fumagalli}, {Lundgren}, {Momcheva},
  {Nelson}, {Skelton}, {van Dokkum}, {Whitaker}, \& {Wuyts}}]{Price2014}
{Price}, S.~H., {Kriek}, M., {Brammer}, G.~B., {et~al.} 2014, \apj, 788, 86

\bibitem[{{Quinn}(1984)}]{Quinn1984}
{Quinn}, P.~J. 1984, \apj, 279, 596

\bibitem[{{Robertson} {et~al.}(2007){Robertson}, {Li}, {Cox}, {Hernquist}, \&
  {Hopkins}}]{Robertson2007}
{Robertson}, B., {Li}, Y., {Cox}, T.~J., {Hernquist}, L., \& {Hopkins}, P.~F.
  2007, \apj, 667, 60

\bibitem[{{Sales} {et~al.}(2012){Sales}, {Navarro}, {Theuns}, {Schaye},
  {White}, {Frenk}, {Crain}, \& {Dalla Vecchia}}]{Sales2012}
{Sales}, L.~V., {Navarro}, J.~F., {Theuns}, T., {et~al.} 2012, \mnras, 423,
  1544

\bibitem[{{Scannapieco} {et~al.}(2010){Scannapieco}, {Gadotti}, {Jonsson}, \&
  {White}}]{Scannapieco2010}
{Scannapieco}, C., {Gadotti}, D.~A., {Jonsson}, P., \& {White}, S.~D.~M. 2010,
  \mnras, 407, L41

\bibitem[{{Schaerer} \& {de Barros}(2009)}]{Schaerer2009}
{Schaerer}, D., \& {de Barros}, S. 2009, \aap, 502, 423

\bibitem[{{Schaye} {et~al.}(2010){Schaye}, {Dalla Vecchia}, {Booth}, {Wiersma},
  {Theuns}, {Haas}, {Bertone}, {Duffy}, {McCarthy}, \& {van de
  Voort}}]{SchayeCSFR}
{Schaye}, J., {Dalla Vecchia}, C., {Booth}, C.~M., {et~al.} 2010, \mnras, 402,
  1536

\bibitem[{{Shen} {et~al.}(2013){Shen}, {Madau}, {Guedes}, {Mayer}, {Prochaska},
  \& {Wadsley}}]{Shen13}
{Shen}, S., {Madau}, P., {Guedes}, J., {et~al.} 2013, \apj, 765, 89

\bibitem[{{Sijacki} \& {Springel}(2006)}]{Sijacki2006}
{Sijacki}, D., \& {Springel}, V. 2006, \mnras, 366, 397

\bibitem[{{Sijacki} {et~al.}(2007){Sijacki}, {Springel}, {Di Matteo}, \&
  {Hernquist}}]{Sijacki2007}
{Sijacki}, D., {Springel}, V., {Di Matteo}, T., \& {Hernquist}, L. 2007,
  \mnras, 380, 877

\bibitem[{{Snyder} {et~al.}(2011){Snyder}, {Cox}, {Hayward}, {Hernquist}, \&
  {Jonsson}}]{Snyder2011}
{Snyder}, G.~F., {Cox}, T.~J., {Hayward}, C.~C., {Hernquist}, L., \& {Jonsson},
  P. 2011, \apj, 741, 77

\bibitem[{{Snyder} {et~al.}(2013){Snyder}, {Hayward}, {Sajina}, {Jonsson},
  {Cox}, {Hernquist}, {Hopkins}, \& {Yan}}]{Snyder2013}
{Snyder}, G.~F., {Hayward}, C.~C., {Sajina}, A., {et~al.} 2013, \apj, 768, 168

\bibitem[{{Snyder} {et~al.}(2014){Snyder}, {Lotz}, {Moody}, {Peth}, {Freeman},
  {Ceverino}, {Primack}, \& {Dekel}}]{Snyder2014}
{Snyder}, G.~F., {Lotz}, J., {Moody}, C., {et~al.} 2014, ArXiv e-prints

\bibitem[{{Somerville} {et~al.}(2012){Somerville}, {Gilmore}, {Primack}, \&
  {Dom{\'{\i}}nguez}}]{Somerville2012}
{Somerville}, R.~S., {Gilmore}, R.~C., {Primack}, J.~R., \& {Dom{\'{\i}}nguez},
  A. 2012, \mnras, 423, 1992

\bibitem[{{Springel}(2005)}]{GADGET}
{Springel}, V. 2005, \mnras, 364, 1105

\bibitem[{{Springel}(2010)}]{AREPO}
---. 2010, \mnras, 401, 791

\bibitem[{{Springel} {et~al.}(2005{\natexlab{a}}){Springel}, {Di Matteo}, \&
  {Hernquist}}]{Springel2005}
{Springel}, V., {Di Matteo}, T., \& {Hernquist}, L. 2005{\natexlab{a}}, \mnras,
  361, 776

\bibitem[{{Springel} \& {Hernquist}(2003{\natexlab{a}})}]{SH03}
{Springel}, V., \& {Hernquist}, L. 2003{\natexlab{a}}, \mnras, 339, 289

\bibitem[{{Springel} \& {Hernquist}(2003{\natexlab{b}})}]{SH03b}
---. 2003{\natexlab{b}}, \mnras, 339, 312

\bibitem[{{Springel} {et~al.}(2001){Springel}, {White}, \&
  {Hernquist}}]{SUBFIND}
{Springel}, V., {White}, M., \& {Hernquist}, L. 2001, \apj, 549, 681

\bibitem[{{Springel} {et~al.}(2005{\natexlab{b}}){Springel}, {White},
  {Jenkins}, {Frenk}, {Yoshida}, {Gao}, {Navarro}, {Thacker}, {Croton},
  {Helly}, {Peacock}, {Cole}, {Thomas}, {Couchman}, {Evrard}, {Colberg}, \&
  {Pearce}}]{Millennium}
{Springel}, V., {White}, S.~D.~M., {Jenkins}, A., {et~al.} 2005{\natexlab{b}},
  \nat, 435, 629

\bibitem[{{Stark} {et~al.}(2013){Stark}, {Schenker}, {Ellis}, {Robertson},
  {McLure}, \& {Dunlop}}]{Stark2013}
{Stark}, D.~P., {Schenker}, M.~A., {Ellis}, R., {et~al.} 2013, \apj, 763, 129

\bibitem[{{Stinson} {et~al.}(2013){Stinson}, {Brook}, {Macci{\`o}}, {Wadsley},
  {Quinn}, \& {Couchman}}]{Stinson2013}
{Stinson}, G.~S., {Brook}, C., {Macci{\`o}}, A.~V., {et~al.} 2013, \mnras, 428,
  129

\bibitem[{{Sutherland} \& {Dopita}(1993)}]{SutherlandDopita}
{Sutherland}, R.~S., \& {Dopita}, M.~A. 1993, \apjs, 88, 253

\bibitem[{{Teyssier} {et~al.}(2009){Teyssier}, {Pires}, {Prunet}, {Aubert},
  {Pichon}, {Amara}, {Benabed}, {Colombi}, {Refregier}, \&
  {Starck}}]{Teyssier2009}
{Teyssier}, R., {Pires}, S., {Prunet}, S., {et~al.} 2009, \aap, 497, 335

\bibitem[{{Theuns} {et~al.}(1998){Theuns}, {Leonard}, {Efstathiou}, {Pearce},
  \& {Thomas}}]{Theuns98}
{Theuns}, T., {Leonard}, A., {Efstathiou}, G., {Pearce}, F.~R., \& {Thomas},
  P.~A. 1998, \mnras, 301, 478

\bibitem[{{Thomas} {et~al.}(2003){Thomas}, {Maraston}, \&
  {Bender}}]{Thomas2003}
{Thomas}, D., {Maraston}, C., \& {Bender}, R. 2003, \mnras, 339, 897

\bibitem[{{Tinsley}(1972)}]{Tinsley1972}
{Tinsley}, B.~M. 1972, \apj, 178, 319

\bibitem[{{Torrey} {et~al.}(2014){Torrey}, {Vogelsberger}, {Genel}, {Sijacki},
  {Springel}, \& {Hernquist}}]{Torrey2014}
{Torrey}, P., {Vogelsberger}, M., {Genel}, S., {et~al.} 2014, \mnras, 438, 1985

\bibitem[{{Torrey} {et~al.}(2012){Torrey}, {Vogelsberger}, {Sijacki},
  {Springel}, \& {Hernquist}}]{Torrey2012}
{Torrey}, P., {Vogelsberger}, M., {Sijacki}, D., {Springel}, V., \&
  {Hernquist}, L. 2012, \mnras, 427, 2224

\bibitem[{{van de Voort} \& {Schaye}(2012)}]{vandeVoort2012a}
{van de Voort}, F., \& {Schaye}, J. 2012, \mnras, 423, 2991

\bibitem[{{van de Voort} {et~al.}(2011){van de Voort}, {Schaye}, {Booth},
  {Haas}, \& {Dalla Vecchia}}]{vandeVoort2011a}
{van de Voort}, F., {Schaye}, J., {Booth}, C.~M., {Haas}, M.~R., \& {Dalla
  Vecchia}, C. 2011, \mnras, 414, 2458

\bibitem[{{V{\'a}zquez} \& {Leitherer}(2005)}]{SB99_2}
{V{\'a}zquez}, G.~A., \& {Leitherer}, C. 2005, \apj, 621, 695

\bibitem[{{Vogelsberger} {et~al.}(2013){Vogelsberger}, {Genel}, {Sijacki},
  {Torrey}, {Springel}, \& {Hernquist}}]{Vogelsberger2013}
{Vogelsberger}, M., {Genel}, S., {Sijacki}, D., {et~al.} 2013, \mnras, 436,
  3031

\bibitem[{{Vogelsberger} {et~al.}(2012){Vogelsberger}, {Sijacki}, {Kere{\v s}},
  {Springel}, \& {Hernquist}}]{Vogelsberger2012}
{Vogelsberger}, M., {Sijacki}, D., {Kere{\v s}}, D., {Springel}, V., \&
  {Hernquist}, L. 2012, \mnras, 425, 3024

\bibitem[{{Vogelsberger} {et~al.}(2014{\natexlab{a}}){Vogelsberger}, {Genel},
  {Springel}, {Torrey}, {Sijacki}, {Xu}, {Snyder}, {Nelson}, \&
  {Hernquist}}]{Vogelsberger2014b}
{Vogelsberger}, M., {Genel}, S., {Springel}, V., {et~al.} 2014{\natexlab{a}},
  ArXiv e-prints 1405.2921

\bibitem[{{Vogelsberger} {et~al.}(2014{\natexlab{b}}){Vogelsberger}, {Genel},
  {Springel}, {Torrey}, {Sijacki}, {Xu}, {Snyder}, {Bird}, {Nelson}, \&
  {Hernquist}}]{Vogelsberger2014a}
---. 2014{\natexlab{b}}, \nat, 509, 177

\bibitem[{{Weinberg} {et~al.}(1997){Weinberg}, {Hernquist}, \&
  {Katz}}]{Weinberg1997}
{Weinberg}, D.~H., {Hernquist}, L., \& {Katz}, N. 1997, \apj, 477, 8

\bibitem[{{Wiersma} {et~al.}(2009{\natexlab{a}}){Wiersma}, {Schaye}, \&
  {Smith}}]{WiersmaCooling}
{Wiersma}, R.~P.~C., {Schaye}, J., \& {Smith}, B.~D. 2009{\natexlab{a}},
  \mnras, 393, 99

\bibitem[{{Wiersma} {et~al.}(2009{\natexlab{b}}){Wiersma}, {Schaye}, {Theuns},
  {Dalla Vecchia}, \& {Tornatore}}]{WiersmaGasReturn}
{Wiersma}, R.~P.~C., {Schaye}, J., {Theuns}, T., {Dalla Vecchia}, C., \&
  {Tornatore}, L. 2009{\natexlab{b}}, \mnras, 399, 574

\bibitem[{{Wild} {et~al.}(2009){Wild}, {Walcher}, {Johansson}, {Tresse},
  {Charlot}, {Pollo}, {Le F{\`e}vre}, \& {de Ravel}}]{Wild2009}
{Wild}, V., {Walcher}, C.~J., {Johansson}, P.~H., {et~al.} 2009, \mnras, 395,
  144

\bibitem[{{Worthey}(1994)}]{Worthey1994}
{Worthey}, G. 1994, \apjs, 95, 107

\bibitem[{Wuyts {et~al.}(2010)Wuyts, Cox, Hayward, Franx, Hernquist, Hopkins,
  Jonsson, \& van Dokkum}]{Wuyts2010}
Wuyts, S., Cox, T.~J., Hayward, C.~C., {et~al.} 2010, \apj, 722, 1666

\bibitem[{Wuyts {et~al.}(2009{\natexlab{a}})Wuyts, Franx, Cox, Hernquist,
  Hopkins, Robertson, \& van Dokkum}]{Wuyts2009a}
Wuyts, S., Franx, M., Cox, T.~J., {et~al.} 2009{\natexlab{a}}, \apj, 696, 348

\bibitem[{{Wuyts} {et~al.}(2007){Wuyts}, {Labb{\'e}}, {Franx}, {Rudnick}, {van
  Dokkum}, {Fazio}, {F{\"o}rster Schreiber}, {Huang}, {Moorwood}, {Rix},
  {R{\"o}ttgering}, \& {van der Werf}}]{Wuyts2007}
{Wuyts}, S., {Labb{\'e}}, I., {Franx}, M., {et~al.} 2007, \apj, 655, 51

\bibitem[{Wuyts {et~al.}(2009{\natexlab{b}})Wuyts, Franx, Cox, {F{\"o}rster
  Schreiber}, Hayward, Hernquist, Hopkins, Labb{\'e}, Marchesini, Robertson,
  Toft, \& van Dokkum}]{Wuyts2009b}
Wuyts, S., Franx, M., Cox, T.~J., {et~al.} 2009{\natexlab{b}}, \apj, 700, 799

\bibitem[{{Wuyts} {et~al.}(2012){Wuyts}, {F{\"o}rster Schreiber}, {Genzel},
  {Guo}, {Barro}, {Bell}, {Dekel}, {Faber}, {Ferguson}, {Giavalisco}, {Grogin},
  {Hathi}, {Huang}, {Kocevski}, {Koekemoer}, {Koo}, {Lotz}, {Lutz}, {McGrath},
  {Newman}, {Rosario}, {Saintonge}, {Tacconi}, {Weiner}, \& {van der
  Wel}}]{Wuyts2012}
{Wuyts}, S., {F{\"o}rster Schreiber}, N.~M., {Genzel}, R., {et~al.} 2012, \apj,
  753, 114

\bibitem[{{Zackrisson} {et~al.}(2008){Zackrisson}, {Bergvall}, \&
  {Leitet}}]{Zackrisson2008}
{Zackrisson}, E., {Bergvall}, N., \& {Leitet}, E. 2008, \apjl, 676, L9

\bibitem[{{Zhang} {et~al.}(1995){Zhang}, {Anninos}, \& {Norman}}]{Zhang95}
{Zhang}, Y., {Anninos}, P., \& {Norman}, M.~L. 1995, \apjl, 453, L57

\end{thebibliography}

\end{document}